\documentclass{iopart}
\usepackage{graphicx,epsfig}
\usepackage{bm}
\usepackage{iopams}
\newcommand{\be}{\begin{equation}}
\newcommand{\ee}{\end{equation}}
\newcommand{\ba}{\begin{eqnarray}}
\newcommand{\ea}{\end{eqnarray}}
\newcommand{\bse}{\numparts}
\newcommand{\ese}{\endnumparts}

\newcommand{\DD}{{\cal {D}}}

\newcommand{\bbq}{\begin{quote}}
\newcommand{\eeq}{\end{quote}}
\newcommand{\RR}{{}^3{\cal{R}}}

\newcommand{\T}{{}^3{\cal{T}}}

\newcommand{\HH}{{\cal{H}}}
\newcommand{\KK}{{\cal{K}}}
\newcommand{\PP}{{\cal{P}}}

\newcommand{\Ome}{\hat\Omega_{e}}

\newcommand{\Omm}{\hat\Omega_{m}}
\newcommand{\Ommi}{\hat\Omega_i}
\newcommand{\Da}{\delta^{(A)}}

\newcommand{\Dj}{\delta^{(J)}}

\newcommand{\Dh}{\delta^{(\HH)}}
\newcommand{\dDh}{\dot\delta^{(\HH)}}

\newcommand{\Dm}{\delta^{(m)}}
\newcommand{\De}{\delta^{(e)}}
\newcommand{\dDm}{\dot\delta^{(\rho)}}
\newcommand{\Dp}{\delta^{(p)}}
\newcommand{\Dk}{\delta^{(\kappa)}}

\newcommand{\Drho}{\delta^{(\rho)}}
\newcommand{\dDrhom}{\dot\delta^{(m)}}
\newcommand{\dDrhoe}{\dot\delta^{(e)}}

\newcommand{\dd}{{\rm{d}}}

\begin{document}
\title[Dynamics of a spherically symmetric inhomogeneous coupled dark energy model]{Dynamics of a spherically symmetric inhomogeneous coupled dark energy model with coupling term proportional to non relatvistic matter.}
\author{Germ\'an Izquierdo${}^\dagger$, Roberto C. Blanquet-Jaramillo${}^\dagger$ and Roberto A. Sussman${}^\ddagger$}
\address{${}^\ddagger$ Facultad de Ciencias, Universidad Aut\'onoma del Estado de M\'exico, Toluca 5000, Instituto literario 100, Edo. Mex.,M\'exico.\\
${}^\ddagger$ Instituto de Ciencias Nucleares, Universidad Nacional Aut\'onoma de M\'exico (ICN-UNAM),A. P. 70--543, 04510 M\'exico D. F., M\'exico.
 }
\ead{gizquierdos@uaemex.mx}
\date{\today}
\begin{abstract}
Quasi--local scalar variables approach is applied to a spherically symmetric inhomogeneous Lema\^\i tre--Tolman--Bondi metric containing a mixture of non-relativistic cold dark matter and coupled dark energy with constant equation of state. The quasi--local coupling term considered is proportional to the quasi--local cold dark matter energy density and a quasi--local Hubble factor-like scalar via a coupling constant $\alpha$. The autonomous numerical system obtained from the evolution equations is classified for different choices of the free parameters: the adiabatic constant of the dark energy $w$ and $\alpha$. The presence of a past attractor in a non-physical region of the energy densities phase-space of the system makes the coupling term non physical when the energy flows from the matter to the dark energy in order to avoid negative values of the dark energy density in the past. On the other hand, if the energy flux goes from dark energy to dark matter, the past attractor lays in a physical region. The system is also numerically solved for some interesting initial profiles leading to different configurations: an ever expanding mixture, a scenario where the dark energy is completely consumed by the non-relativistic matter by means of the coupling term, a scenario where the dark energy disappears in the inner layers while the outer layers expand as a mixture of both sources, and, finally, a structure formation toy model scenario, where the inner shells containing the mixture collapse while the outer shells expand.
\end{abstract}

\section{Introduction}

The cosmological observational data is explained in a very satisfactory way by assuming that the Universe is a flat homogeneous Friedman-Lema\^\i tre-Robertson-Walker (FLRW) space-time currently undergoing a accelerated expansion ruled out by a mysterious form of energy with negative pressure \cite{copeland, wmap, plank}. This source does not interact with the ordinary non-relativistic matter (except by gravitational interaction) and is known in the literature as dark energy (DE). The data also support the existence of a cold dark matter (CDM) source: a non-relativistic matter source of massive particles which, in turn, is only coupled to ordinary matter through gravity. Both sources are known as the dark sector of the Universe and much effort is made to obtain new information that clarifies the nature of them. Although the simple and most successful model to explain the observational data is the $\Lambda$-CDM model where the DE is a cosmological constant term, a great number of DE models arise. Coupled Dark Energy (CDE) models assume that the DE is coupled to the CDM through a coupling term and are sustained by data obtained from observations of the dynamics of galaxy clusters \cite{abdalla} and the integrated Sachs-Wolfe effect \cite{ol3}. There is a large number of coupling terms used in the literature, motivated by particle physics, phenomenological approach, etc. \cite{copeland,boe08}.

Some sets of the observational data mentioned early can give us direct information on the dynamics of the homogeneous Universe (and, consequently, on the DE and CDM sources), such as the luminosity distance of Supernovae Type Ia \cite{copeland}, the history of Hubble parameter \cite{Riess}, or the expected redshift derivatives data that would come in a future \cite{martins}. Other sets of data give us information on the inhomogeneous part of the Universe, such as the data of anisotropy of the cosmic microwave background (CMB) and the data from Barionic Acoustic Oscillations \cite{copeland},  the redshift drift \cite{gil-marin}, evolution of the growth function \cite{linder}, etc. All the data regarding the inhomogeneous cosmology can give us information on the Universe through the dynamics of the perturbations in Cosmology. Perturbation theory is very well understood to linear order but higher order corrections are still very difficult to work with. The linear perturbation evolution has been treated in the CDE models for different coupling terms: a coupling proportional to both the Hubble factor and the dark matter energy density \cite{ol, ol2, Maar}, a coupling proportional to the dark energy \cite{Maar}, a coupling proportional to both the Hubble factor and the coupled dark energy density {Maar, Gavela}, etc. In all the cases the linear perturbation equations are solved numerically, the power spectrum of the energy density fluctuations is computed by means of numerical codes such as CAMB \cite{camb} and, finally, the results are compared with the data obtaining interesting bounds on the free parameters of the theory.

Inhomogeneous exact solutions of the Einstein equation can give us very useful information about the perturbation dynamics that can complement (or be used as an alternative approach to) those of the linear order inhomogeneous Cosmology. In particular, the spherically symmetric Lema\^\i tre--Tolman--Bondi (LTB) metric Quasi-Local (QL) scalar approach \cite{suss13} can be used to study a local spherical exact solution that matches the homogeneous FLRW at larger scales and the linear perturbations of the FLRW metric can be related to the LTB fluctuations with respect to the QL scalars defined \cite{suss15}. LTB metrics are well known in the context of pure dust solutions (see \cite{kras1} for a review), and also in the context of dust plus a cosmological constant \cite{izsuss10, suss15}, but not much work is done in the context of more exotic sources such as DE or CDE \cite{suss08}. In this sense, we believe that an understanding of the dynamics of the inhomogeneous metrics containing those sources is necessary.

In this work we will consider a LTB metric containing a mixture of CDM and CDE with a interaction term proportional to CDM energy density. We will make use of the QL scalar variables approach \cite{suss15,izsuss10, suss08, sussPRD} to obtain a set of autonomous evolution equations and its critical points in terms of the free parameters of the model. We will also use some initial conditions to solve the evolution equations and get a better understanding of some interesting scenarios.

The plan of the article is the following. In section \ref{genLTB}, we apply the QL scalar variables approach to the evolution equations in order to obtain an autonomous dynamical system. In section \ref{criticalpointsLTB}, we study the dynamical system and the behavior of the critical points for different ranges of the free parameters. In section \ref{numerical}, we numerically solve the evolution equations for some interesting sets of initial conditions. Finally, in section \ref{conclusions}, we summarize the findings. From now on, we assume units for which $c=1$.

\section{LTB spacetimes and quasi--local (QL) scalar variables}\label{genLTB}
We shall follow the methodology described in \cite{suss13,izsuss10,suss08}, which we will summarize briefly in order to clarify the notation used in this work. We consider spherically symmetric space-times whose source is an anisotropic fluid
\begin{equation}T^{ab} = \rho\,u^a u^b + p\,h^{ab}+\Pi^{ab},\label{Tab}\end{equation}
where $u^a=\delta^a_0$, $\rho=\rho(ct,r)$ and $p=p(ct,r)$ are the energy density and isotropic pressure in the comoving frame, respectively, $\Pi^{ab}$ is the anisotropic pressure and $h_{ab}=u_au_b+g_{ab}=\delta_a^i\delta_b^jg_{ij}$, where $i,j = r,\theta,\phi$. We shall use the term ``LTB spacetimes'' to denote all solutions of Einstein's equations for the source that are described by the spherically symmetric Lema\^\i tre--Tolman--Bondi metric in a comoving frame \cite{kras1}
\begin{equation} ds^2 = -dt^2 +\frac{R'^2\,dr^2}{1+E}+R^2[d\theta^2+\sin^2\theta\,d\phi^2],\label{LTB}\end{equation}
where $R=R(t,r)$, \, $R'=\partial R/\partial r$,\, $E=E(r)$. The comoving geodesic 4--velocity field defines a foliation of spacelike hypersurfaces, $\T(t)$, orthogonal to $u^a$, marked by $t$ constant, and having an induced metric $h_{ab}$.

Besides $u^a,\,\rho$, $p$, the remaining covariant objects associated with LTB spacetimes are two scalars: the expansion, $\Theta=\nabla_a u^a=3\HH$, and $\RR$, the Ricci scalar of hypersurfaces $\T(t)=6 \KK$. As in all spherically symmetric spacetimes, the spacelike traceless tensor $\Pi^{ab} = \left[h^{(a}_ch^{b)}_d-\frac{1}{3}h^{ab}h_{cd}\right]\, T^{cd}$ can be completely and covariantly determined in terms of a single scalar function $\PP$ as \cite{sussPRD,sussLTB1,sussLTB2}
\bse\be
\Pi^{ab} = \PP\,\left(h^{ab}-3\chi^{a}\chi^b\right),\label{PPsc} \ee\ese
where $\chi^a=\sqrt{h^{rr}}\,\delta^a_r$ is the unit vector orthogonal to $u^a$ and to the 2-spheres, orbits of SO(3).

Following \cite{izsuss10,suss08}, we will use the alternative representation of covariant quasi--local (QL) scalars \cite{sussPRD,sussLTB1,hayward}. For any scalar function $A$\footnote {$A$ being a smooth integrable scalar functions in a comoving regular domain $\DD=\mathbb{S}^2\times\vartheta\subset \T(t)$, where $\mathbb{S}^2$ is the unit 2--sphere parametrized by $(\theta,\phi)$ and $\vartheta =\{x\,|\, 0\leq x\leq r\}$, where $x=0$ marks a symmetry center, see \cite{sussPRD,sussLTB1,sussLTB2}.}, a ``dual'' QL scalar function $A_q$ follows from
\begin{equation}A_q =  \frac{\int_{x=0}^{x=r}{A\,R^2R' dx}}{\int_{x=0}^{x=r}{R^2R' dx}}.\label{QLfunc}\end{equation}
The QL scalar functions $A_q$ depend on the upper integration limit $r$, and generalize to any scalar the QL mass--energy definition of the Misner--Sharp QL mass--energy function.

From their definition and (\ref{QLfunc}), the QL scalars dual to $\HH,\,\KK$ are
\bse\label{HKql}\ba \HH_q &=& \frac{\dot R}{R},\label{theta2}\\
 \KK_q &=&-\frac{E}{R^2}\quad\Rightarrow\quad \frac{\dot\KK_q}{\KK_q}=-2\HH_q.\label{RR2}
 \ea\ese
The equations of above together with the field equations $G^{ab}=\kappa T^{ab}$ ($\kappa =8\pi G$) for (\ref{LTB}) lead to a FLRW Raychaudhuri equation, and its integral, the  Friedman equation \cite{izsuss10,suss08}
\ba \dot\HH_q &=& -\HH_q^2-\frac{\kappa}{6}\,(\rho_q+3p_q),\label{Raych2}\\
\HH_q^2  &=& \frac{\kappa}{3}\,\rho_q-\KK_q,\label{cHam2}\ea
where $\HH_q^2=(\HH_q)^2 \ne (\HH^2)_q$. These equations can be combined to yield the FLRW energy balance equation
\begin{equation}  \dot\rho_q = -3\,(\rho_q+p_q)\,\HH_q.\label{Econs2}\end{equation}
Applying (\ref{QLfunc}) to $p$ and from its definition, $\PP = \frac{1}{2}\,[p-p_q]$.

\subsection{Evolution equations for the QL scalars}\label{eveqs}

The local and QL scalars can be related by means of ``relative deviations''
\begin{equation} \Da \equiv \frac{A-A_q}{A_q},\quad \Rightarrow\quad A = A_q\,\left[1+\Da\right].\label{Da_def}\end{equation}
which allows us to eliminate $\rho,\,p,\,\HH$ in terms of their duals $A_q$ and the corresponding $\Da$. Hence, we have a complete scalar representation of LTB spacetimes given by
\begin{equation} \{\HH_q,\,\rho_q,\,p_q,\,\KK_q,\,\Dh,\,\Drho,\,\Dp,\,\Dk\}.\label{ql_scals}\end{equation}
which is alternative to the local representation. We will denote (\ref{ql_scals}) the ``QL scalar representation''. We can derive now the evolution and constraint equations for the representation (\ref{ql_scals}) from the local field equations $G^{ab}=\kappa T^{ab}$ and the corresponding constraints.

From (\ref{QLfunc}), its is posible to deduce the property
\be A_q{}' = \frac{3R'}{R}\,[A-A_q],\label{prop2} \ee
thus, the radial gradients of $\rho_q,\,p_q$ and $\HH_q$ can be given from the $\delta$ functions
\begin{equation} \frac{\HH_q{}'}{\HH_q} = \frac{3R'}{R}\,\Dh,\qquad \frac{\rho_q{}'}{\rho_q} = \frac{3R'}{R}\Drho,\qquad \frac{p_q{}'}{p_q} = \frac{3R'}{R}\Dp, \label{rad_grads}\end{equation}
while (\ref{Raych2}) and (\ref{Econs2}) are evolution equations for $\rho_q$ and $\HH_q$. Hence, the evolution equations for $\Drho$ and $\Dh$ follow from the consistency condition $\left[A_q{}'\right]\,\dot{}=\left[\dot A_q\right]'$, applied to (\ref{Raych2}), (\ref{Econs2}) and (\ref{rad_grads}) for $A=\HH_q,\,\rho_q$. The result is the following set of autonomous evolution equations for the QL representation (\ref{ql_scals}):
\bse\label{eveqs_ql}\ba
\dot\rho_q &=& -3\,\left[\,1+w\,\right]\,\rho_q\,\HH_q,\label{evmu_ql}\\
\dot\HH_q &=& -\HH_q^2 -\frac{\kappa}{6}\,\left[\,1+3\,w\,\right]\,\rho_q,
\label{evHH_ql}\\
\dDm &=& 3\HH_q\,\left[\left(\Drho-\Dp\right)\,w-\left(1+w+\Drho\right)\Dh\right],
\label{evDmu_ql}\\
\dDh &=& -\HH_q\,\left(1+\Dh\right)\,\Dh\nonumber\\
 &{}&\,\,+ \frac{\kappa\rho_q}{6\,\HH_q}\left[\Dh-\Drho+3w\,\left(\Dh-\Dp\right)\right],
\label{evDth_ql}
\ea\ese
where $w\equiv p_q/\rho_q$ is the adiabatic coefficient defined by the QL energy density and pressure EOS\footnote{In this work we restrict ourselves to constant adiabatic coefficients, which assures that both QL and local fields fulfill the same EOS. A detailed discussion on why the EOS is defined for the QL energy densities and pressures and not necessarily for the local ones in a perturbation scheme such as this is done in \cite{suss08}.}.
The spacelike constraints associated with these evolution equations are simply the spatial  gradients (\ref{rad_grads}), while the Friedman equation (or Hamiltonian constraint) is (\ref{cHam2}). Equations (\ref{evmu_ql})--(\ref{evDth_ql}) become fully determined once an ``equation of state'' that fixes $p_q,\,\Dp$ as functions of $\rho_q,\,\Drho$ is selected.

It is straightforward to  prove that the evolution equations (\ref{evmu_ql})--(\ref{evDth_ql}) and the constraints (\ref{cHam2}) and (\ref{rad_grads}) are wholly equivalent to the local evolution equations and their constraints. Hence, given an EOS, they completely determine the dynamics of LTB spacetimes.

\subsection{The coupled dark energy model}

Coupled dark energy models (CDE) have been introduced in the literature as an attempt to avoid the coincidence problem present in the $\Lambda$CDM model and the fine-tuning problems of the quintessence and phantom models with constant or parametric adiabatic coefficient $w$ in the context of the FLRW metric. In the CDE models the FLRW metric has three sources: the barionic matter, the CDM, both of them pressureless, and a dark energy fluid with EOS. The two dark sources (CDM and CDE) are coupled by means of an interaction term, and a flux of energy flows from one source to another \cite{copeland}.

We will consider now a LTB metric with a source which is a mixture of both CDM and CDE fluids. The energy--momentum tensor for this source reads $T^{ab}=T_{m}^{ab}+T_{e}^{ab}$ where the subindex $m$ and $e$ refers to CDM and CDE sources, respectively. Although the system energy-momentum should be conserved $\nabla_bT^{ab}=0$, it is possible to consider a non null a energy-momentum flux between both components.Then, the conservation laws for  the individual tensors read
\be
\nabla_b T_{m}^{ab}= j^a=-\nabla_b T_{e}^{ab},
\ee
where $j^a$ is the interaction current (or coupling) that characterizes an interactive mixture,  so that if $j^a=0$ the mixture is non--interactive (decoupled).  We take this current as a vector parallel to the 4--velocity, so that $j_a=Ju_a$ and $h_{ca}j^a=0$ hold. The spatially projected conservation equation $h_{ac}\nabla_bT^{ab}=0$ remains as it is, but the projection along $u^a$ becomes
\begin{equation}u_a \nabla_b T_{m}^{ab}= J =- u_a \nabla_b T_{e}^{ab}.
\label{uTab_cons_J}\end{equation}

The energy density and the pressure of the LTB metric can be decomposed as

\bse\label{Tabdec}\ba \rho(ct,r) &=& \rho_{m}(ct,r) + \rho_{e}(ct,r),\\ p(ct,r) &=& p_{m}(ct,r)+p_{e}(ct,r).\ea\ese
The scalars $\rho_e$, $\rho_m$ can be used to define their QL counterparts as $\rho_{mq}$, $\rho_{eq}$ and their respective delta functions $\Dm$ and $\De$. Additionally we need the EOS for the QL densities and pressures of both fluids: we consider $p_{mq}=0$ for the CDM source (a dust source), while $p_{eq}=w\rho_{eq}$ with constant $w<-1/3$ is considered for the CDE. With the EOS of both sources defined,  it follows that $\Dp_m=0$ and $\Dp_e=\De$.

The evolution equation (\ref{evmu_ql}) is split in two coupled evolution equations and the delta functions evolution equations can be obtained from them by derivation with respect to $r$, and the system reads
\bse\label{eveqs_q2}\ba
\dot\HH_q &=& -\HH_q^2 -\frac{\kappa}{6}\,\left[\rho_{mq}+(1+3\,w\,)\rho_{eq}\right]\,,\label{evHH_q2}\\
\dot{\rho}_{mq} &=& -3{\HH}_q\,\rho_{mq}+J_q,\label{evm_q2}\\
\dot{\rho}_{eq} &=& -3{\HH}_q\left( 1+w\right)\rho_{eq}-J_q,\label{eve_q2}\\
\dDrhom &=& -3\HH_q\,\left(1+\Dm\right)\Dh-\frac{J_q}{\rho_{mq}}\left(\Dm-\Dj\right),\label{evDm_q2}\\
\dDrhoe &=& 3\HH_q\,[w\,\De-\left(1+w+\De\right)\Dh] -\frac{J_q}{\rho_{eq}}\left(\De-\Dj\right),\label{evDe_q2}\\
\dDh &=& -\HH_q\Dh\left(1+3\Dh\right)+\nonumber\\
&&\frac{\kappa}{6\HH_q}
\left[\rho_{mq}\,\left(\Dh-\Dm\right)+(1+3w)\rho_{eq}\,\left(\Dh-\De\right)\right],\label{evDh_q2}
\ea\ese
where $J_q$ is the QL energy density flux defined from $J$ or defined by itself and $\Dj=(J-J_q)/J_q$. Additionally, the constraint (\ref{cHam2}) reads
\be
\HH_q^2=\frac{\kappa}{3}\rho_{mq}+\frac{\kappa}{3}\rho_{eq}-\KK_q.
\ee

The system (\ref{evHH_q2}--\ref{evDh_q2}) can be solved for a determined adiabatic coefficient $w$ once the local energy density flux $J=J(ct,r)$ (or its QL counterpart, $J_q$) is defined, as the scalar $\Dj$ can be obtained from
\be
\Dj=\frac{R}{3R'}\frac{J_q'}{J_q}=\frac{R}{3R'}\left(ln(J_q)\right)'.
\ee

There is an extensive literature in cosmology for the interaction term[Ref copeland,etc]. In this work we consider an interaction term, and, consequently, a $\Dj$ function, of the form
\be
J_{q} = 3\,\alpha\,\HH_{q}\,\rho_{mq},\qquad \Dj=\Dh+\Dm, \label{intj_m}
\ee
where $\alpha$ is an dimensionless constant. The QL energy density flux is considered to match the coupling term of different cosmological models of CDE in a FLRW scheme \cite{copeland, ol, Maar}. If $\alpha>0$ the energy flows from the CDE to the CDM. On the other hand, $\alpha<0$ means the energy flux goes from the CDM to the CDE. This coupling is deduced in the literature from phenomenological grounds (although a microscopic description of the quantum field theory could be obtained from it).

Solving the system (\ref{evHH_q2}--\ref{evDh_q2}) allows us to represent the local $J$ corresponding to the QL one, as $J=J_q\left(1+\Dh+\Dm\right)$ for every shell $r=r_i$ and as a funtion of time. In this way, the LTB model allows us to obtain interesting information about the local interaction between CDE and CDM that might occur in galaxies and clusters of galaxies and that is lost in the FLRW model. In FLRW models, the perturbations evolution is introduced as a first order perturbation correction of the background dynamics and the local $J$ can be obtained perturbatively as well by numerical methods. On the other hand, LTB model is an exact solution of Einstein equations with easy to compute equations provided the spherical symmetry that can a match FLRW background connecting the density fluctuations with respect to the QL counterpart as linear FLRW perturbations with a given set of conditions (see \cite{suss15} for a detailed description of how this identification can be done in a $\Lambda$--CDM LTB metric).

\subsection{Dimensionless dynamical system and critical points}

At this point it is convenient to define dimensionless functions that allow us to transform the system (\ref{evHH_q2}--\ref{evDh_q2}) in order to use the convenient dynamical system methods, to find the critical points of the system.

For the QL densities, we can make use of the partial energy density $\Omega$ functions of Cosmology. We can define then
\be
\Omm=\frac{\kappa}{3\HH_q^2}\rho_{mq}, \qquad \Ome=\frac{\kappa}{3\HH_q^2}\rho_{eq}.
\ee
It is straightforward to obtain evolution equations of $\Omm$ and $\Omm$ in terms of equations (\ref{evHH_q2}), (\ref{evm_q2}) and (\ref{eve_q2}) as
\be
\frac{1}{\HH_q}\dot{\hat\Omega_A}=\frac{\kappa}{3\HH_q^3}\dot{\rho}_{Aq}-\hat{\Omega}_A\frac{\dot{\HH_q}}{\HH_q},
\ee
where $A=m,e$. The constraint (\ref{cHam2}) reads, then,
\be
\Omm+\Ome+\hat{\Omega}_{\KK}=1,
\ee
where $\hat{\Omega}_{\KK}=-\KK_q/ \HH^2_q$.

We define a dimensionless coordinate $\xi(t,r)$ that, for all the comoving curves $r=r_i$%
\begin{equation}\frac{\partial}{\partial\xi}=\frac{1}{\HH_q}\frac{\partial}{\partial c t}
=\frac{3}{\Theta_q}\frac{\partial}{\partial t}.\label{xidef}\end{equation}
In terms of the new dimensionless $\xi$ and using (\ref{intj_m}), the system (\ref{evHH_q2}--\ref{evDh_q2}) is transformed into
\bse\label{eveqs_q3}
\ba
\frac{\partial{\Omm}}{\partial{\xi}} &=& \Omm\,\left[ -1+\Omega_m+\left( 1+3\,w\right)\,\Ome+3\,\alpha \right], \label{sistdinint1a}\\
\frac{\partial{\Ome}}{\partial{\xi}} &=& \Ome\,\left[ \left( 1+3\,w \right) \left( -1+\Ome \right)+\Omega_m \right]-3\,\alpha\,\Omm, \label{sistdinint1b}\\
\frac{\partial{\Dm}}{\partial{\xi}} &=& -3\,\Dh\,\left(1+\Dm-\alpha \right), \label{sistdinint1c}\\
\frac{\partial{\De}}{\partial{\xi}} &=&-3\Dh\left(1+w+\De+\frac{\alpha\Omm}{\Ome} \right)-\frac{3\alpha\Omm\left( \Dm-\De \right)}{\Ome}, \label{sistdinint1d}\\
\frac{\partial{\Dh}}{\partial{\xi}} &=& -\Dh\left( 1+3\Dh \right)+\frac{\Omm\,\left( \Dh-\Dm \right)}{2}\nonumber\\
&&\qquad+\frac{\left( 1+3\,w \right)\,\Ome\,\left(\Dh-\De \right)}{2}. \label{sistdinint1e}
\ea
\ese
The system (\ref{sistdinint1a}-\ref{sistdinint1e}) is 5-dimensional and can be computed for a set of initial conditions for every shell $r=r_i$ once we fix the parameters $w$ and $\alpha$. From the solution, it is possible to compute $\HH_q(\xi,r_i)$ provided that
\be
\frac{\partial{\HH_q}}{\partial{\xi}}=\frac{1}{\HH_q}\dot{\HH}_q=-\HH_q\left(1+\frac{1}{2}\Omm+\frac{1+3w}{2}\Ome\right).
\ee
The LTB metric can then be fully solved as the QL energy densities are evaluated as $\rho_{aq}=(3\HH_q^2 \Ommi)/\kappa$ with $a=m,e$ and
\ba
\KK_q&=&\HH^2_q\left(-1+\Omm+\Ome\right),\nonumber\\
\Dk&=&\frac{ \HH_q^2}{ \KK_q}\left(-2\Dh +\Omm\Dm+\Ome\De\right)\nonumber
\ea
Additionally, the local quantities can be obtained from the definition (\ref{Da_def}). For every shell $r=r_i$, it is possible to implicitly recover the instant $t$ corresponding to the variable $\xi(t ,r_i)$ as
\be
t= \int_0^{\xi(t,r_i)} {\frac{d \xi'}{\HH_q(\xi',r_i)}}.
\ee

The scalar functions $\Omm$ and $\Ome$ form an independent subsystem that is formally identical to that of the $\Omega$-functions of the FLRW model. As their evolution do not depend upon the $\delta$-functions, we will refer to this subsystem as the homogeneous projection. On the other hand, we will chose constant values for $\Omm$ and $\Ome$ in order to represent the $\delta$-functions evolution and we will refer to this three dimensional projection as the inhomogeneous projection. With both projections we have a complete representation of the system.

\begin{table}[]
\centering
\caption{The critical points and their respective eigenvalues of the system (\ref{sistdinint1a}-\ref{sistdinint1e}).}
\label{criticalpoints}
\begin{tabular}{|l|l|l|}
\hline
\begin{tabular}[c]{@{}l@{}}Critical\\  points\end{tabular} & $\left(\Omm,\Ome,\Dm,\De,\Dh\right)^T$ & Eigenvalues \\ \hline
PC1 & $\left(0,\,1,\, \Dm arbitrary, \, 0,\,0\right)^T$ & \begin{tabular}[c]{@{}l@{}}$ \lambda_1= \lambda_2 = 1+3\,w,\,\lambda_3 = -\frac{3}{2}(1+w),$\\ $\lambda_4 = 3\,(w+\,\alpha),\,\lambda_5 = 0.$\end{tabular} \\ \hline
PC2 & $\left(1+\frac{\alpha}{w},\,\frac{-\alpha}{w},\,0,\,0,\,0\right)^T$ & \begin{tabular}[c]{@{}l@{}}$ \lambda_1 = \lambda_2 = -3\,(\alpha+\,w),\,\lambda_3 = -\frac{3}{2}(1+\alpha),$\\ $\lambda_4 =\lambda_5 = 1-3\,\alpha.$\end{tabular} \\ \hline
PC3 & $\left(0,\,1,\, -1+\alpha,\, -(1+w),\,-\frac{1}{2}\left(1+ w\right)\right)^T$ & \begin{tabular}[c]{@{}l@{}}$\lambda_1 = \frac{9w+5}{2},\,\lambda_2 = 1+3\,w,$\\ $\lambda_3 = \lambda_4 = \frac{3}{2}(1+w),\,\lambda_5 = 3\,(w+\,\alpha) $\end{tabular} \\ \hline
PC4 & $\left(0,\,1,\, -1+\alpha,\,-(1+w),\,\frac{1}{3}+w\right)^T$ & \begin{tabular}[c]{@{}l@{}}$ \lambda_1 = \lambda_2 =-\lambda_3 = -1-3\,w,\,$\\ $\lambda_4 = -\frac{9\,w+5}{2},\,\lambda_5 = 3\,(w+\alpha)$\end{tabular} \\ \hline
PC5 & $\left(1+\frac{\alpha}{w},\,-\frac{\alpha}{w},\,-1+\alpha,-1+\alpha,\,\frac{1}{3}-\alpha\right)^T$ & \begin{tabular}[c]{@{}l@{}}$ \lambda_1 = -\lambda_4= 1-3\,\alpha,\,\lambda_2 = -3(\,\alpha+\,w),$\\ $\lambda_3 = -(1+3\,w),\,\lambda_5 = \frac{9\,\alpha-5}{2}$\end{tabular} \\ \hline
PC6 & \begin{tabular}[c]{@{}l@{}}$\left(1+\frac{\alpha}{w},\,-\frac{\alpha}{w},\, -1+\alpha,\right.$\\ $\qquad\left.\, -1+\alpha,\,\frac{1}{2}\left( -1+\alpha \right)\right)^T$\end{tabular} & \begin{tabular}[c]{@{}l@{}}$\lambda_1 = -3\,(\alpha+\,w),\,\lambda_2 = \frac{3}{2}(1- \,\alpha),\,$\\ $\lambda_3 = \frac{3}{2}(1-3\alpha-2\,w),\,$\\ $\lambda_4 = \frac{5- 9\,\alpha}{2},\,\lambda_5 = 1-3\,\alpha$\end{tabular} \\ \hline
PC7 & \begin{tabular}[c]{@{}l@{}}$\left(1+\frac{\alpha}{w},\,-\frac{\alpha}{w},\, -1+\alpha,\,\right.$\\ $\qquad\left. \frac{2\,w^2-w+3\,\alpha\,w-\alpha+\alpha^2}{\alpha}, -(\alpha+w)\right)^T$\end{tabular} & \begin{tabular}[c]{@{}l@{}}$ \lambda_1 = -\lambda_4=3\,(\alpha+ w),\,\lambda_2 = 1-3\,\alpha,\,$\\ $\lambda_3 = 1+3\,w,\lambda_5 =- \frac{3}{2}(1-3\,\alpha-2\,w)$\end{tabular} \\ \hline
\end{tabular}
\end{table}

The critical points of the system (\ref{sistdinint1a}-\ref{sistdinint1e}) and their respective eigenvalues are shown in table \ref{criticalpoints} and depend on the parameters $w$ and $\alpha$ except for $PC1$. The critical point $PC1$ is in fact a line parallel to the $\Dm$ axis. The eigenvalue $\lambda_5$ of $PC1$ is null and corresponds to a eigenvector that is also parallel to the $\Dm$ axis, indicating that near the line there is no evolution of the space phase trajectory in that direction. For some choices of the free parameters, critical points $PC2$, $PC5$, $PC6$ and $PC7$ can be non physical as $\Ome<0$, which means the CDE energy density is negative. We next examine the homogeneous for $\alpha>0$ subspace closely.

\subsection{Homogeneous subspace for $\alpha>0$}

As we have stated before the equations (\ref{sistdinint1a}-\ref{sistdinint1b}) form a subsystem independent of the $\delta$-functions, the homogeneous subsystem. The subsystem has a future attractor $PCA=({\Omm}^{PCA}=0,{\Ome}^{PCA}=1)^T$, a past attractor  $PCR=({\Omm}^{PCR}=1+\frac{\alpha}{w},{\Ome}^{PCR}=-\frac{\alpha}{w})^T$ and a saddle point $PCS=({\Omm}^{PCS}=0,{\Ome}^{PCS}=0)^T$. Both, $PCA$ and $PCR$ can be considered as critical points of the FLRW homogeneous scheme, or as a projection of the $PC1-PC7$ points over the $\Omm-\Ome$ subspace in a full five-dimensional representation. In the former case, the trajectories in the phase-space are computed for a given set of initial conditions with $\Dm=\De=\Dh=0$ and live completely in the homogeneous space, while in the later case the trajectories are computed with a general choice of $\Dm,\,\De$ and $\Dh$ and are represented in the homogeneous subspace as projections of the five-space trajectories over the $\Omm-\Ome$ subspace.

Additionally, the line of the homogeneous subspace that contains both the saddle point and the past attractor, i.e.,
\be
\Ome=-\frac{\alpha}{w+\alpha}\Omm,\label{invline}
\ee
is an invariant subspace of the homogeneous system as, from eqs. (\ref{sistdinint1a}-\ref{sistdinint1b}),
\be
\frac{d}{d \xi}\left(\Ome+\frac{\alpha}{w+\alpha}\Omm\right)=0.
\ee
Over the invariant line, the system can evolve from the past attractor to the saddle point (for initial conditions $(\Omm(\xi=0),\Ome(\xi=0))^T$ on the line with ${\Omm}(0)<1+\frac{\alpha}{w}$) or from the attractor to infinity (for $\Omm(0)>1+\frac{\alpha}{w}$). The trajectories of the homogeneous phase space cannot cross the invariant line, and the $\Omm-\Ome$ plane is divided in two: the region where trajectories evolve to the $\Ome=0$ axis and the region where $\Ome \neq 0$ at any instant. The later region contains the attraction basin of $PCA$, where trajectories evolve to the future attractor, but it is also possible to find some trajectories in it that evolve to infinity.

It is not clear if the trajectories evolving to  the to the $\Ome=0$ axis are physical or not as we have not a complete microscopical description of the coupling term $J_q$. On one hand, we can argue that once the trajectory reaches the $\Ome=0$ point, the QL energy density $\rho_{eq}$ of the LTB shell is null, (and the local energy density is also null provided that $\rho_{e}=\rho_{eq}\left(1+\De\right)$), and , consequently, the energy flux from the CDE to the CDM should end (provided that $\alpha>0$ which indicates that the CDE is ceding its energy density to the CDM). In this case, once $\Ome=0$ in a given shell, it will keep its evolution as a pure dust scenario. On the other hand, we can argue that the coupling term $J_{q}$ is independent of the CDE density and, consequently, those initial conditions lead to a non-physical scenario with negative values of CDE density. In this case the initial conditions should be avoided in any physical context, giving us a theoretical limit on the parameters of the model and the initial conditions. Both approaches should be not discarded beforehand, but we will consider the pure dust shell scenario more closely in the numerical examples as we believe it is physically more interesting and leads to exotic profiles such as pure CDM spheres surrounded by a mixture of CDE and CDM background.

\subsection{Initial value formulation, scaling laws and singularities.}
It is useful to introduce a initial value formulation of the CDE LTB model. In order to integrate (\ref{sistdinint1a})--(\ref{sistdinint1e}) we need to specify initial conditions given at the hypersurface $t=t_{in}$. We can rephrase the functions $R$ and $R'$ from (\ref{LTB}) as dimensionless scalar factors
\be
L=\frac{R}{R_{in}}, \qquad \Gamma=\frac{R'/R}{R'_{in}/R_{in}}=1+\frac{L'/L}{R'_{in}/R_{in}},
\ee
where the subindex ${}_{in}$ denote evaluation at $t=t_{in}$. In terms of the new scalar factors, $E=-\KK_{qi}R_{in}^2$ and, consequently, the LTB metric can be written as
\begin{equation}\dd s^2 = -\dd t^2 +L^2\,\left[\frac{\Gamma^2\,R_{in}'{}^2 \dd r^2}{1-\KK_{qi} R_{in}^2 }+R_{in}^2\,(\dd\theta^2+\sin^2\theta\dd\phi^2)\right],\label{LTB2}\end{equation}
which highlights the role of $L$ as a FLRW--like scale factor, while $\Gamma$ can be understood as a scale factor associated with the anisotropy of the LTB metric. Since the LTB metric, in either form (\ref{LTB}) or (\ref{LTB2}), admits an arbitrary rescaling of the radial coordinate, the initial value function $R_{in}$ can be used to define a specific radial coordinate. Additionally, $L=0$ is related to a central singularity, while $\Gamma=0$ is related to a shell crossing singularity \cite{izsuss10}.

It is straightforward that $\HH_q=\dot{L}/L$ and, from (\ref{RR2}), (\ref{evm_q2}) and (\ref{eve_q2}), the QL functions scale as in the FLRW case, i.e.,
\ba
\KK_q&=&\KK_{q\,in}L^{-2},\\
\rho_{mq}&=&\rho_{mq\,in}L^{-3(1-\alpha)}, \label{scalinglawrhomq} \\ \rho_{eq}&=&\rho_{eq\,in}L^{-3(1-w)}+\rho_{mq\,in}\frac{\alpha}{w+\alpha}\left(L^{-3(1+w)}-L^{-3(1-\alpha)}\right).\\
\ea
Additionally, $\HH_q$ follows the Hubble-like equation
\ba
\HH_q^2=\left(\frac{\dot{L}}{L}\right)^2&=&\frac{\kappa}{3}
\left[\rho_{mq\,in}\left(\frac{w}{w+\alpha}L^{-3(1-\alpha)}+\frac{\alpha}{w+\alpha}L^{-3(1+w)}\right)\right.\\
&&\left.+\rho_{eq\,in}L^{-3(1-w)}\right]-\KK_{q\,in}L^{-2}\label{EvL}.
\ea
The initial QL profiles $\rho_{mq\,in},\,\rho_{eq\,in},$ and $\KK_{q\,in},$ can be computed from a set of given initial local profiles $\rho_m(t_{in},r), \rho_e(t_{in},r)$ and $\KK(t_{in},r)$ using an arbitrary choice for the function $R_{in}(r)$. Defining the variable $\xi_{in}=\xi(t=t_in,r_i)=0$ for each shell $r=r_i$, the initial conditions for the system (\ref{sistdinint1a})--(\ref{sistdinint1e}), i.e., $\Omm(\xi=0)$, $\Ome(\xi=0)$, $\Dm(\xi=0)$, $\De(\xi=0)$ and $\Dh(\xi=0)$, can be evaluated from their respective definitions. Also, note that from the definition of $d\xi=\HH_q dt$, it is straightforward that $\xi=\ln(L)$.

From (\ref{EvL}), we can define ${\dot{L}}^2=L^{-1}Q(L)$ where
\ba
Q(L)&=&L^3\HH_q^2 =\HH_{q\,in}^2\left[a L^{3\alpha}+b L^{-3w}+c L\right],\\
a&=&\Omm(0)\frac{w}{w+\alpha}=\Omm(0)/{\Omm}^{PCR},\\
b&=&\Ome(0)+\Omm(0)\frac{\alpha}{w+\alpha}=\Ome(0)-\Omm(0)({\Ome}^{PCR}/{\Omm}^{PCR}),\\
c&=&1-\Ome(0)-\Omm(0)={\Omega}_{\KK}(0).
\ea
The initial conditions $\Omm(0)$ and $\Ome(0)$ for the shell $r=r_i$ determines whether $Q(L)$ has roots or not. If $Q(L_*)=0$ for some shell at $L_*$, the corresponding shell experiments a bounce at the instant where the coordinate $\xi=ln(L_*)$ (i.e., the shell stops the expanding (collapsing) evolution and starts collapsing (expanding)). If only the shell $r_i$ experiments a bouncing evolution while the neighbor shells keep their expanding (collapsing) behavior, the configuration will experiment a shell cross singularity.

Given that $L'/L=R'_{in}(\Gamma-1)/R_{in}$, deriving respect to the radius coordinate the scaling law (\ref{scalinglawrhomq}) and using the property $\Dm=(R'/R)(ln(\rho_{mq}))'$, it is straightforward that
\be
\Dm=-1+\alpha+\frac{\Dm(0)+3(1-\alpha)}{3\Gamma},\qquad \Rightarrow \qquad \Gamma=\frac{\Dm(0)+3(1-\alpha)}{3\Dm+3(1-\alpha)}.
\ee
This relation can be useful to evaluate the $\Gamma$ function at any shell and any instant except when $\Dm \rightarrow \infty$, that can be related to a shell cross singularity. Although similar relations can be obtained for $\De$ and $\Dh$, the corresponding scaling laws lead to a more complicated relations.

\section{Critical points clasification in terms of the free parameters $w$ and $\alpha$}\label{criticalpointsLTB}

In this section we study the critical points of above considering different possibilities of the free parameters $w$ and $\alpha$. As both parameters are widely used in a cosmological frame in the FLRW model, we will restrain ourselves to a range of parameters that is usefull in Cosmology.

First, the adiabatic parameter $w$ is assumed to be constant in this work. The observational data seems to favor  the $\Lambda$--CDM model for which $w=-1$, although small variations from it are still possible\cite{plank}. Dark energy with $w>-1$ is referred to as quintessence models in the literature, while dark energy with $w<-1$ are called phantom models of dark energy. The latter models present several theoretical problems, as the violation of the second law of thermodynamics once we assign a entropy to the phantom fluid, or the presence of a negative kinetic energy of the phantom field term \cite{copeland}. In this work we will assume that the dark energy can behave as cosmological constant ($w=-1$), quintessence ($w>-1$) or phantom ($w<-1$). In the latter two cases the adiabatic coefficient value will be close to $-1$.

Regarding the parameter $\alpha$, in \cite{db}, the authors state that the second law of thermodynamics regarding the entropy of the CDE field gets violated if $\alpha<0$ and the CDE is an effective field, while the entropy is null for a scalar field in a pure quantum state. Assuming that $\alpha>0$, the coupling parameter must be smaller than $0.1$ in order to reproduce the observed values of BAO and CMB anisotropy \cite{ol,ol2}. On the other hand, in \cite{Maar,Gavela}, the evolution of the linear perturbations in a FLRW approach of both CDM and CDE are considered for a coupling term of the same kind of $J_q$ concluding that when $\alpha>0$ and $w$ is constant, early non-adiabatic large-scale instabilities are present (a non constant adiabatic coefficient $w=w(a)$ could lead to avoid the instabilities). Those results are specially interesting in this work as our LTB approach is a full perturbation scheme, closely related to the linear perturbation FLRW scheme. We can study the evolution of perturbations with the addition of having a local representation of the energy densities and the coupling. In this work we will assume positive and negative values of $\alpha$.

\begin{figure}[tbp]
\includegraphics*[scale=0.30]{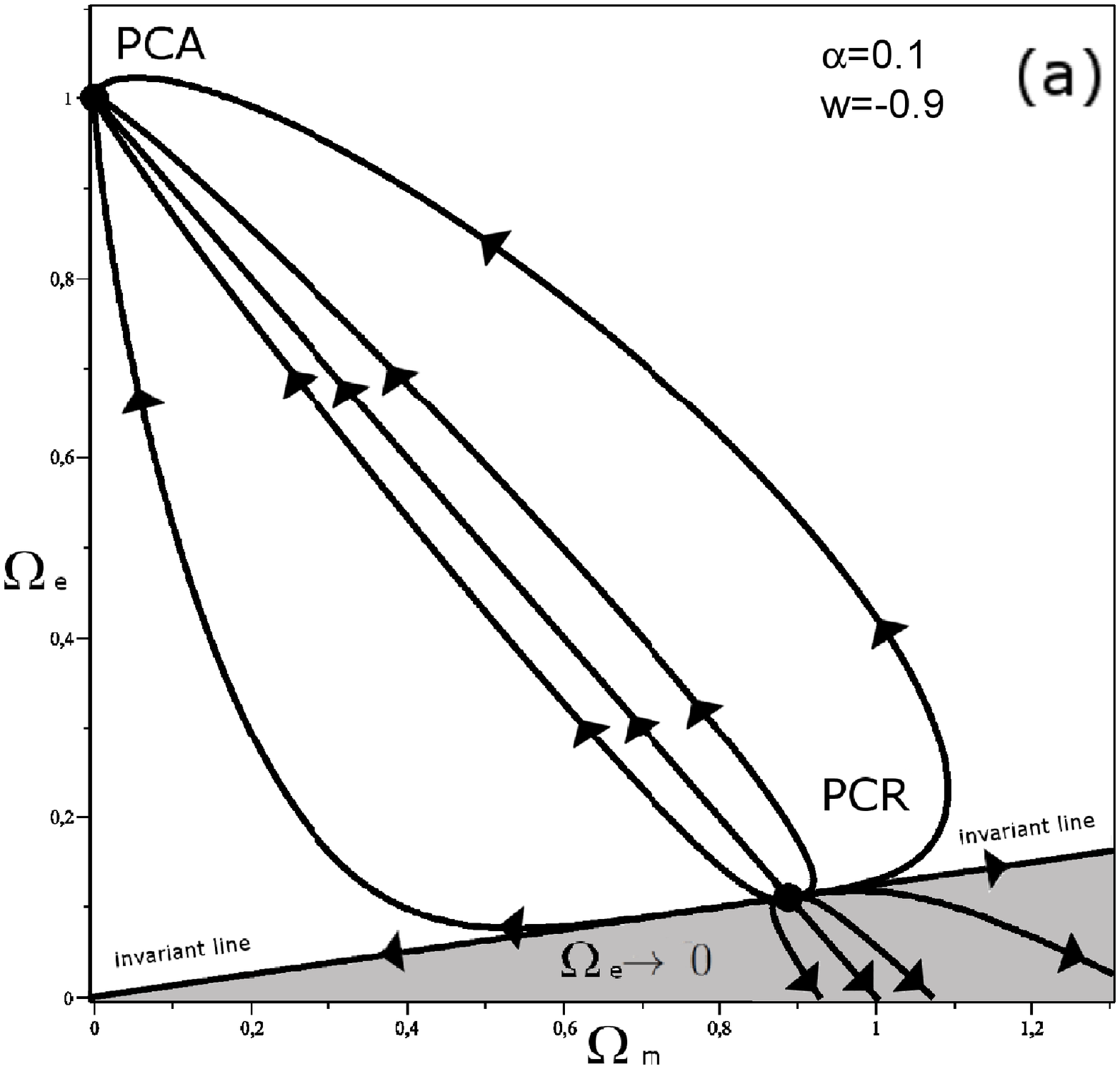}
\includegraphics*[scale=0.30]{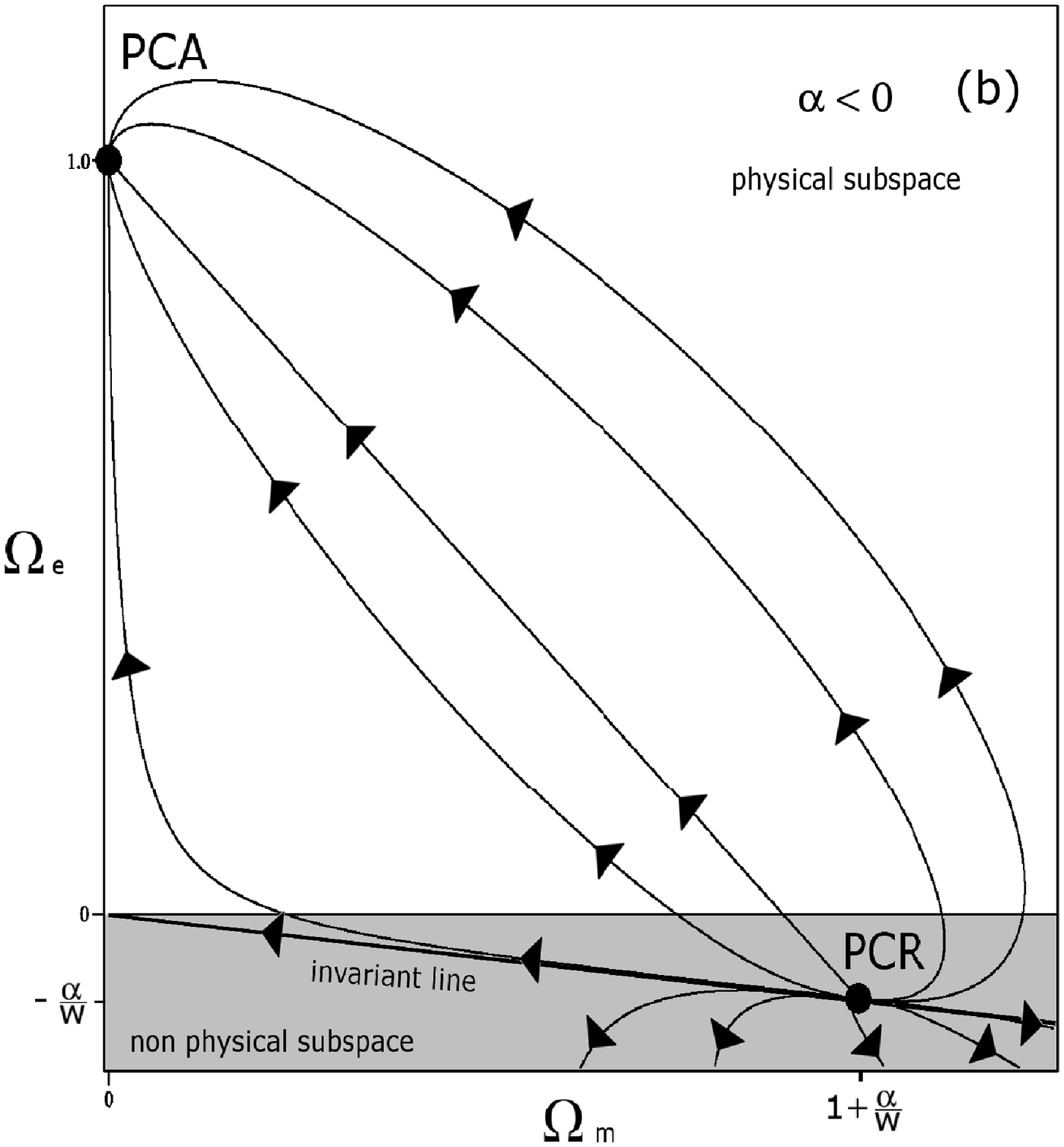}
\caption{Panel (1a): Critical points and numerical trajectories of the dynamical system (\ref{sistdinint1a}-\ref{sistdinint1e}) in the homogeneous projection for $\alpha=0.1$ and $w=-0.9$. For other choices of the parameters with $\alpha>0$ the point $PCR$ will be in a diferent position, and, consequently, the invariant line will have a different slope.  Panel (1b): Critical points and numerical trajectories of the dynamical system (\ref{sistdinint1a}-\ref{sistdinint1e}) in the homogeneous projection for $\alpha<0$. For any initial conditions choice the trajectory evolves to negative $\Ome$ in the past.} \label{fig1}\end{figure}

Figure \ref{fig1} shows the homogeneous subspace together with the critical points $PCR$ and $PCA$ and the invariant line for both cases: $\alpha>0$ in panel (a), and $\alpha<0$ in panel (b). Some numerically computed trajectories are shown for illustration purposes only.

\begin{figure}[tbp]
\includegraphics*[scale=0.30]{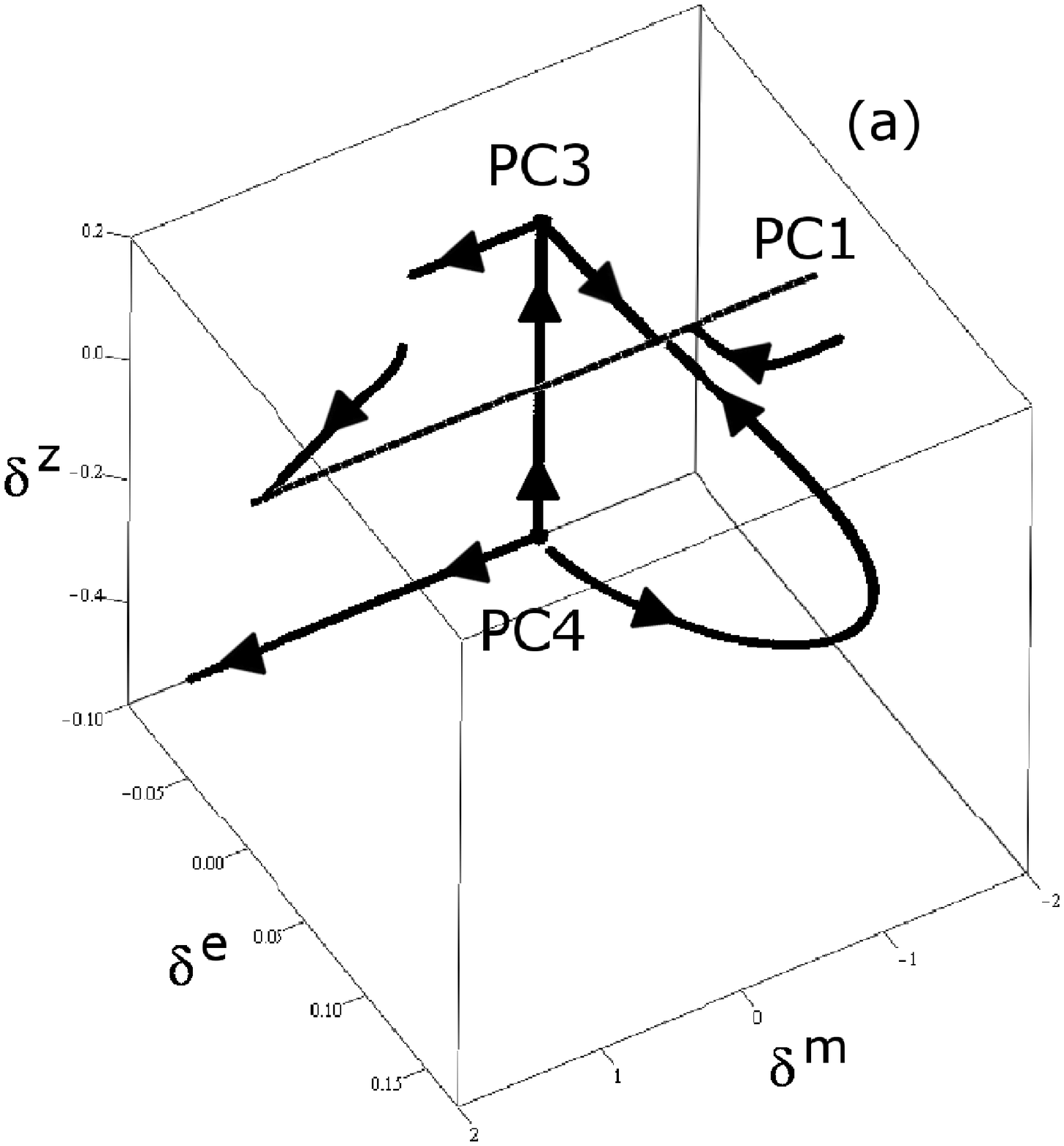}
\includegraphics*[scale=0.3]{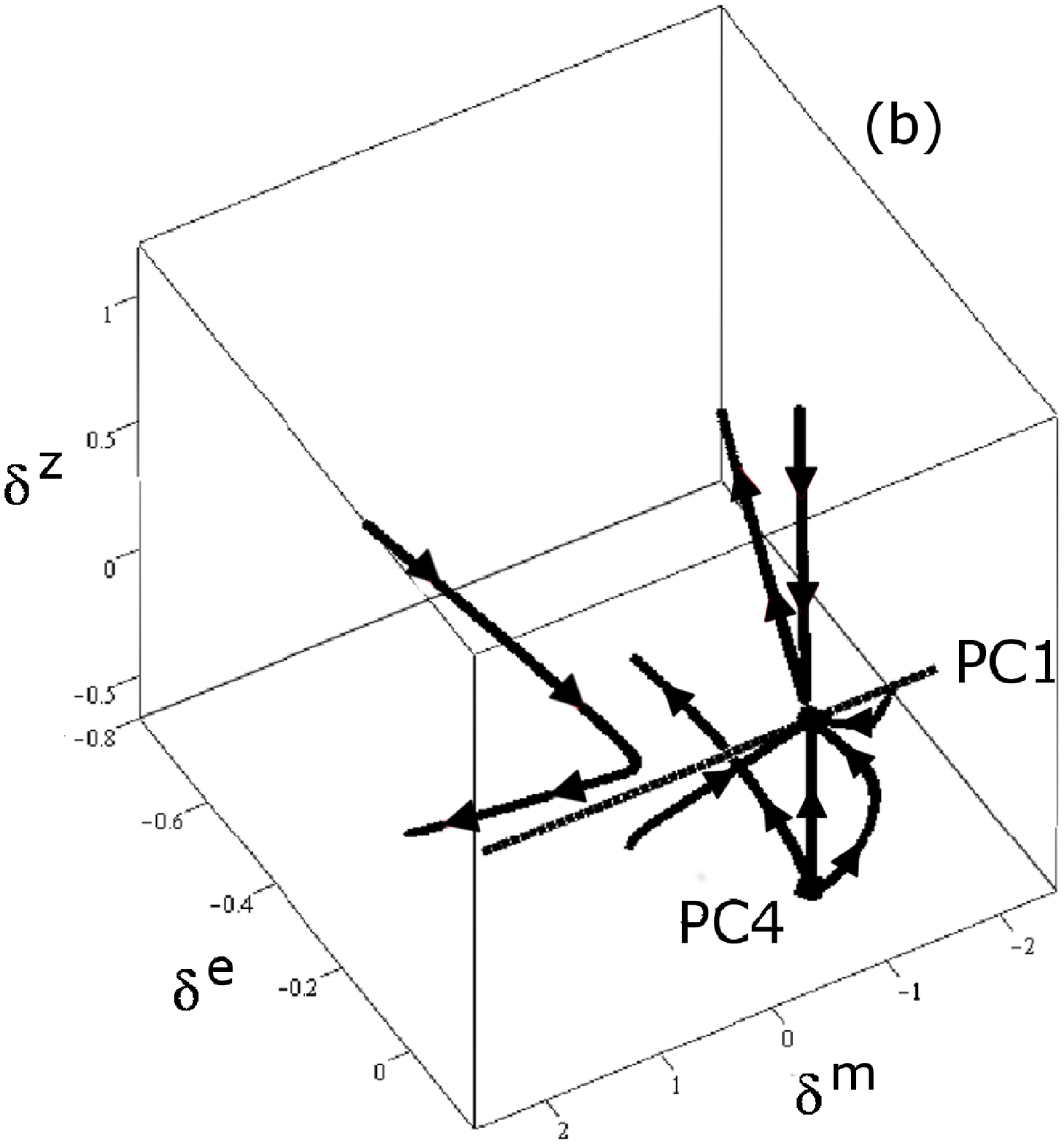}
\includegraphics*[scale=0.3]{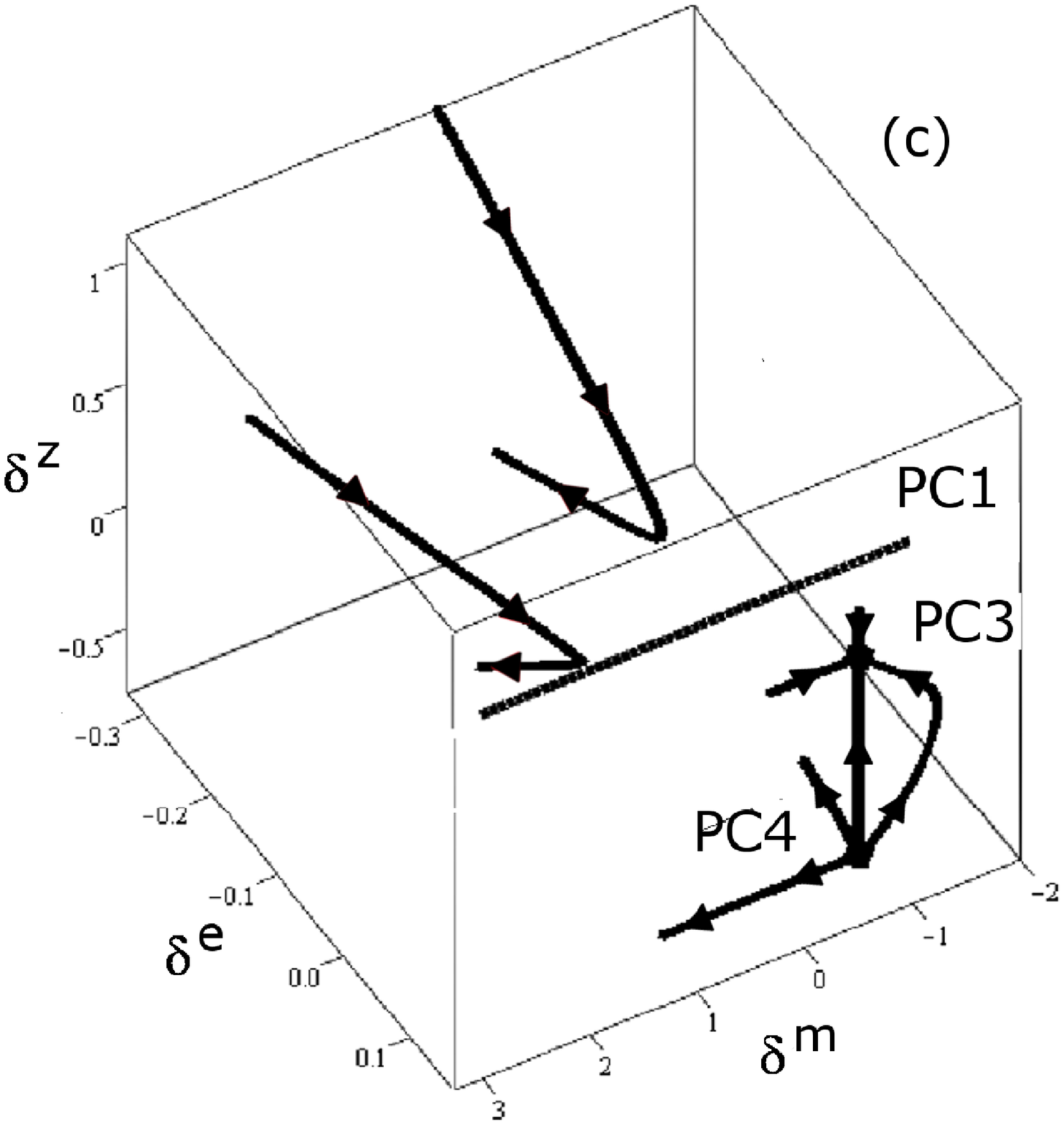}
\includegraphics*[scale=0.3]{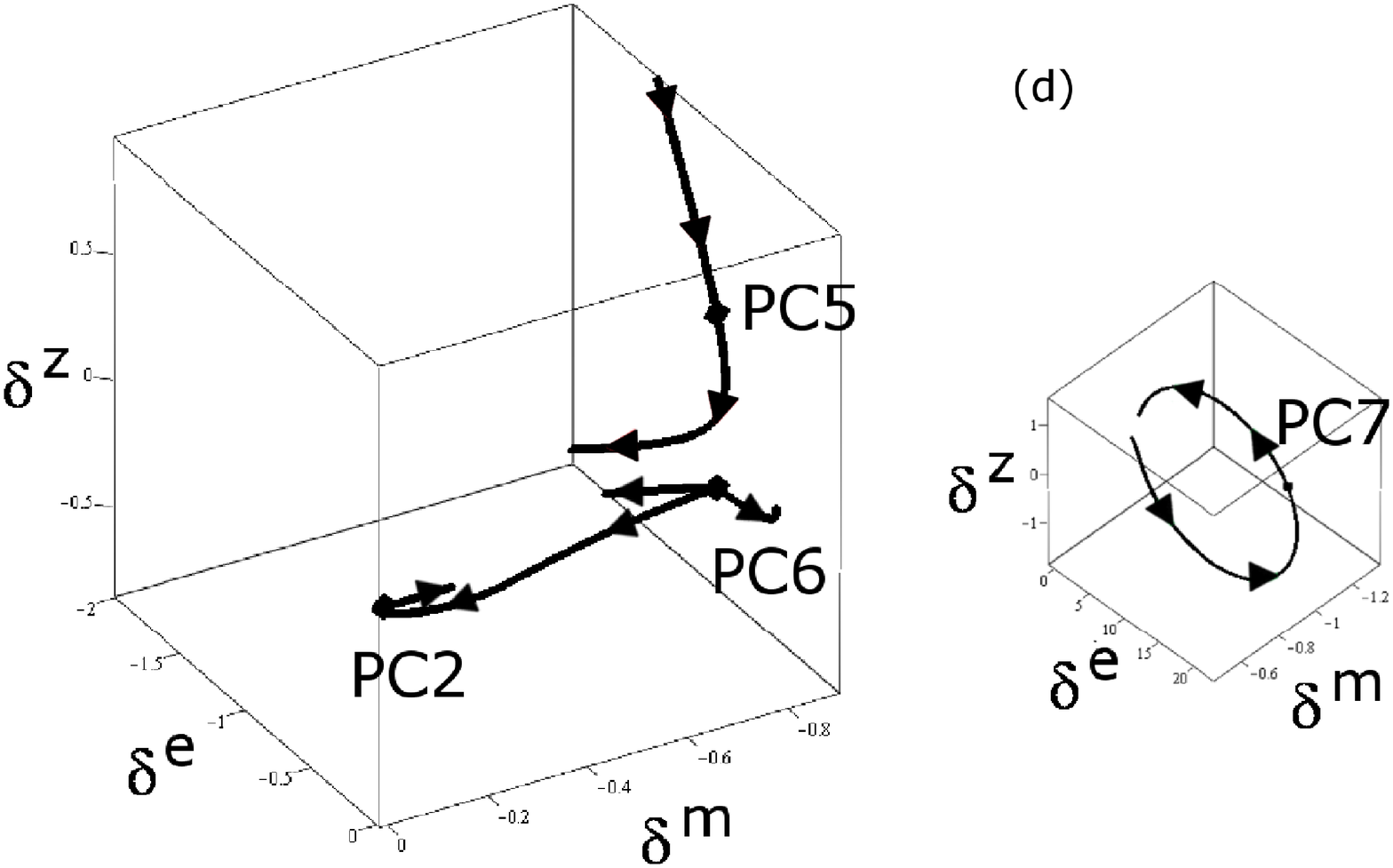}
\caption{Critical points and numerical trajectories of the dynamical system (\ref{sistdinint1a}-\ref{sistdinint1e}) in the inhomogeneous projections. Panel (2a): Inhomogeneous subspace $\Dm$ vs. $\De$ vs. $\Dh$ with $\Omm=0$ y $\Ome=1$ for $\alpha=0.1$ and $w=-0.9$. Panel (2b): Inhomogeneous subspace $\Dm$ vs. $\De$ vs. $\Dh$ with $\Omm=0$ y $\Ome=1$ for $\alpha=0.1$ and $w=-1.0$; the critical points $PC1$ and $PC4$ are shown, the point $PC3$ disappears in this case. Panel (2c): Inhomogeneous subspace $\Dm$ vs. $\De$ vs. $\Dh$ with $\Omm=0$ y $\Ome=1$ for $\alpha=0.1$ and $w=-1.1$; the point $PC3$ has different coordinates and behavior than in the $w=-0.9$ case as it is now the future atractor. Panel (2d): Inhomogeneous subspace  $\Dm$ vs. $\De$ vs. $\Dh$ with $\Omm=1+\alpha/w=8/9$ y $\Ome=-\alpha/w=1/9$ for $\alpha=0.1$ and $w=-0.9$, the critical points $PC2$, $PC5-PC7$ are represented.} \label{fig2}\end{figure}

Figure \ref{fig2} shows the two inhomogeneous projections of the system (\ref{sistdinint1a}-\ref{sistdinint1e}). Panel (a), (b) and (c) represent the projection with $\Omm=0$ y $\Ome=1$ for different choices of $w$ and $\alpha>0$. Panel (d) shows the projection $\Omm=1+\alpha/w=0.9$ y $\Ome=-\alpha/w=0.1$ for $\alpha=0.1$ and $w=-0.9$, although choosing a different value of $w$ will not change the general behavior of the points or the trajectories. A similar figure to figure \ref{fig2} would be obtained when plotting panels (a-c) and $\alpha<0$. Panel 2(d), on the other hand, is not physical in the $\alpha<0$ case.

\subsection{Energy density flux from CDE to CDM ($\alpha>0$)\label{subsecalphapos}}
In this case all seven critical points are physical, as $\Ome>0$ for them. The attractor is a different point for the different choices of $w$.

\subsubsection{Quintessence}
When $w>-1$, the line $PC1$ is a future attractor as the trajectories near the critical point evolve to converge with the line at a fixed point of it in the phase space (the convergence point will have a different constant value of $\Dm$ for each trajectory). On the other hand, the critical point $PC6$ is a past attractor as all the eigenvalues of the jacobian matrix of the system computed at $PC6$ have positive values. the rest of critical points are saddle points with their own attraction subspace generated by the corresponding eigenvectors.

Panel \ref{fig1}a shows the homogeneous subspace, the future attractor $PCA$, the past attractor $PCR$, the invariant line and the two kind of trajectories evolving to the future attractor or to the $\Ome=0$ axis, respectively. In the panel \ref{fig2}a, the inhomogeneous projection $\Omm=0,\,\Ome=1$ is represented. The attractor $PC1$ is shown and also some trajectories in its vicinity that evolve to it at different values of $\Dm$. Also the saddle points $PC3$ and $PC4$ are represented. In this projection, the $\Dm-\De-\Dh$ subspace, the trajectories near $PC4$ can only evolve away from it as the eigenvectors with negative eigenvalues at $PC4$ are orthogonal to the subspace. Finally, the panel \ref{fig2}d the inhomogeneous subspace with $\Omm=(w+\alpha)/w=8/9,\,\Ome=-\alpha/w=1/9$ is plotted. In this projection, the past attractor $PC6$ and the saddle points $PC2,\, PC5,\, PC6,\, PC7$ are represented.

\subsubsection{Cosmological constant}

When $w=-1$, the critical point $PC3$ is superposed with the line $PC1$. Additionally, $PC1$ behavior is no longer as a future attractor but a non hyperbolical point. For some trajectories, $PC1$ still acts as an attractor as it has three negative eigenvalues while for other trajectories it is no longer a future attractor. The rest of the points show a phenomenologically identical behaviour to the $w>-1$ case.

The homogeneous subspace has a similar behavior than the one shown in panel \ref{fig1}a. The only difference is the position of the $PCR$ point and the slope of the invariant line. In the panel \ref{fig2}b, the inhomogeneous projection $\Omm=0,\,\Ome=1$ is represented for the $w=-1$ case. The point $PC1$ is represented and we appreciate some trajectories evolving to it while other trajectories in its vicinity evolve away from it, this is due to the non hyperbolic behavior of $PC1$ in this case in contrast with the $w>-1$ case. Critical point $PC4$ is also represented and is phenomenologically identical to the $w<-1$ case.  Finally, the inhomogeneous subspace with $\Omm=(w+\alpha)/w=0.90,\,\Ome=-\alpha/w=0.10$ is very similar to the one represented in panel \ref{fig2}d.

\subsubsection{Phantom dark energy}
When $w<-1$, the critical points $PC3$ and $PC4$ have positive values of its $\De$ coordinate. Additionally, $PC1$ behaves as a saddle point while the attractor is $PC3$. Both the $w>-1$ and the $w<-1$ cases present an attractor but in the $w>-1$ case the future attractor allow different values of $\Dm$ while in the $w<-1$ the future attractor allow a single possibility for $\Dm=-1+\alpha$. The rest of the points show a phenomenologically identical behavior to the $w<-1$ and $w=-1$ cases.

The homogeneous subspace is identical to that of panel \ref{fig1}a except for the position of the $PCR$ point and the slope of the invariant line. In the panel \ref{fig2}c, the inhomogeneous projection $\Omm=0,\,\Ome=1$ is represented. The point $PC1$ is a saddle point and the point $PC3$ is now the attractor of the system in contrast with the $w<-1$ and $w=-1$ cases. Critical point $PC4$ is also represented and is phenomenologically identical to the previous cases.  Finally, the inhomogeneous subspace with $\Omm=(w+\alpha)/w=10/11,\,\Ome=-\alpha/w=1/11$ is as in the previous case similar to that in \ref{fig2}d.

\subsection{Energy density flux from CDM to CDE ($\alpha<0$).}

When $\alpha<0$, the energy flows from the CDM to CDE. In this case only $PC1$, $PC3$ and $PC4$ have physical meaning while the rest of the points present values with $\Ome<0$. In the homogeneous subsystem, the past attractor $PCR$ is no longer physical and consequently the invariant line is also non physical. Thus, the attraction basin of $PCA$ is the physical space and all the trajectories computed for any physical initial condition lead to the attractor $PCA$. On the other hand, the trajectories evolve from $PCR$ which is non physical in this scenario. The fact that $PCR$ present negative values of the CDE energy density is stated in several works regarding FLRW scenarios in Cosmology \cite{GZun14} and makes the coupling (\ref{intj_m}) with $\alpha<0$ model very unlikely. A possible solution to this problem is considered in \cite{bespro} where the coupling is activated at a concrete instant in the past previous to the point where the CDE present negative energy, avoiding in this way the negative values of $\Ome$. But, as the authors state, this activation mechanism is purely defined on phenomenological grounds and present a fine tuning problem similar to the problem that CDE models try to solve on the first place. Again, given the lack of microscopical description of the particle interaction leading to a coupling like (\ref{intj_m}), this option should be not discarded beforehand. In this sense, we consider the coupling (\ref{intj_m}) with $\alpha<0$ restricting ourselves to the physical space of parameters, i.e., the region for which $\Ome\ge 0$.

There is no significative difference between the homogeneous space for the different possibilities of the parameter $w$. Although the trajectories follow a different curve for every choice of $w$, they all lead to the $PCA$. Panel \ref{fig1}b shows schematically the homogeneous subspace for $\alpha<0$. The behavior of $PC1$, $PC3$ and $PC4$ in the inhomogeneous subspace with $\Omm=1, \Ome=0$ is identical to the $\alpha>0$ case (the reader should refer to subsection \ref{subsecalphapos} to read a detailed description of them for different choices of $w$).

\section{Numerical examples and its evolution.}\label{numerical}
In this section we will chose some initial profiles for the local scalars $\rho_m(t_{in},r)$, $\rho_e(t_{in},r)$ and $\KK(t_{in},r)$ and an arbitrary initial function $R_{in}(r)$. From those functions, it is possible to do a partition of the $r$ variable defining the number of shells $n$ we will use, and evaluate the initial conditions for the system (\ref{sistdinint1a}-\ref{sistdinint1e}). Then, after fixing the parameters $w$ and $\alpha$, we will compute numerical solutions for every shell $r_i$. For simplicity we will set the initial time $t_{in}$ to zero, so that $\xi(t_{in}=0,r_i)=0$ at any shell. We next compute the local quantities profiles at a fixed instant of time $t$, by evaluating the QL quantities at the variable $\xi$ corresponding to $t$ at every shell $r=r_i$.

Given that the LTB metric is scale invariant, it possible to define dimensionless coordinate ${\textsf{t}}=H_s t$ where $H_s$ is an arbitrary constant with dimensions of inverse of time. In this case, in terms of the new variable, $\HH_q=H_s(\dot{R}/R)$. The constant $H_s^{-1}$ will set the time scale of the metric LTB (or the length scale of it as $l_s=H_s^{-1}$). Additionally the energy densities will be rescaled as $\rho_a= H_s^{-2}\rho_a$ (similarly to what is done in Cosmology with the current Hubble factor $H_0$). For the sake of simplicity, we will set the arbitrary scale as $H_s=1$ for the numerical work, and we will use $t$ to denote the dimensionless time coordinate.

\subsection{Expanding mixture of CDM and CDE.}
In this configuration we set the free parameters as $w=-0.9$ and $\alpha=0.1$. We consider the initial local profiles
\ba
\rho_{m\,in}&=&{ m_{10}}+{\frac {{ m_{11}}-{ m_{10}}}{1+\tan(r)^{2}}},{ m_{10}
}= 0.01,{ m_{11}}= 20.0;\nonumber\\
\rho_{e\,in}&=&20.75;\label{p8}\\
\KK_{in}&=&k_{10}+\frac{k_{11}-k_{10}}{1+\tan(r)^2}, k_{10}=-4.1, k_{11}=35.5;\nonumber
\ea
and the scalar $R_{in}(r)=\tan(r)$. The variable $r$ goes from $0$ to $\pi/2$ and we made a partition of the interval of $n=20$.

\begin{figure}[tbp]
\includegraphics*[scale=0.30]{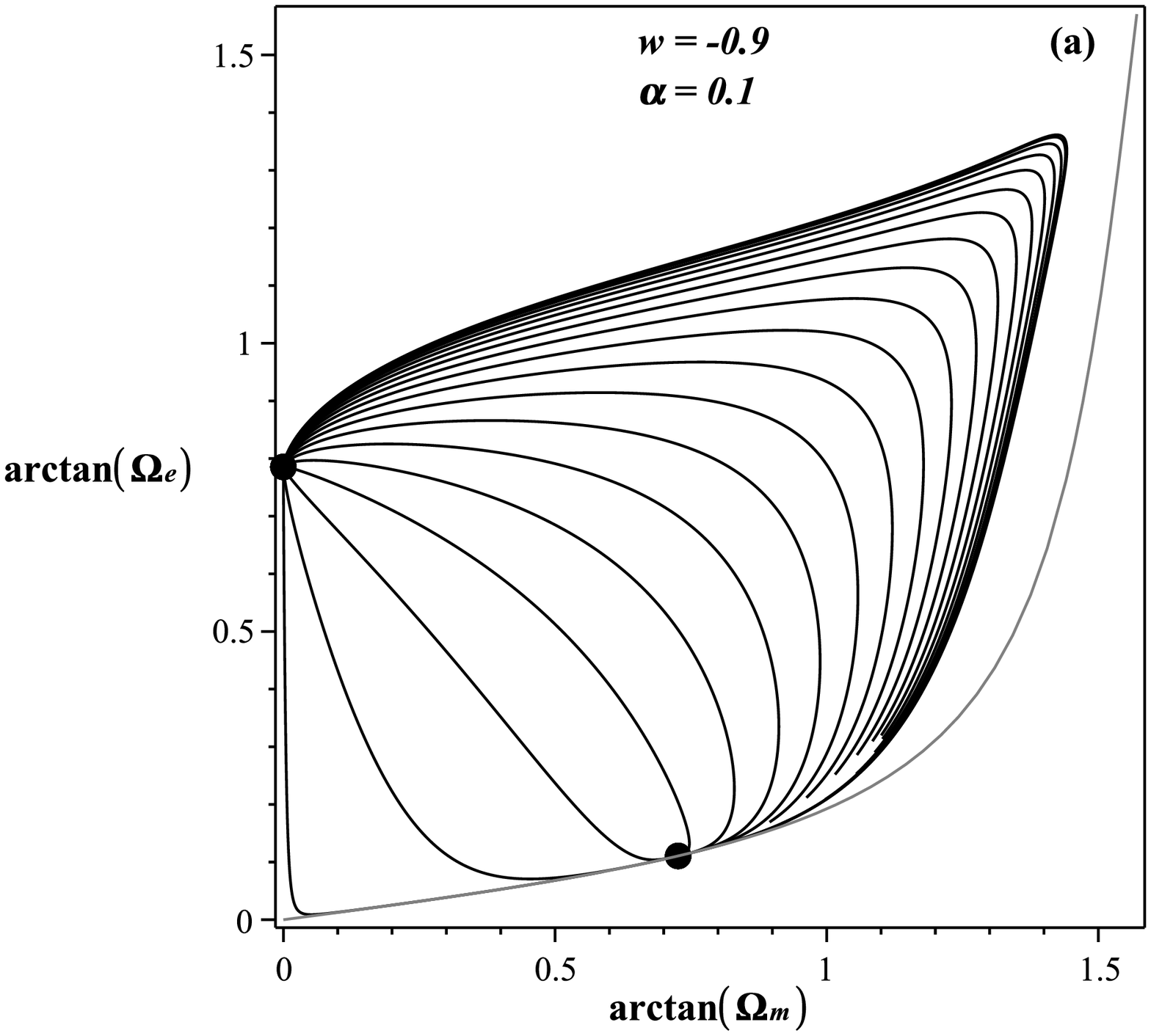}
\includegraphics*[scale=0.4]{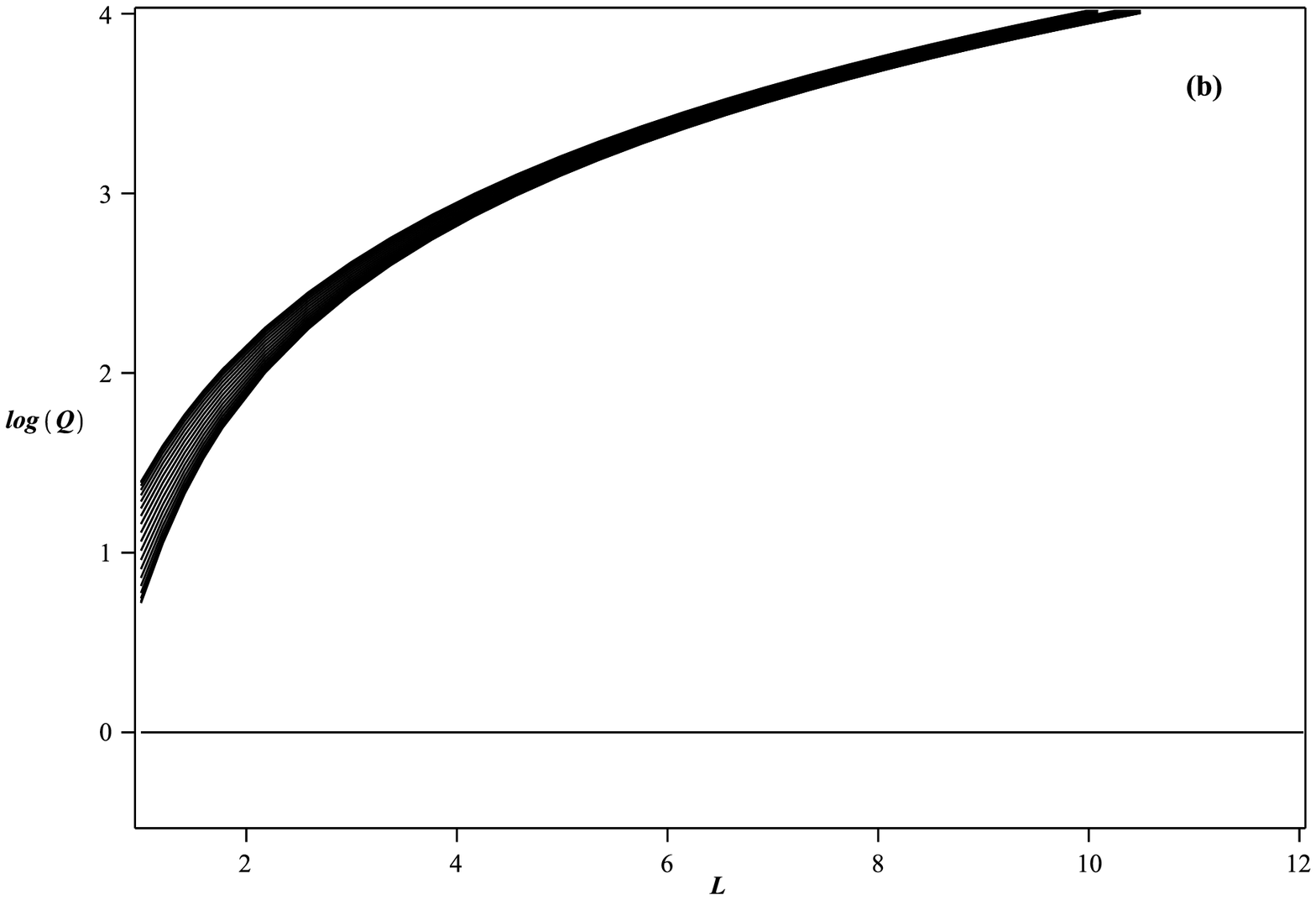}
\caption{Panel (a): Homogeneous projection of the trajectories of the system (\ref{sistdinint1a}-\ref{sistdinint1e}) for the different shells of the configuration with initial conditions given by (\ref{p8}) and $w=-0.9$, $\alpha=0.1$. Panel(b): Evolution of $\log(Q(L))$ vs. $L$ for the different shells of the configuration with initial conditions given by (\ref{p8}) and $w=-0.9$, $\alpha=0.1$. Refer to the text for a detailed discussion of the panels.} \label{fig3}\end{figure}

Panel (a)of fig. \ref{fig3} shows the homogeneous projection of the trajectories of every shell. The critical points and the invariant line are also represented as a grey line. The initial conditions for all the shells are in the $PCA$ attraction basin, consequently the shells evolve to the future attractor for a long time. In panel (b)of fig. \ref{fig5}, the scalar $Q(L)$ is computed for the different shells, and no shell experiments a bounce at any point.

\begin{figure}[tbp]
\includegraphics*[scale=0.3]{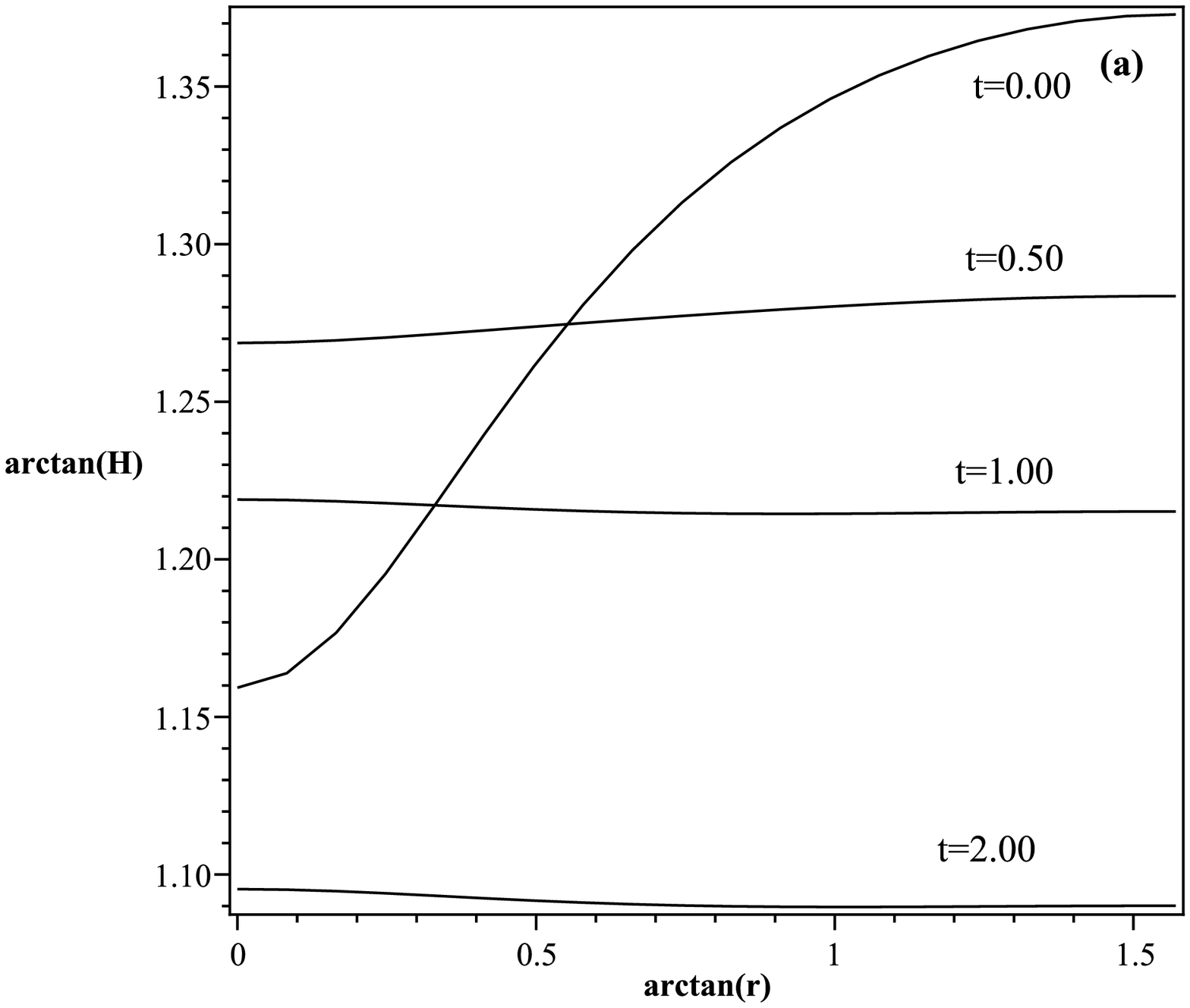}
\includegraphics*[scale=0.3]{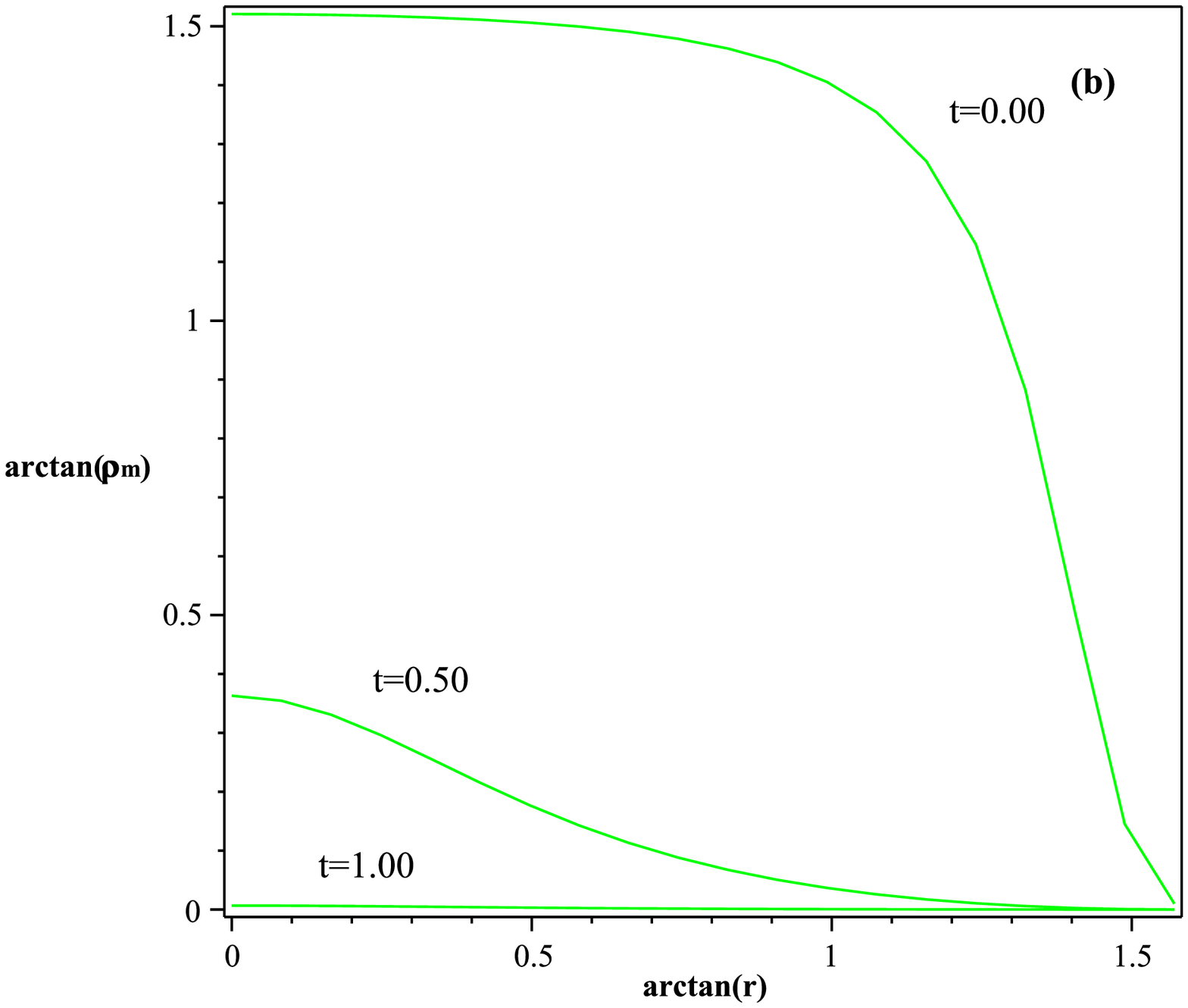}
\includegraphics*[scale=0.3]{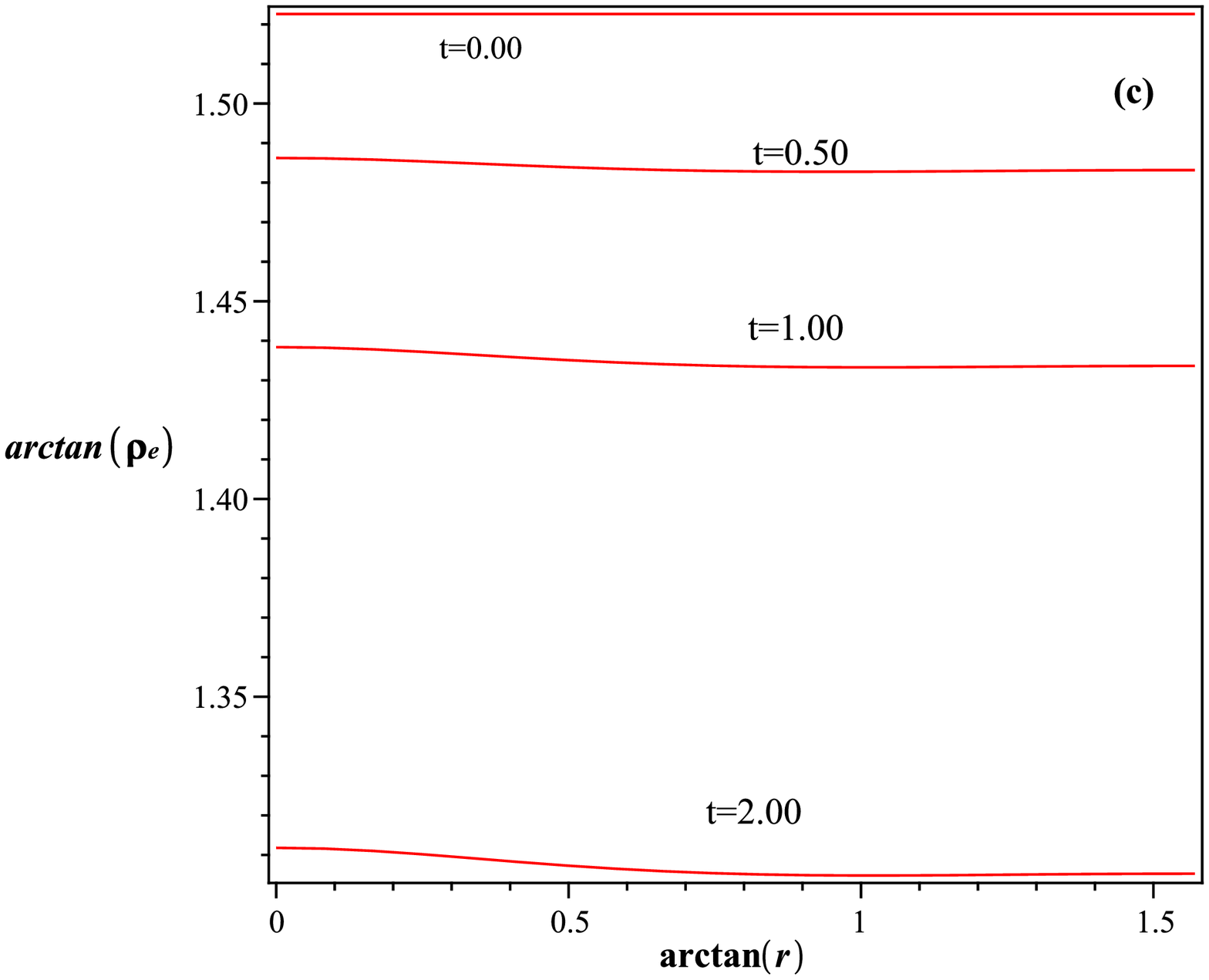}
\includegraphics*[scale=0.3]{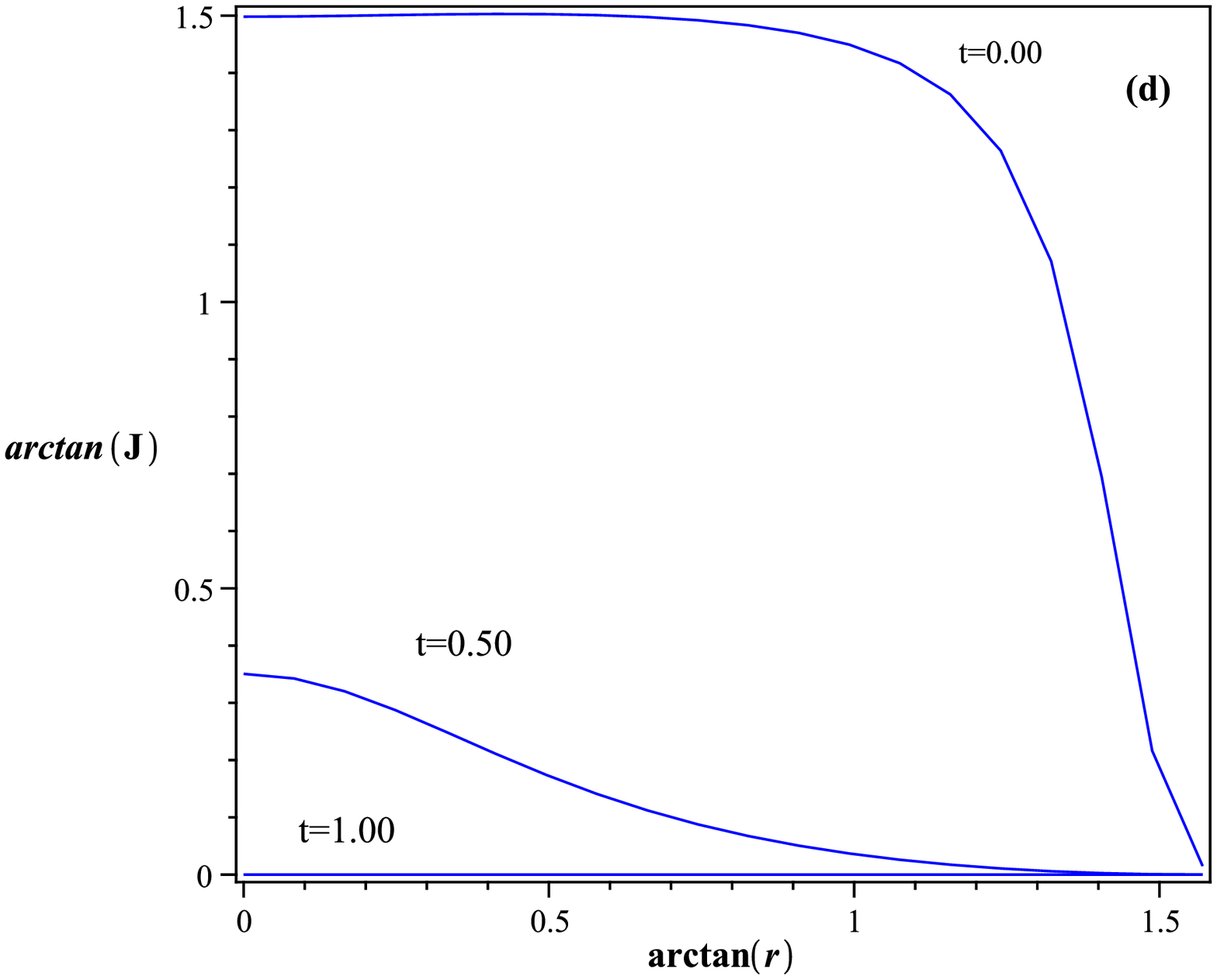}
\caption{Panel (a): Local profile of the scalar $\HH$ for different instants of time for the configuration with initial conditions given by (\ref{p8}) and $w=-0.9$, $\alpha=0.1$. Panel (b): Local profile of the scalar $\rho_m$ for different instants of time for the configuration with initial conditions given by (\ref{p8}) and $w=-0.9$, $\alpha=0.1$. Panel (c): Local profile of the scalar $\rho_e$ for different instants of time for the configuration with initial conditions given by (\ref{p8}) and $w=-0.9$, $\alpha=0.1$. Panel (d): Local profile of the scalar $J$ for different instants of time for the configuration with initial conditions given by (\ref{p8}) and $w=-0.9$, $\alpha=0.1$. Refer to the text for a detailed discussion of the panels.} \label{fig4}\end{figure}

In figure \ref{fig4}, the evolution of the local profiles of $\HH$, $\rho_m$, $\rho_e$ and $J$ are plotted in panels (a), (b), (c), and (d) respectively, for different instants of time. As the shells evolve to the $PCA$, the local profiles of $\rho_m$ and $\rho_e$ decrease with the expansion.

For this configuration, when changing the values of the parameters $w$ and $\alpha$, the initial conditions of all the shells will lay in the attraction basin of $PCA$, so the general behavior of the trajectories and the local profiles will be phenomenologically identical. On a side note, when solving the system for negative times, i.e., in the past, some shells experiment a bounce at a definite instant. The trajectories near this instant of bounce in the past evolve to the past attractor.

Considering any local profile with $\alpha<0$, the evolution of the trajectories will be very similar to that scenario, as all the shells will evolve to the future attractor. Those scenarios will have problems in the past as they evolve to non physical negative values of the CDE energy density.

\subsection{Ever-expanding mixture scenario evolving to pure CDM.}
For the shells with initial conditions under the invariant line, the evolution leads to $\Ome=0$ as stated before. In this case for the numerical work, we will assume that the coupling term from this point on is null and the shell evolve as a pure dust LTB scenario.The pure dust shells will follow an evolution determined by the equations
\ba
\frac{\partial{\Omm}}{\partial{\xi}} &=& \Omm\,\left[ -1+\Omega_m\right], \label{sistdinint1dusta}\\
\frac{\partial{\Dm}}{\partial{\xi}} &=& -3\,\Dh\,\left(1+\Dm \right), \label{sistdinint1dustc}\\
\frac{\partial{\Dh}}{\partial{\xi}} &=& -\Dh\left( 1+3\Dh \right)+\frac{\Omm\,\left( \Dh-\Dm \right)}{2}, \label{sistdinint1duste}
\ea
with initial conditions for $\Omm$, $\Dm$ and $\Dh$ given at the instant where the shell reaches $\Ome=0$. From this point on the scalar $\HH_q$ of the pure dust shell $r_i$ is defined as
\be
\HH_q^2(\xi)=\frac{\kappa}{3}\rho_{mq}-\KK_q.
\ee
The pure dust shells can experiment a bounce following the dust LTB dynamics as $Q(L)=L^{3} \HH_q^2(\xi=\ln(L),r_i)$ can be null at some $L$.
In this example, we choose a mixture of CDE and CDM with initial conditions for the different shells that evolve to the pure dust scenario with no shell crossing singularities.

In this case the choice of free parameters is $w=-1.0$ and $\alpha=0.1$, the initial local profiles
\ba
\rho_{m\,in}&=&{ m_{10}}+{\frac {{ m_{11}}-{ m_{10}}}{1+\tan(r)^{2}}},{ m_{10}
}= 0.00,{ m_{11}}= 13.10;\nonumber\\
\rho_{e\,in}&=&{ e_{10}}+{\frac {{ e_{11}}-{ e_{10}}}{1+\tan(r)^{2}}},{ e_{10}
}= 0.00,{ e_{11}}= 0.65;\label{p15}\\
\KK_{in}&=&k_{10}+\frac{k_{11}-k_{10}}{1+\tan(r)^2}, k_{10}-1.10, k_{11}=-3.50;\nonumber
\ea
and the scalar $R_{in}(r)=\tan(r)$. The variable $r$ goes from $0$ to $\pi/2$ and we made a partition of the interval of $n=20$..

\begin{figure}[tbp]
\includegraphics*[scale=0.30]{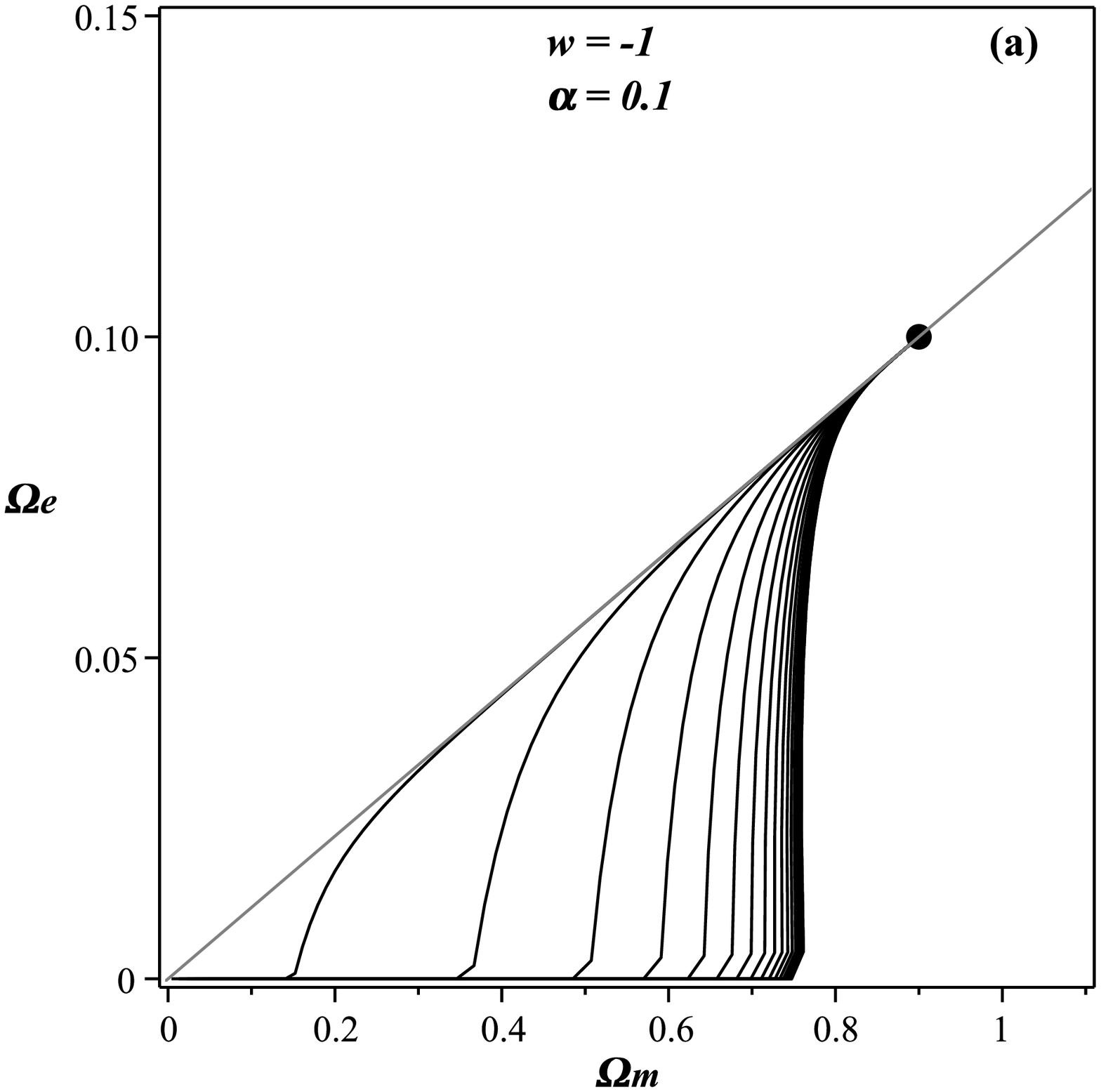}
\includegraphics*[scale=0.4]{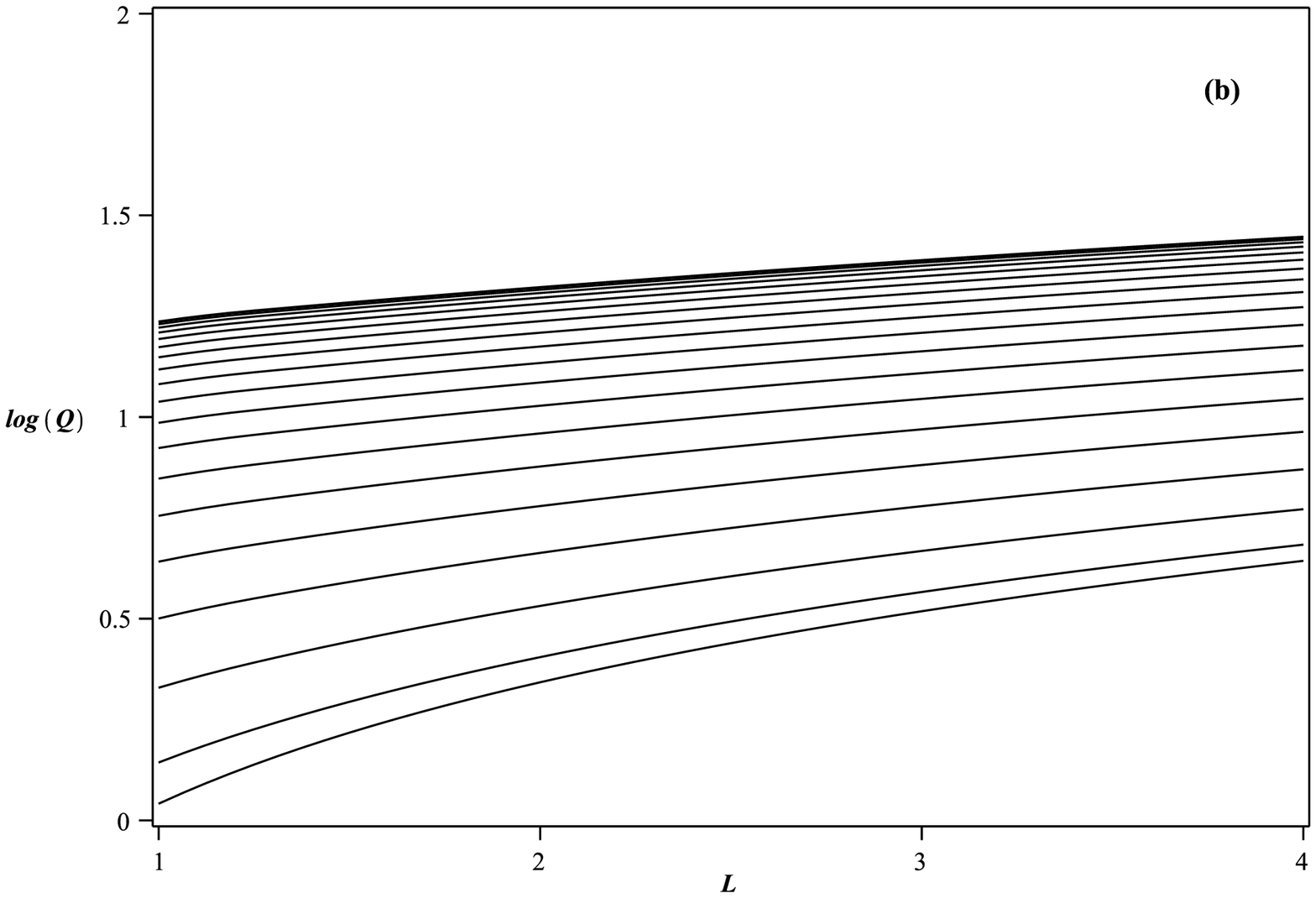}
\caption{Panel (a): Local profile of the scalar $\HH$ for different instants of time for the configuration with initial conditions given by (\ref{p15}) and $w=-1.0$, $\alpha=0.1$. Panel (b): Local profile of the scalar $\rho_m$ for different instants of time for the configuration with initial conditions given by (\ref{p15}) and $w=-1$, $\alpha=0.1$. Panel (c): Local profile of the scalar $\rho_e$ for different instants of time for the configuration with initial conditions given by (\ref{p15}) and $w=-1.0$, $\alpha=0.1$.Refer to the text for a detailed discussion of the panels.} \label{fig5}\end{figure}

Panel (a)of fig. \ref{fig5} shows the homogeneous projection of the trajectories of every shell from negative times where all the trajectories evolve from the past attractor $PCR$. The initial conditions for all the shells are under the invariant line and the shells evolve in the future to the $\Ome=0$ axis. Once the shell reaches the point where no longer has CDE, we consider that the coupling term is null and the shell follows a pure dust LTB evolution. In panel (b)of fig. \ref{fig5}, the scalar $Q(L)$ is computed for the different shells, and no shell experiments a bounce at any point.

\begin{figure}[tbp]
\includegraphics*[scale=0.30]{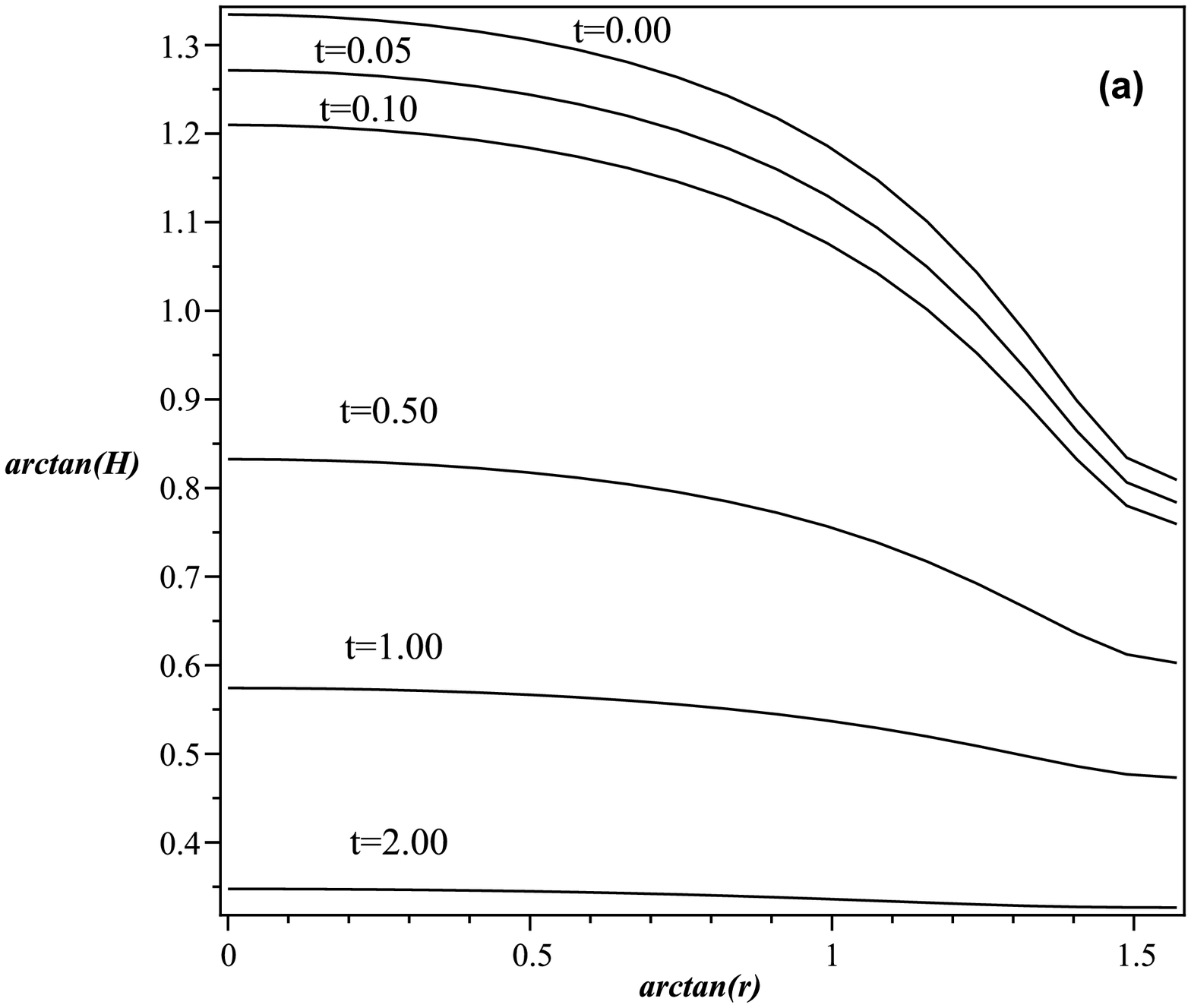}
\includegraphics*[scale=0.3]{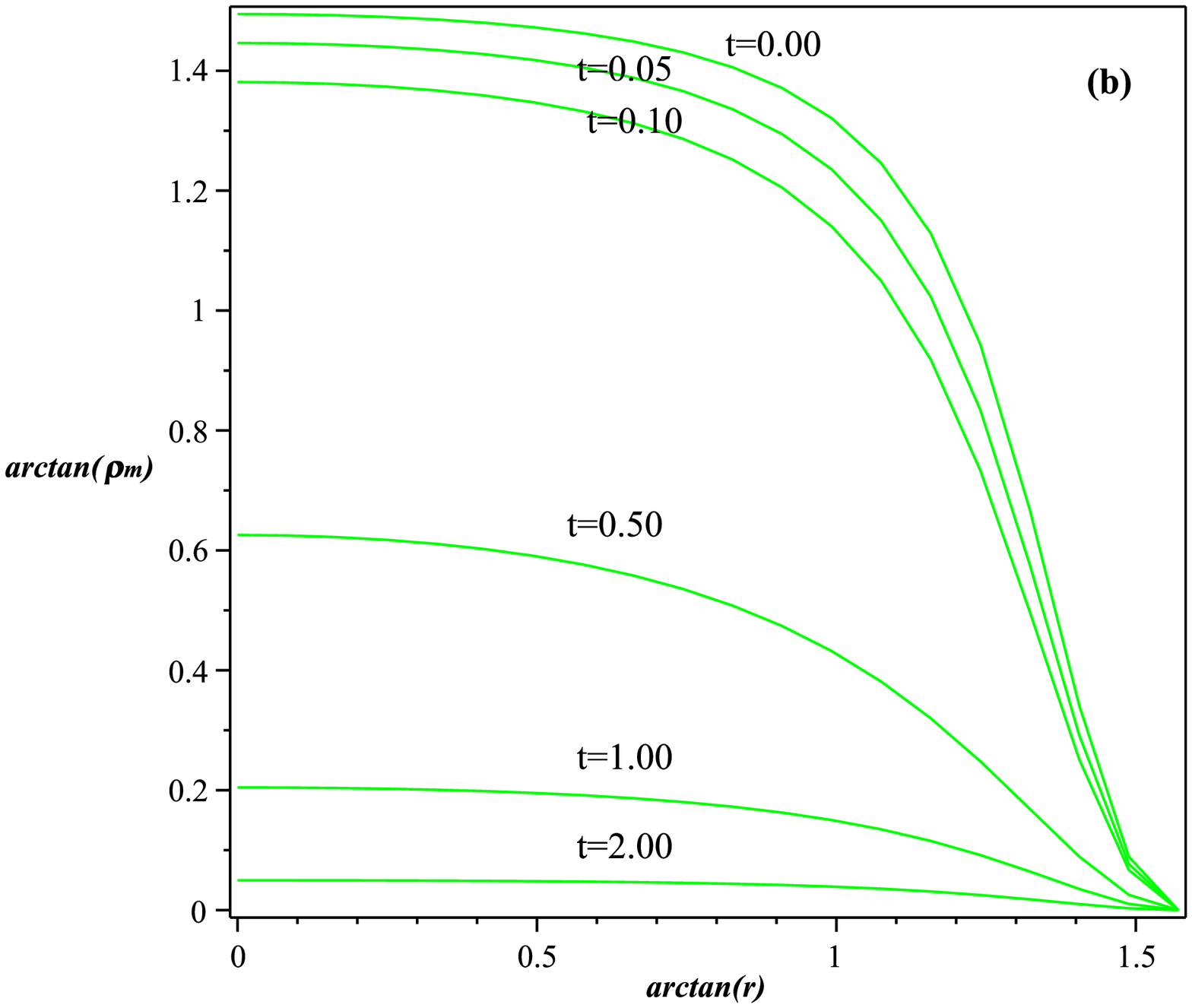}
\includegraphics*[scale=0.3]{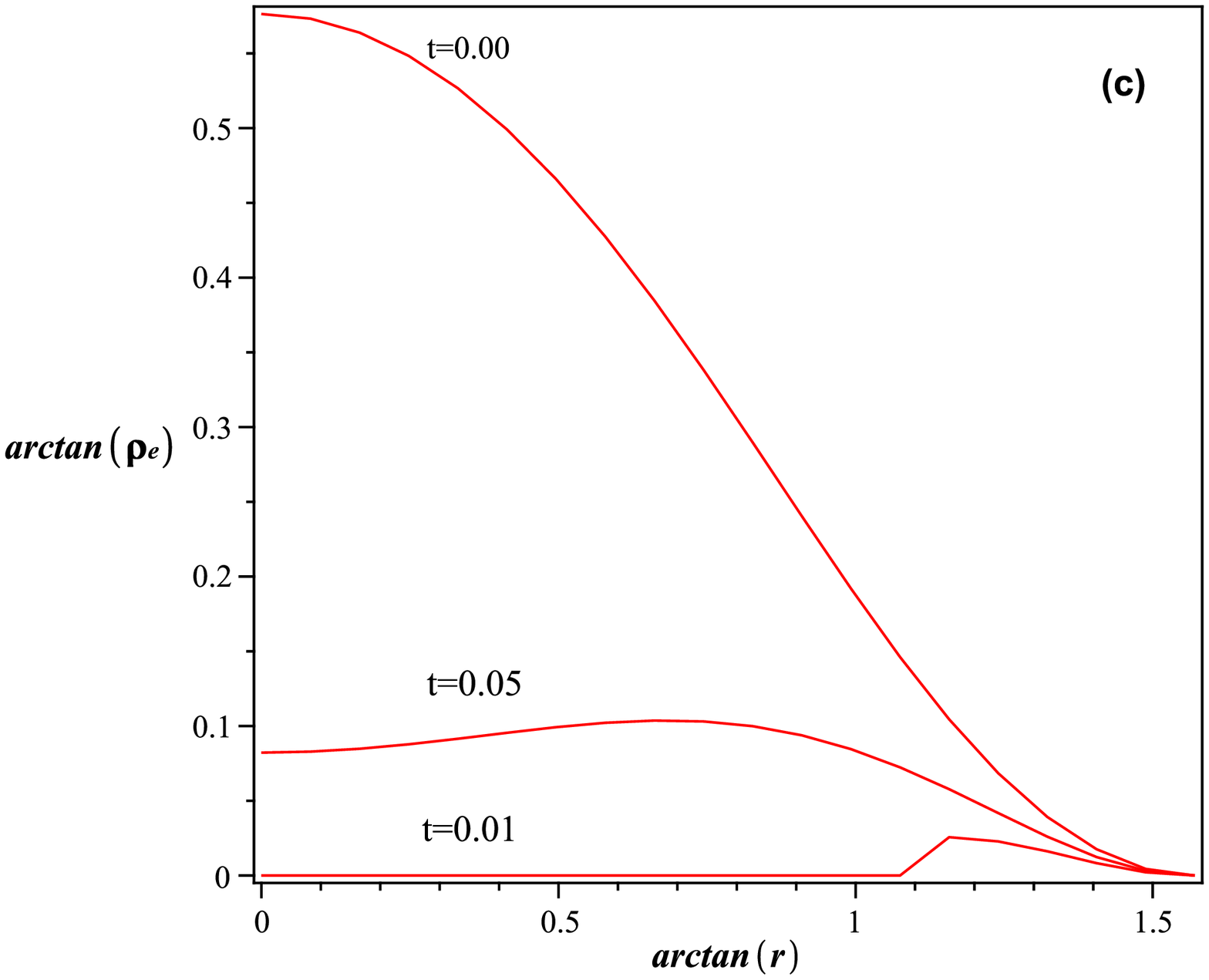}
\includegraphics*[scale=0.3]{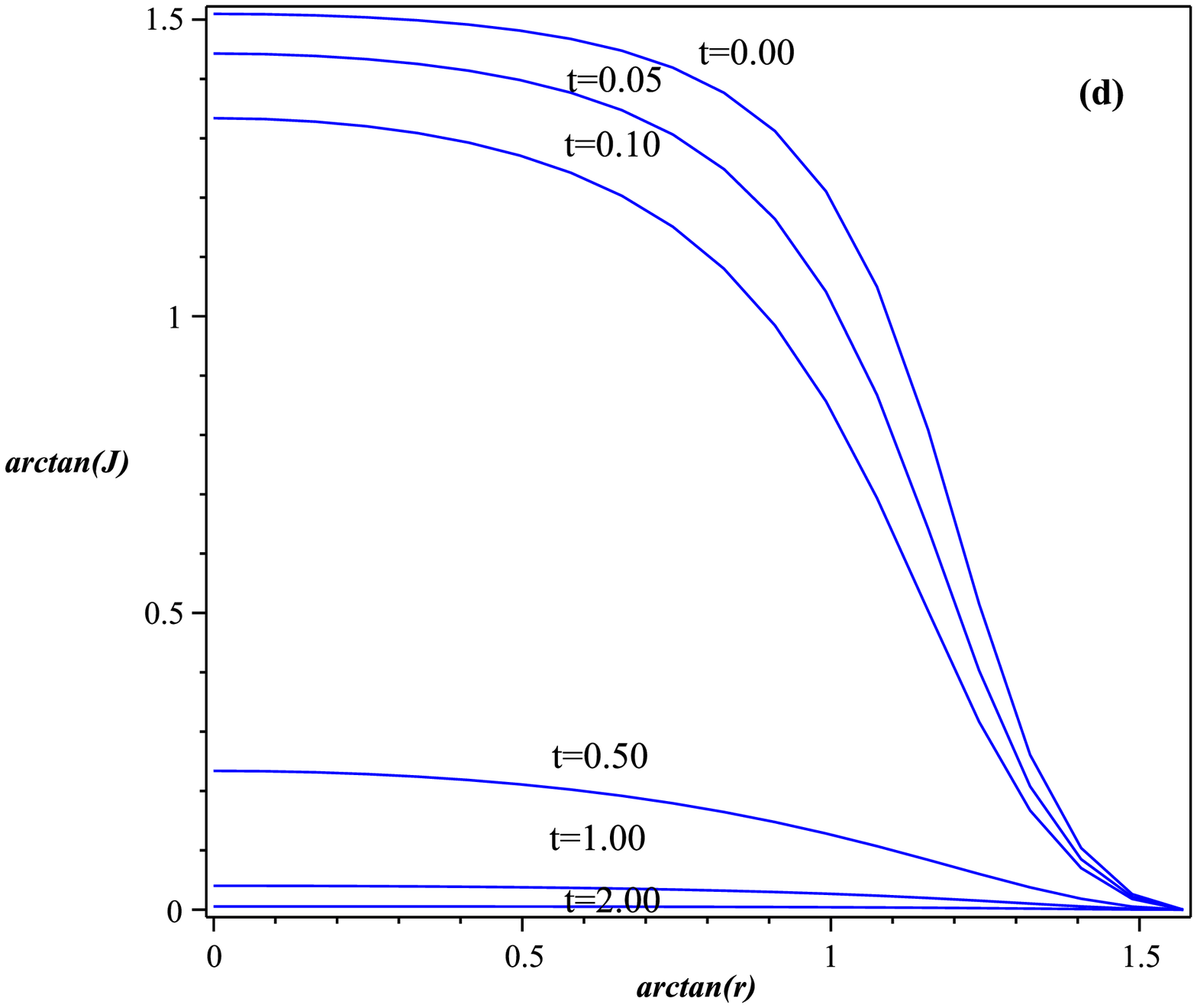}
\caption{Panel (a): Local profile of the scalar $\HH$ for different instants of time for the configuration with initial conditions given by (\ref{p15}) and $w=-1.0$, $\alpha=0.1$. Panel (b): Local profile of the scalar $\rho_m$ for different instants of time for the configuration with initial conditions given by (\ref{p15}) and $w=-1.0$, $\alpha=0.1$. Panel (c): Local profile of the scalar $\rho_e$ for different instants of time for the configuration with initial conditions given by (\ref{p15}) and $w=-1$, $\alpha=0.1$. Panel (d): Local profile of the scalar $J$ for different instants of time for the configuration with initial conditions given by (\ref{p15}) and $w=-1$, $\alpha=0.1$.Refer to the text for a detailed discussion of the panels.} \label{fig6}\end{figure}

The evolution of the local profiles of $\HH$, $\rho_m$, $\rho_e$ and $J$ are plotted in panels (a), (b), (c) and (d) of figure \ref{fig6}, respectively, for different instants of time. Note that the CDE is consumed at a very fast rate in the different shells and, at the instant $t=0.10$, no CDE is present.

In this scenario, changing the values of the free parameters will alter the slope of the invariant line. Thus, the initial conditions of some shells will be over the invariant line and evolving to the future attractor while other shells still evolve to the $\Ome=0$ axis. In the next subsection two of those mixed configurations are studied.

\subsection{Mixed configurations: CDM sphere sorrounded by a mixture of CDM and CDE.}
In the next two examples, we chose initial profiles so that the inner shells evolve to a pure dust evolution when the CDE is consumed, while the external shells evolve to the future attractor. For the former shells, we follow the same procedure as in the previous subsection. In the first example a shell crossing singularity is found during the pure dust evolution stage of the inner shells, as some shells bounce while the neighbor shells keep its expanding configuration.
\subsubsection{Configuration leading to a shell cross singularity.}
This configuration corresponds to $w=-1$ and $\alpha=0.1$ and
\ba
\rho_{m\,in}&=&{ m_{10}}+{\frac {{ m_{11}}-{ m_{10}}}{1+{r}^{3}}},{ m_{10}
}= 0.0,{ m_{11}}= 15.3;\nonumber\\
\rho_{e\,in}&=&0.7; \label{p7}\\
\KK_{in}&=&k_{10}+\frac{k_{11}-k_{10}}{1+r^4}, k_{10}=+1.2, k_{11}=-0.1.\nonumber
\ea
The scalar $R_{in}(r)=r$. For the numerical work we assume that $0<r<2$, and we do a partition of $n=20$ (then $r_j=j\cdot0.1$ with $j\in[0,20]$). We compute the evolution of the system (\ref{sistdinint1a}-\ref{sistdinint1e}). The choice of $r$ is totally arbitrary.

\begin{figure}[tbp]
\includegraphics*[scale=0.30]{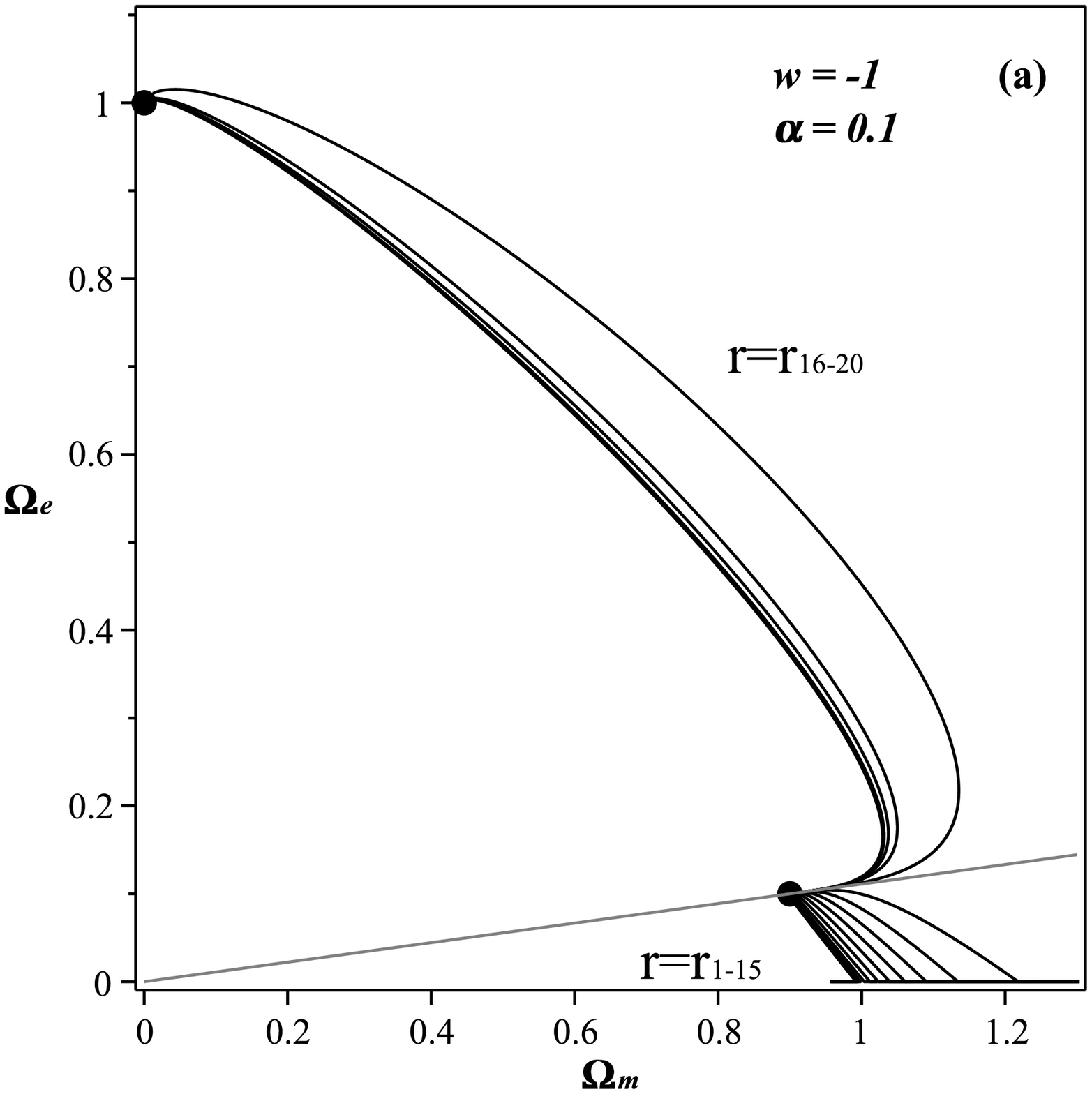}
\includegraphics*[scale=0.4]{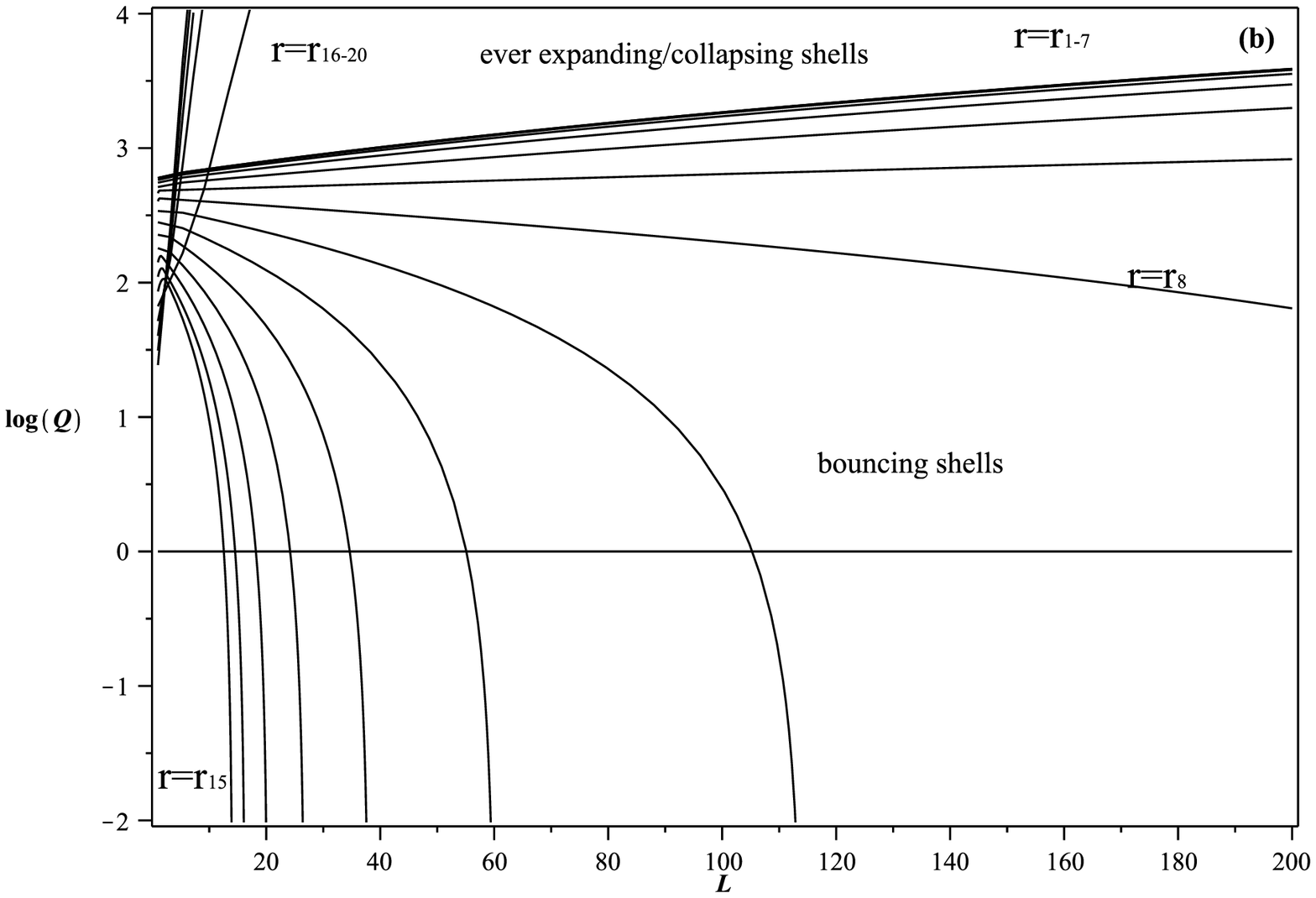}

\caption{Panel (a): Homogeneous projection of the trajectories of the system (\ref{sistdinint1a}-\ref{sistdinint1e}) for the different shells of the configuration with initial conditions given by (\ref{p7}) and $w=-1$, $\alpha=0.1$. Panel(b): Evolution of $\log(Q(L))$ vs. $L$ for the different shells of the configuration with initial conditions given by (\ref{p7}) and $w=-1$, $\alpha=0.1$. Refer to the text for a detailed discussion of the panels.} \label{fig7}\end{figure}

Panel (a)of fig. \ref{fig7} shows the homogeneous projection of the trajectories of every shell. The critical points and the invariant line are also represented. The inner shells $r=r_j$ with $j\in[0,15]$ evolve to the $\Omm$ axis, the outer shells $r=r_j$ with $j=[16,20]$ evolve to the future attractor. In this sense, we can assume that a sphere of pure CDM surrounded by a mixture of CDM and CDE is formed. In panel (b)of fig. \ref{fig7}, the scalar $Q(L)$ is computed for the different shells. The inner shells with $with r=r_{8-15}$ experiment a bounce at different instants of time while the inner shells with $r=r_{1-7}$ and the outer shells expand/collapse forever. The first shell to experiment a collapse is the $r=r_{15}=1.5$ one, followed by the $r=r_{14}=1.4$ shell, etc. With this in mind, we conclude that this configuration leads to a shell-cross singularity at the instant the shell $r=r_{15}=1.5$ bounces while the shells next to it still experiment expansion/contraction.

\begin{figure}[tbp]
\includegraphics*[scale=0.30]{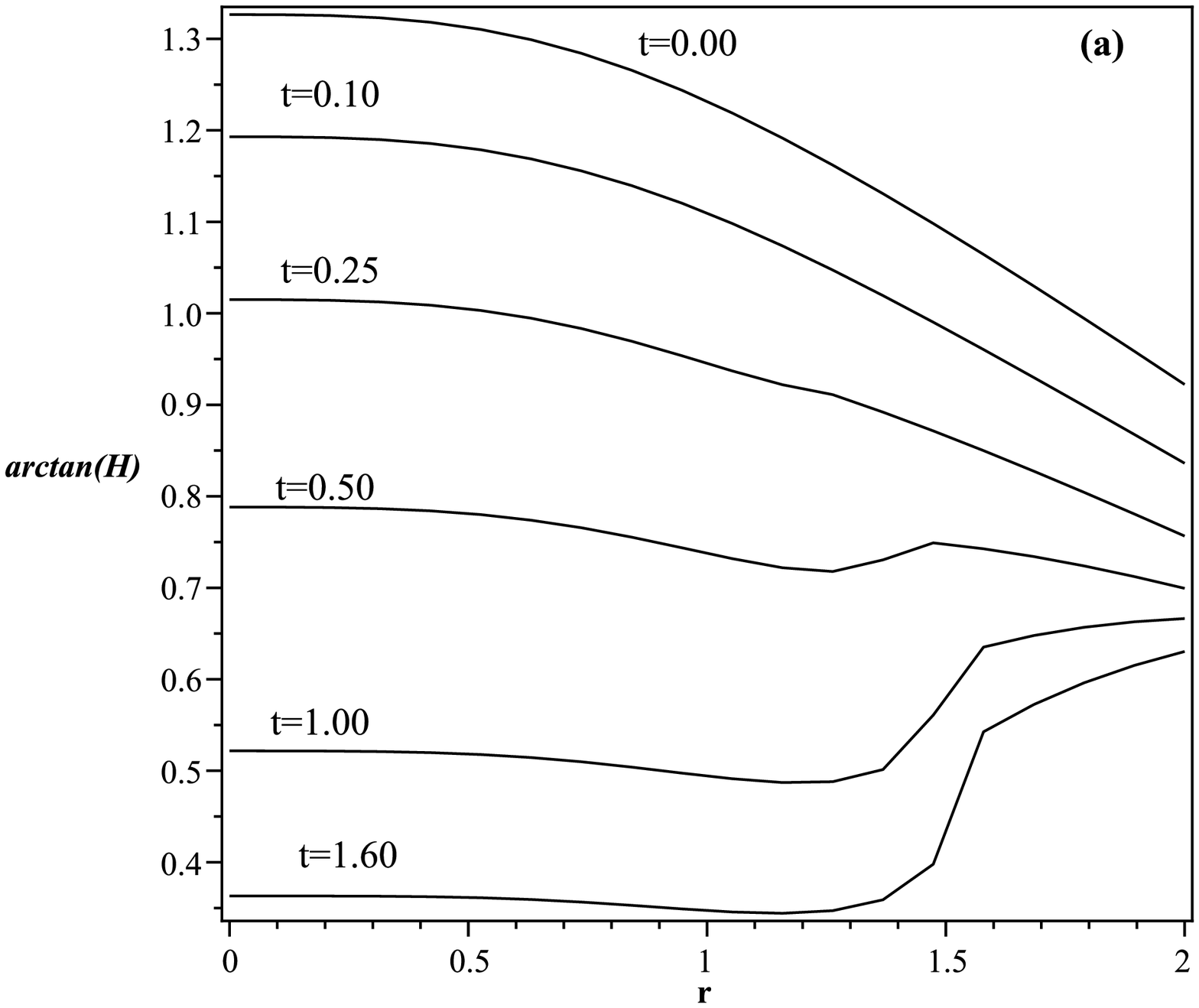}
\includegraphics*[scale=0.3]{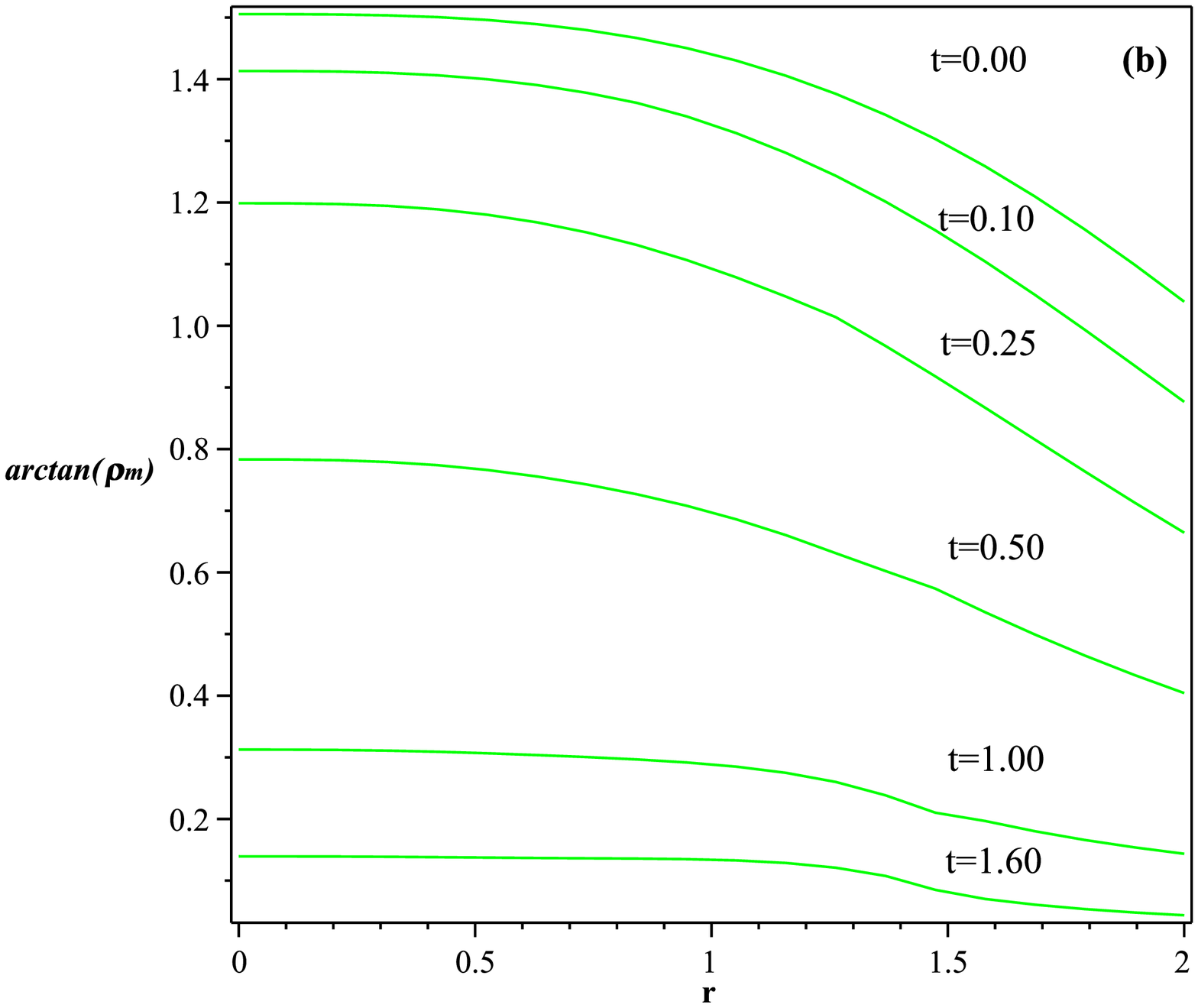}
\includegraphics*[scale=0.3]{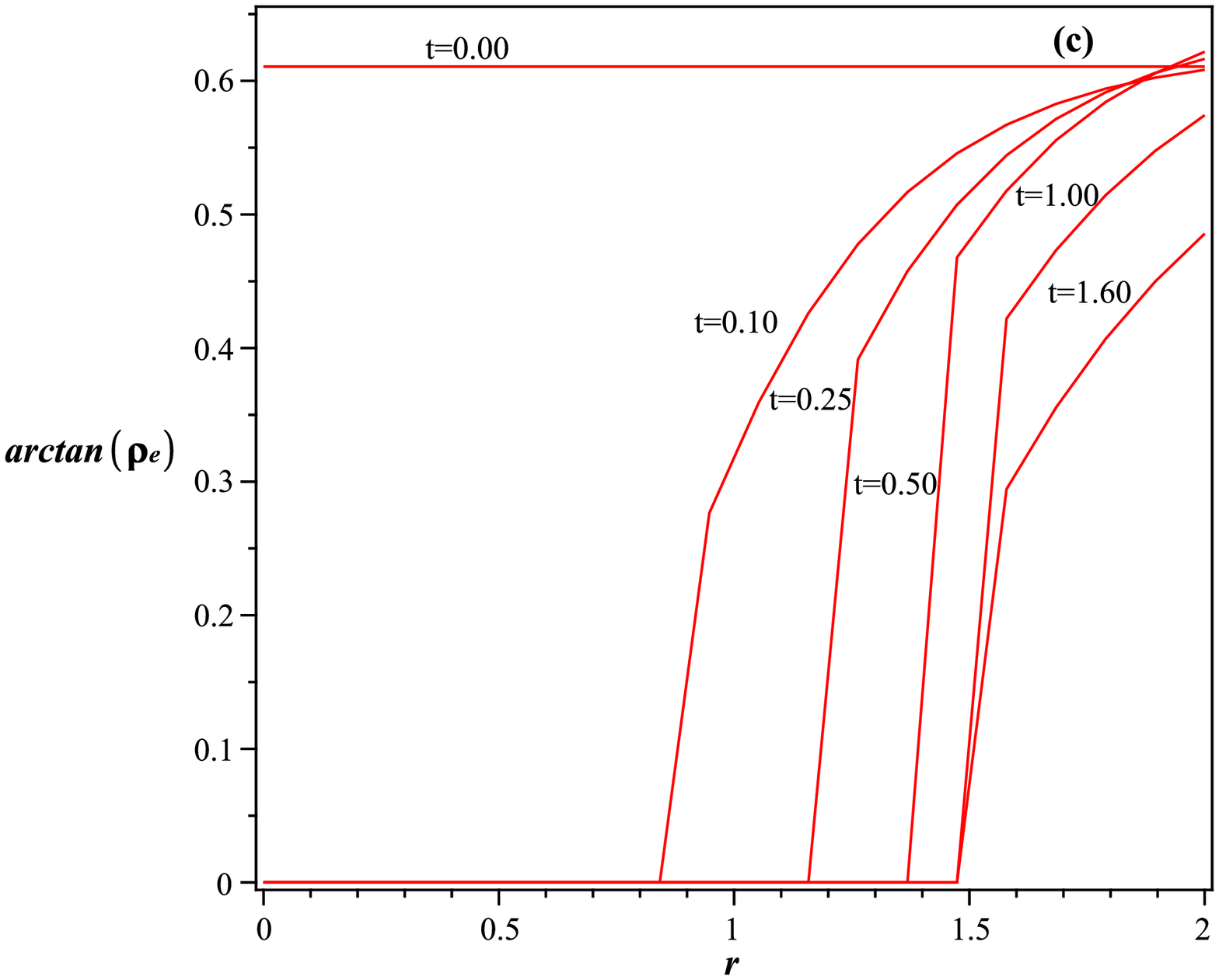}
\includegraphics*[scale=0.3]{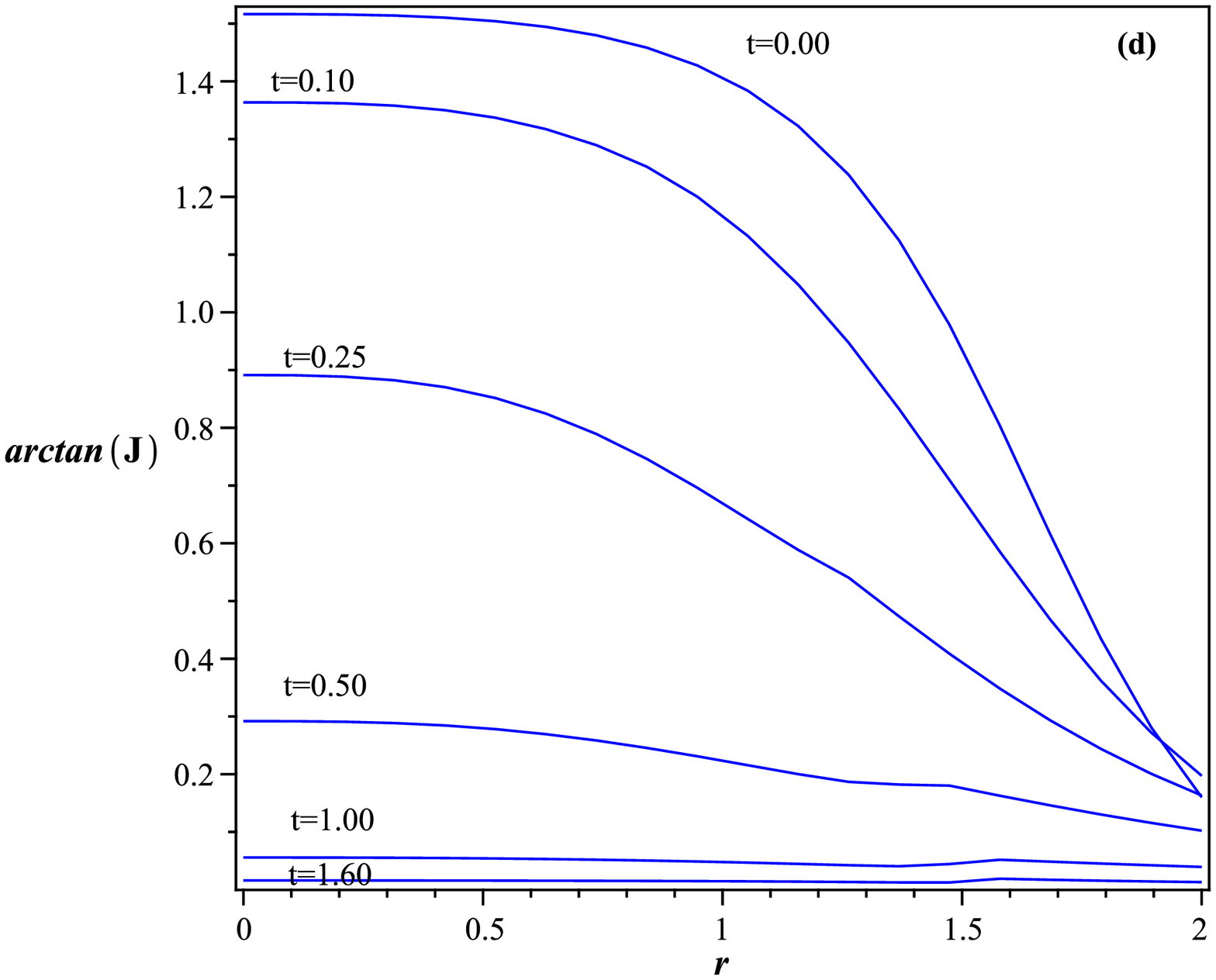}
\caption{Panel (a): Local profile of the scalar $\HH$ for different instants of time for the configuration with initial conditions given by (\ref{p7}) and $w=-1$, $\alpha=0.1$. Panel (b): Local profile of the scalar $\rho_m$ for different instants of time for the configuration with initial conditions given by (\ref{p7}) and $w=-1$, $\alpha=0.1$. Panel (c): Local profile of the scalar $\rho_e$ for different instants of time for the configuration with initial conditions given by (\ref{p7}) and $w=-1$, $\alpha=0.1$. Panel (c): Local profile of the scalar $J$ for different instants of time for the configuration with initial conditions given by (\ref{p7}) and $w=-1$, $\alpha=0.1$. Refer to the text for a detailed discussion of the panels.} \label{fig8}\end{figure}

In figure \ref{fig8}, the evolution of the local profiles of $\HH$, $\rho_m$, $\rho_e$ and $J$ are plotted in panels (a), (b), (c), and (d) respectively. In the panel (a), we can appreciate how the scalar $\HH$ of the shell $r_{15}=1.5$ tends to zero with time while the neighbour shells remain with a positive value. At any instant $t>1.6$ the shell $r_{15}=1.5$ will bounce while the other shells still experiment expansion/collapse, leading to a shell-cross singularity. In panel (c), we can appreciate how the inner shells consume the CDE. From the initial instant to the instant $t=0.1$, the shells with $r<0.9$ have consumed their CDE, and from the instant $t=0.1$ to the $t=1$ the following shells are consume the CDE until the only shells that keep its CDE are the outer shells with $r\geq 1.6$. In this sense we can conclude that a pure CDM sphere is formed surrounded by a background with a mixture of CDM and CDE before the shell-crossing singularity occurs.

Changing the values of the free parameters $w$ and $\alpha$ in this configuration will change the position of the past attractor and the slope of the invariant line. The number of inner shells that evolve to the $\Omm$ axis and bounce is changed as well (e.g., assuming $w=-0.90$ and $\alpha=0.1$ the shells $r=r_{16,17}$ also consume the initial CDE and bounce at a certain instant, while, for $w-1.10$ and $\alpha=0.1$, the shells $r_{14-15}$ evolve to the future attractor and expand/collapse forever).

\subsubsection{Configuration ever expanding.}
In this configuration we set the free parameters as $w=-1$ and $\alpha=0.1$ and  the initial local profiles
\ba
\rho_{m\,in}&=&{ m_{10}}+{\frac {{ m_{11}}-{ m_{10}}}{1+{r}^{3}}},{ m_{10}
}= 0.0,{ m_{11}}= 15.3;\nonumber\\
\rho_{e\,in}&=&0.7;\label{p2}\\
\KK_{in}&=&k_{10}+\frac{k_{11}-k_{10}}{1+r^4}, k_{10}=-1.2, k_{11}=-0.1;\nonumber
\ea
with the scalar function $R_{in}(r)=r$. For the numerical work we assume that $0<r<2$, and we do a partition of $n=20$ (then $r_j=j\cdot0.1$ with $j\in[0,20]$). This configuration is very similar to the previous one but with negative initial curvature.

\begin{figure}[tbp]
\includegraphics*[scale=0.30]{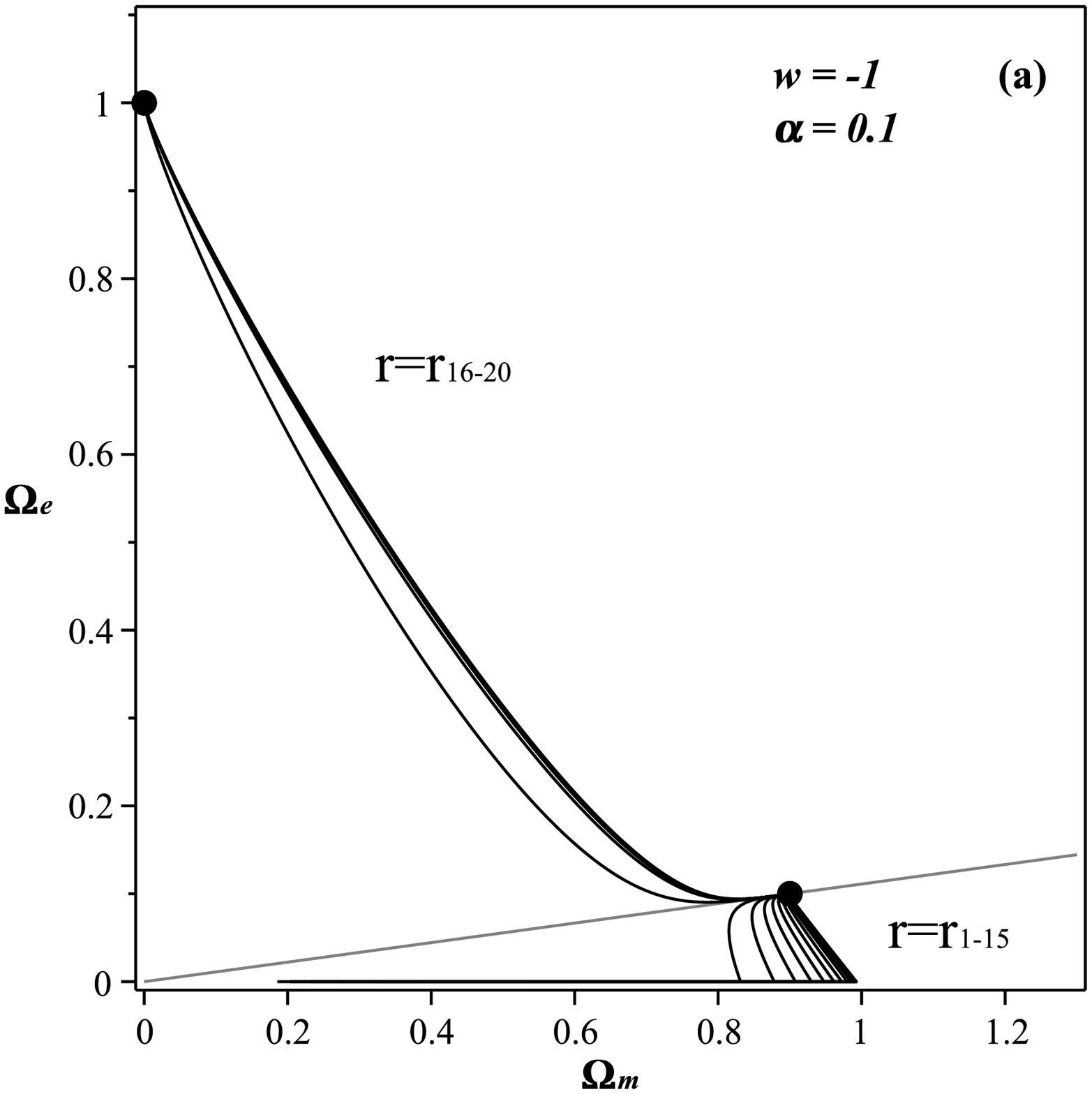}
\includegraphics*[scale=0.4]{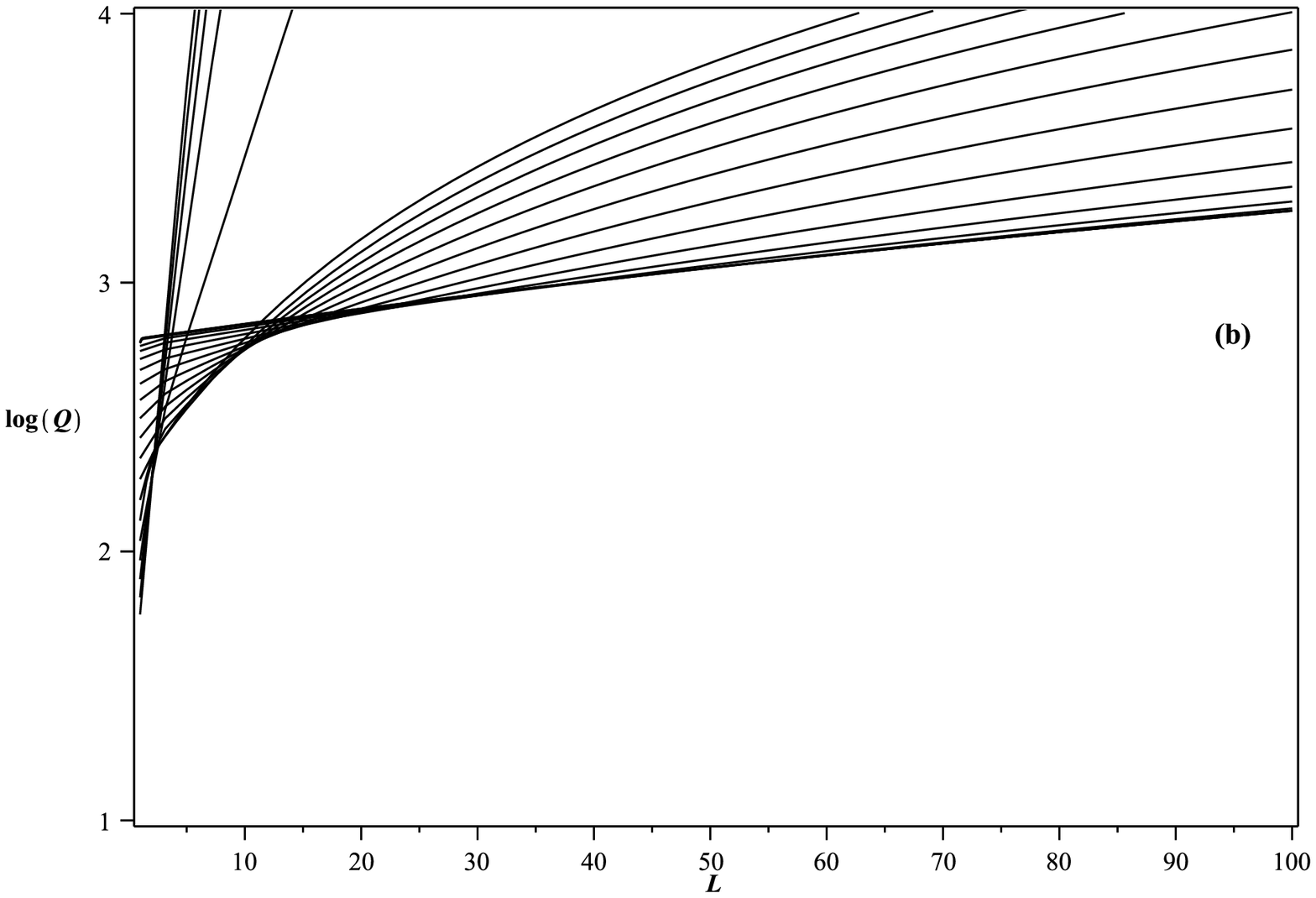}

\caption{Panel (a): Homogeneous projection of the trajectories of the system (\ref{sistdinint1a}-\ref{sistdinint1e}) for the different shells of the configuration with initial conditions given by (\ref{p2}). Panel(b): Evolution of $\log(Q(L))$ vs. $L$ for the different shells of the configuration with initial conditions given by (\ref{p2}). Refer to the text for a detailed discussion of the panels.} \label{fig9}\end{figure}

Panel (a)of fig. \ref{fig9} shows the homogeneous projection of the trajectories of every shell. The critical points and the invariant line are also represented. The inner shells $r=r_j$ with $j\in[0,15]$ evolve to the $\Omm$ axis, the outer shells $r=r_j$ with $j=[16,20]$ evolve to the future attractor. In panel (b)of fig. \ref{fig9}, the scalar $Q(L)$ is computed for the different shells. From it, we conclude that no shell experiment bouncing in this case and all the shells expand/collapse forever.

\begin{figure}[tbp]
\includegraphics*[scale=0.30]{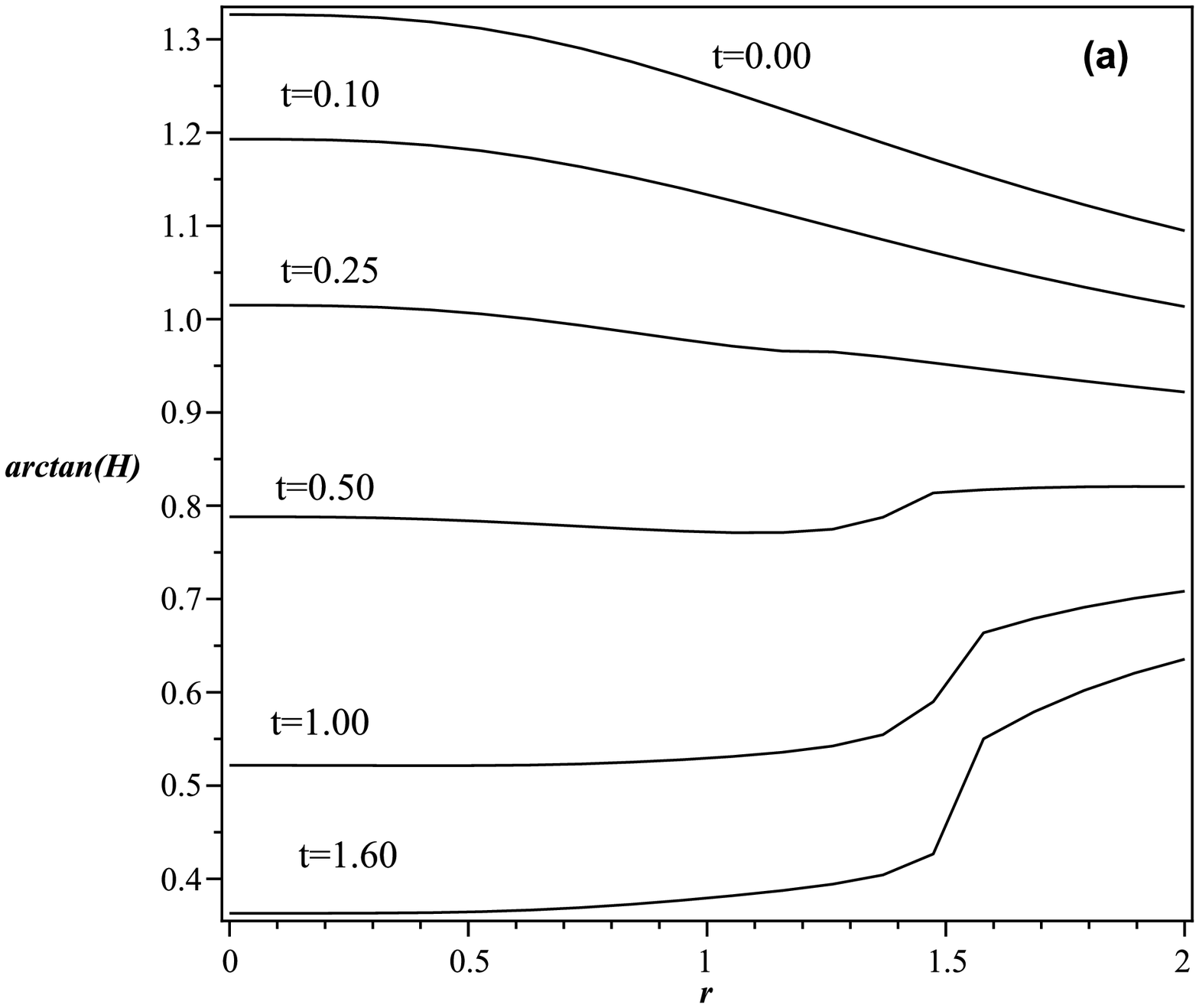}
\includegraphics*[scale=0.3]{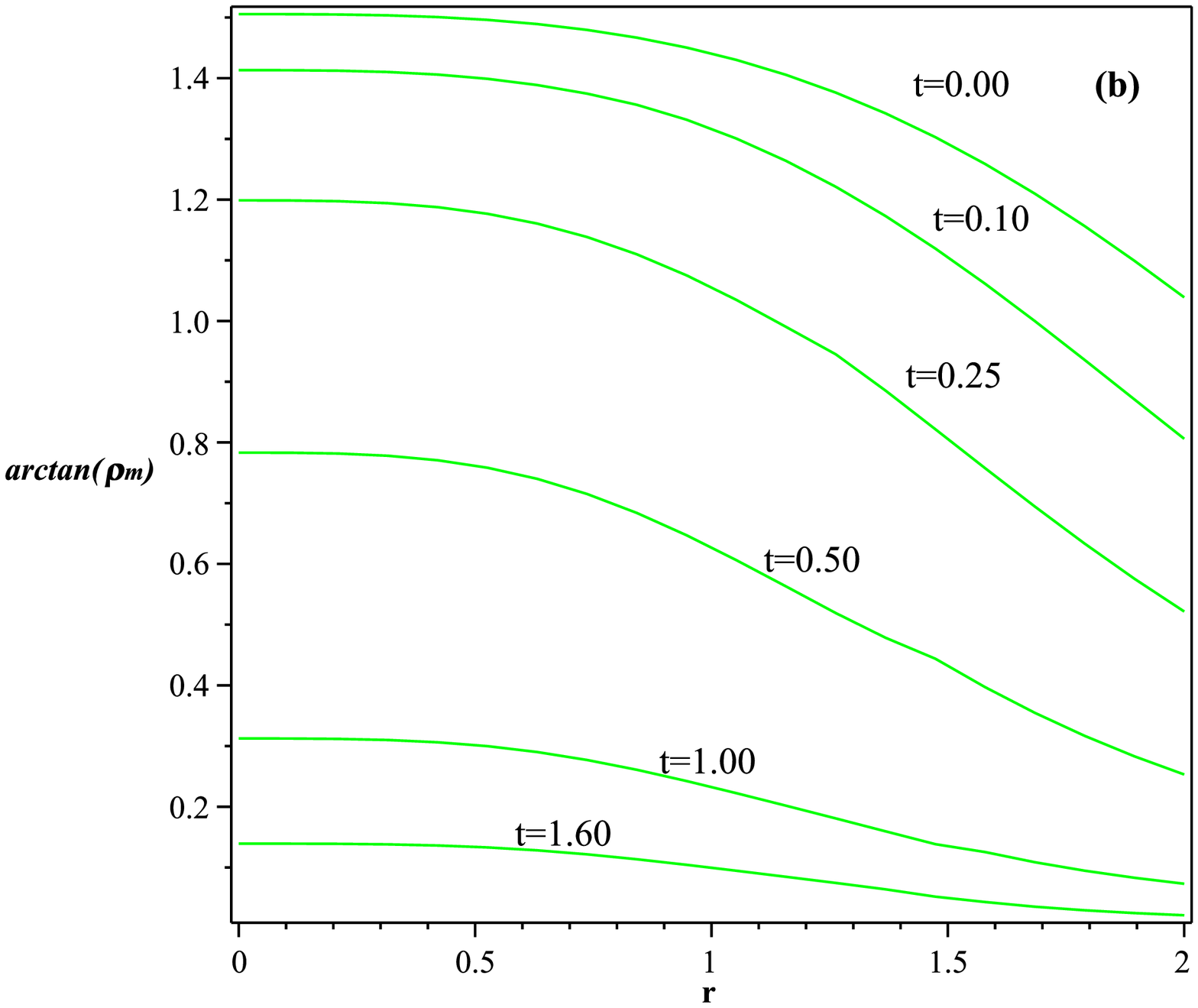}
\includegraphics*[scale=0.3]{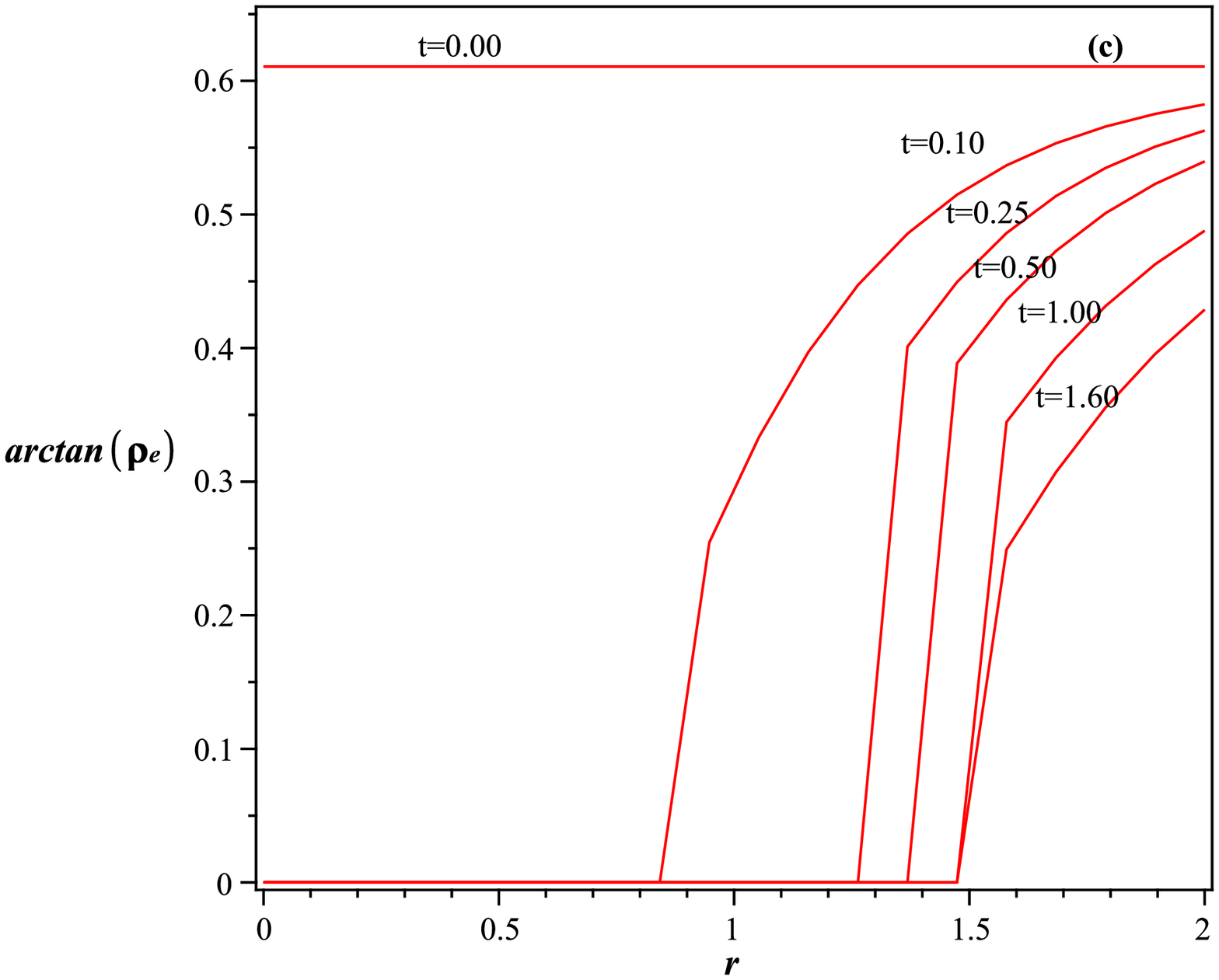}
\includegraphics*[scale=0.3]{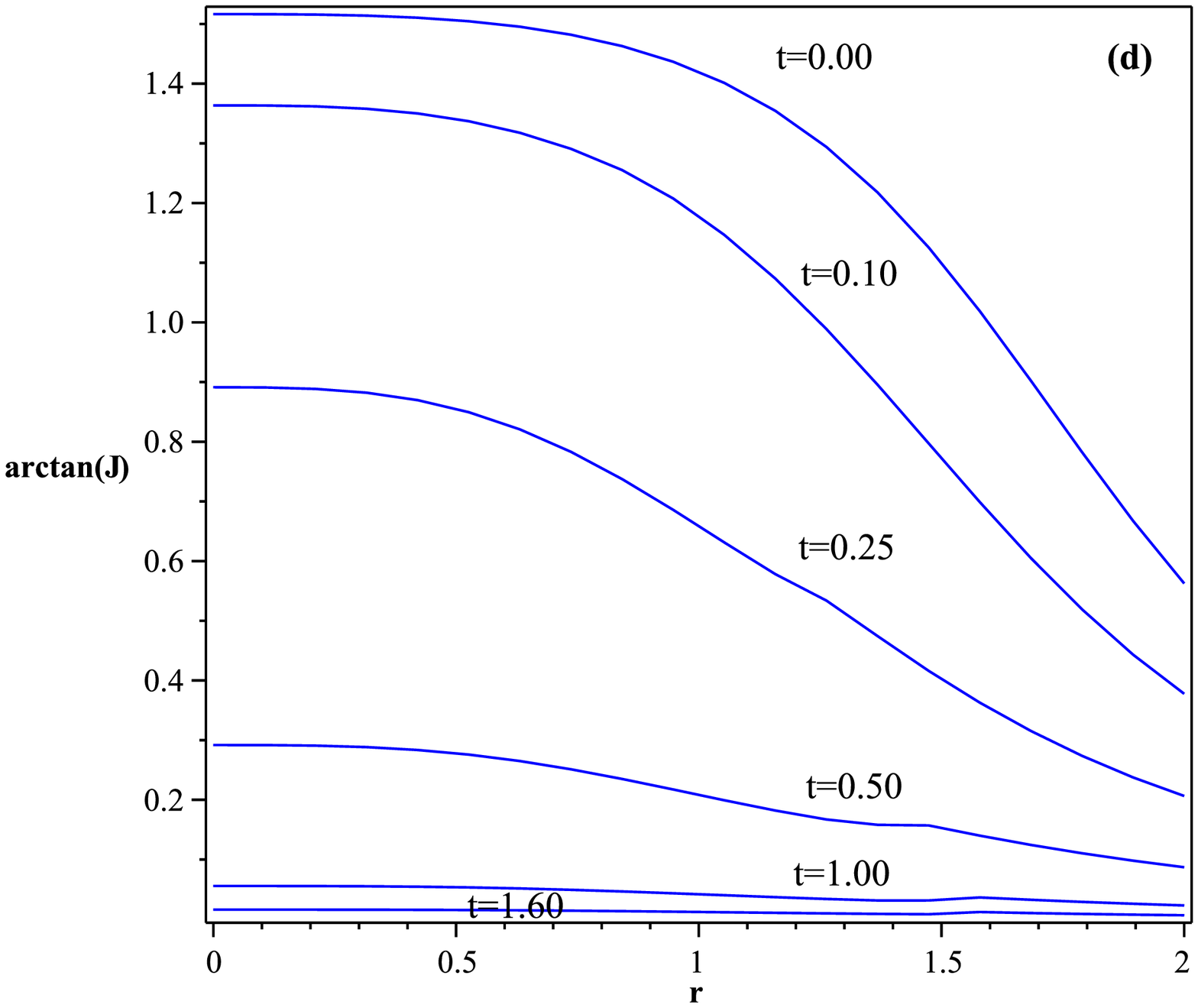}
\caption{Panel (a): Local profile of the scalar $\HH$ for different instants of time for the configuration with initial conditions given by (\ref{p2}) and $w=-1$, $\alpha=0.1$. Panel (b): Local profile of the scalar $\rho_m$ for different instants of time for the configuration with initial conditions given by (\ref{p2}) and $w=-1$, $\alpha=0.1$. Panel (c): Local profile of the scalar $\rho_e$ for different instants of time for the configuration with initial conditions given by (\ref{p2}) and $w=-1$, $\alpha=0.1$. Panel (d): Local profile of the scalar $J$ for different instants of time for the configuration with initial conditions given by (\ref{p2}) and $w=-1$, $\alpha=0.1$. Refer to the text for a detailed discussion of the panels.} \label{fig10}\end{figure}

In figure \ref{fig10}, the evolution of the local profiles of $\HH$, $\rho_m$, $\rho_e$ and $J$ are plotted in panels (a), (b), (c), and (d), respectively. Although the  $\rho_m$ and $\rho_e$ profiles are similar to the previous configuration, the $\HH$ profile is different specially in the inner shells. In this case, the scalar $\HH$ of the inner shells with $0.5<r<1.5$ decreases ar a slower rate than in the previous example. As a result of this behavior, the $\HH$ profile at $t=1.6$ is a increasing function of $r$ while in the previous example $\HH$ presents a minimum at the shell $r=1.5$. In this case we can compute the profile at longer times than $t=1.6$ as no shell-crossing is found.

\subsection{Mixed configuration: Structure formation scenario.}
In this example $w=-1.0$ and $\alpha=0.1$, the initial local profiles read
\ba
\rho_{m\,in}&=&{ m_{10}}+{\frac {{ m_{11}}-{ m_{10}}}{1+\tan(r)^{2}}},{ m_{10}
}= 0.00,{ m_{11}}= 13.10; \nonumber\\
\rho_{e\,in}&=&{ e_{10}}+{\frac {{ e_{11}}-{ e_{10}}}{1+\tan(r)^{2}}},{ e_{10}
}= 0.00,{ e_{11}}= 1.47;\label{p14}\\
\KK_{in}&=&k_{10}+\frac{k_{11}-k_{10}}{1+\tan(r)^2}, k_{10}=-4.10, k_{11}=-5.50;\nonumber
\ea
and the scalar $R_{in}(r)=\tan(r)$. The variable $r$ goes from $0$ to $\pi/2$ and we made also a partition of $n=20$. We assume that the configuration is initially in expansion.

\begin{figure}[tbp]
\includegraphics*[scale=0.30]{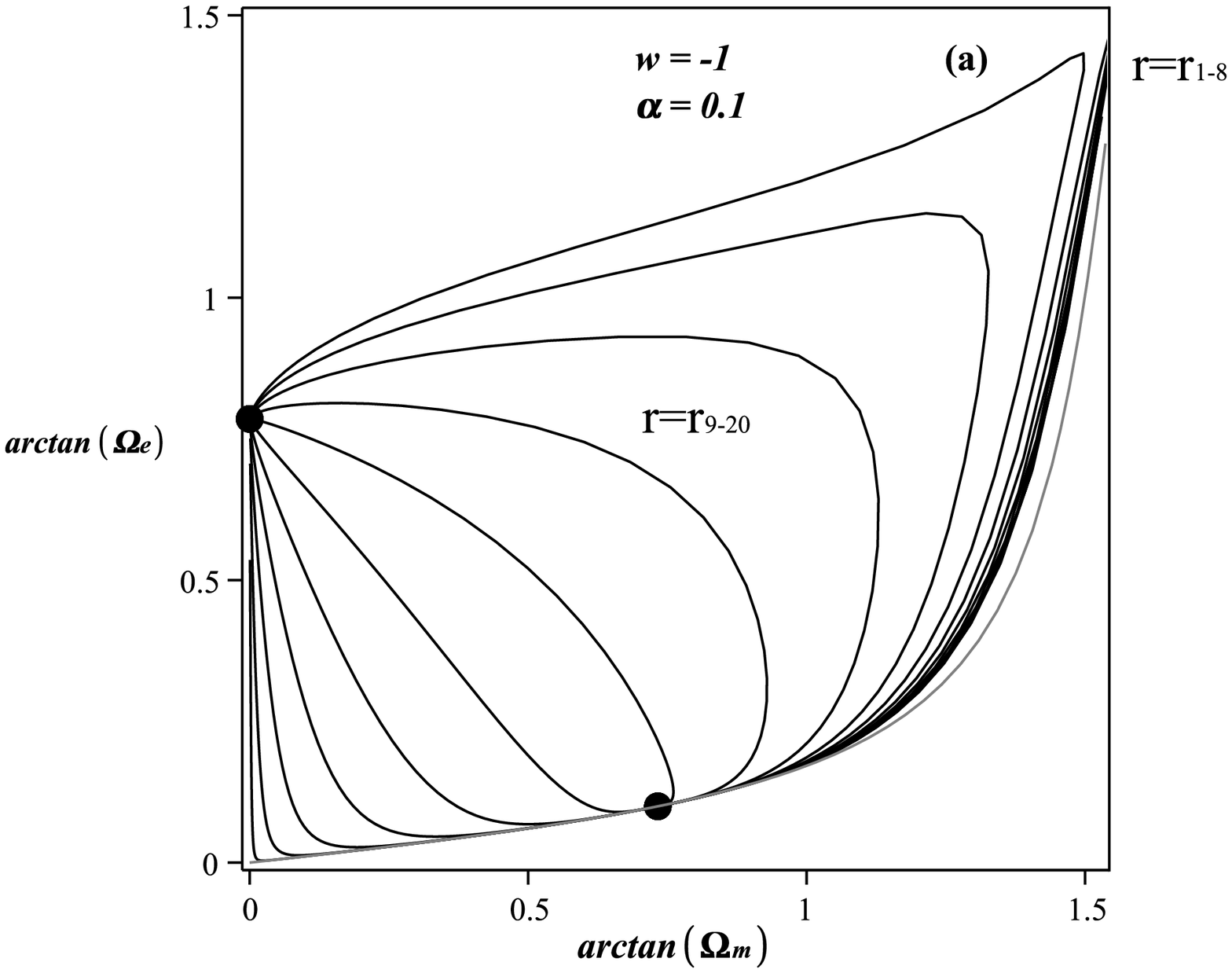}
\includegraphics*[scale=0.4]{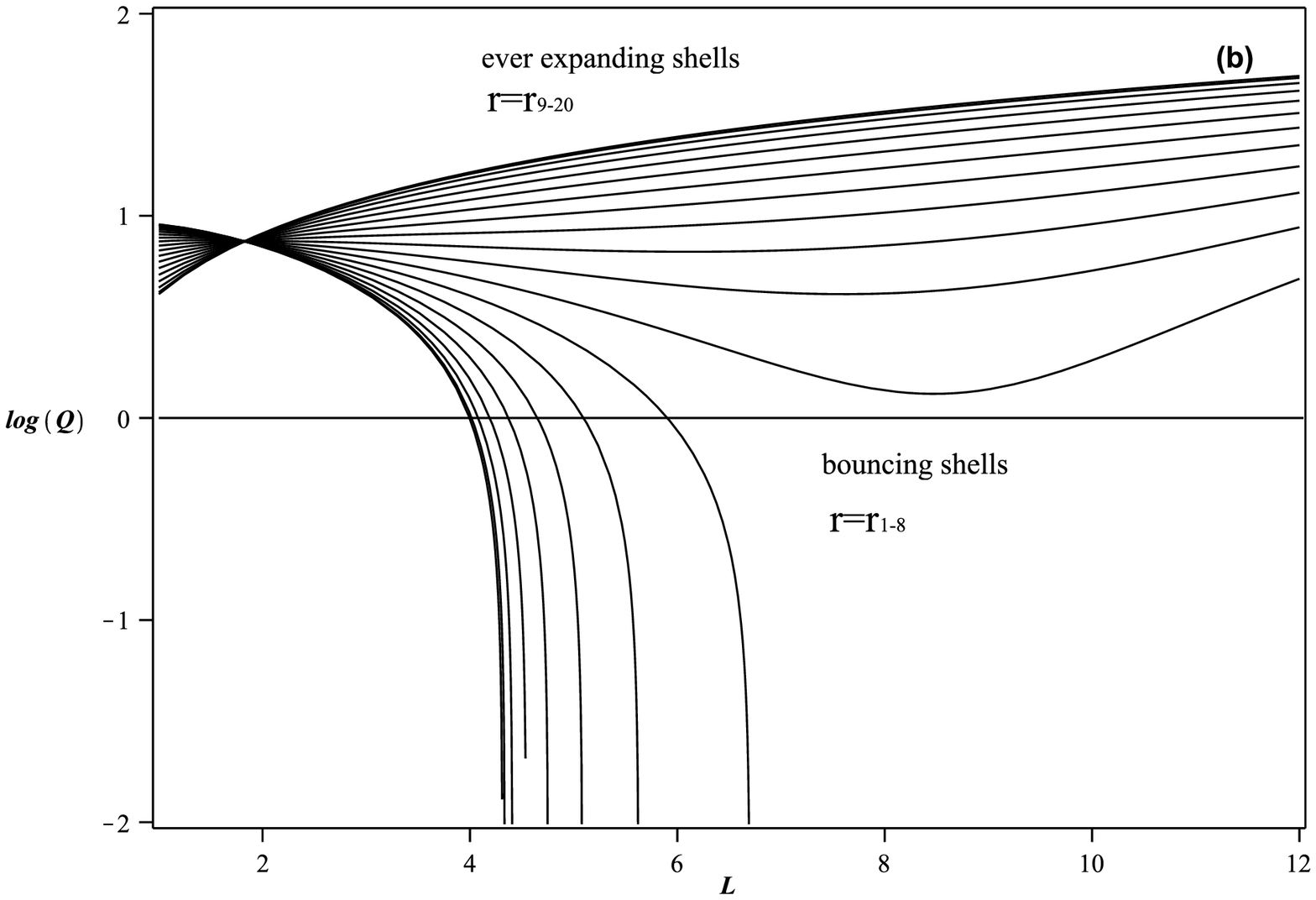}

\caption{Panel (a): Homogeneous projection of the trajectories of the system (\ref{sistdinint1a}-\ref{sistdinint1e}) for the different shells of the configuration with initial conditions given by (\ref{p14}). For this plot we have used $\arctan(\Ome)$ vs. $\arctan(\Omm)$ plot, as the shells $r_{1-8}$ do evolve to infinity. The invariant line is plotted as a grey curve as well. Panel(b): Evolution of $\log(Q(L))$ vs. $L$ for the different shells of the configuration with initial conditions given by (\ref{p14}). Refer to the text for a detailed discussion of the panels.} \label{fig11}\end{figure}

From panel (a) of figure \ref{fig11}, we notice that the inner shells $r_{1-8}$ have initial conditions over the invariant line but their homogeneous trajectories in the phase-space evolve to infinity, while the rest of the shells have initial conditions in the attraction basin of $PCA$. Additionally, from panel (b) of figure \ref{fig11}, we also know that the inner shells experiment a bounce: first the $r_1$ shell collapses, followed by the $r_2$ shell, etc. The outer shells on the other hand keep their expanding behavior. In this sense, we can conclude that this configuration leads to a structure formation toy model scenario as the inner shells collapse with no shell crossing while the outer shells can be interpreted as a expanding background. All the shells evolve to $PCR$ in the far past, that can be considered the 'big bang' instant in this structure formation toy model scenario.

\begin{figure}[tbp]
\includegraphics*[scale=0.30]{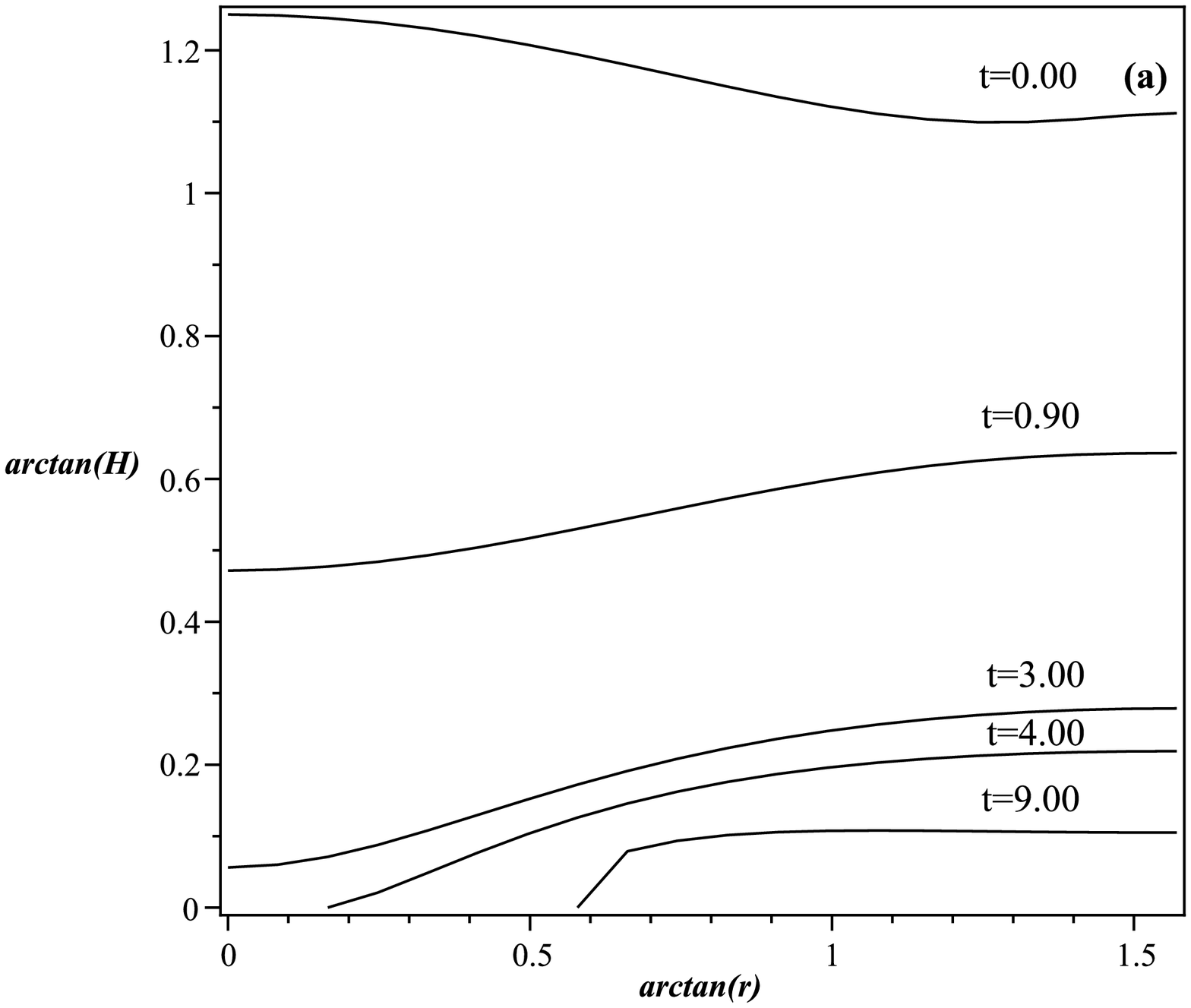}
\includegraphics*[scale=0.3]{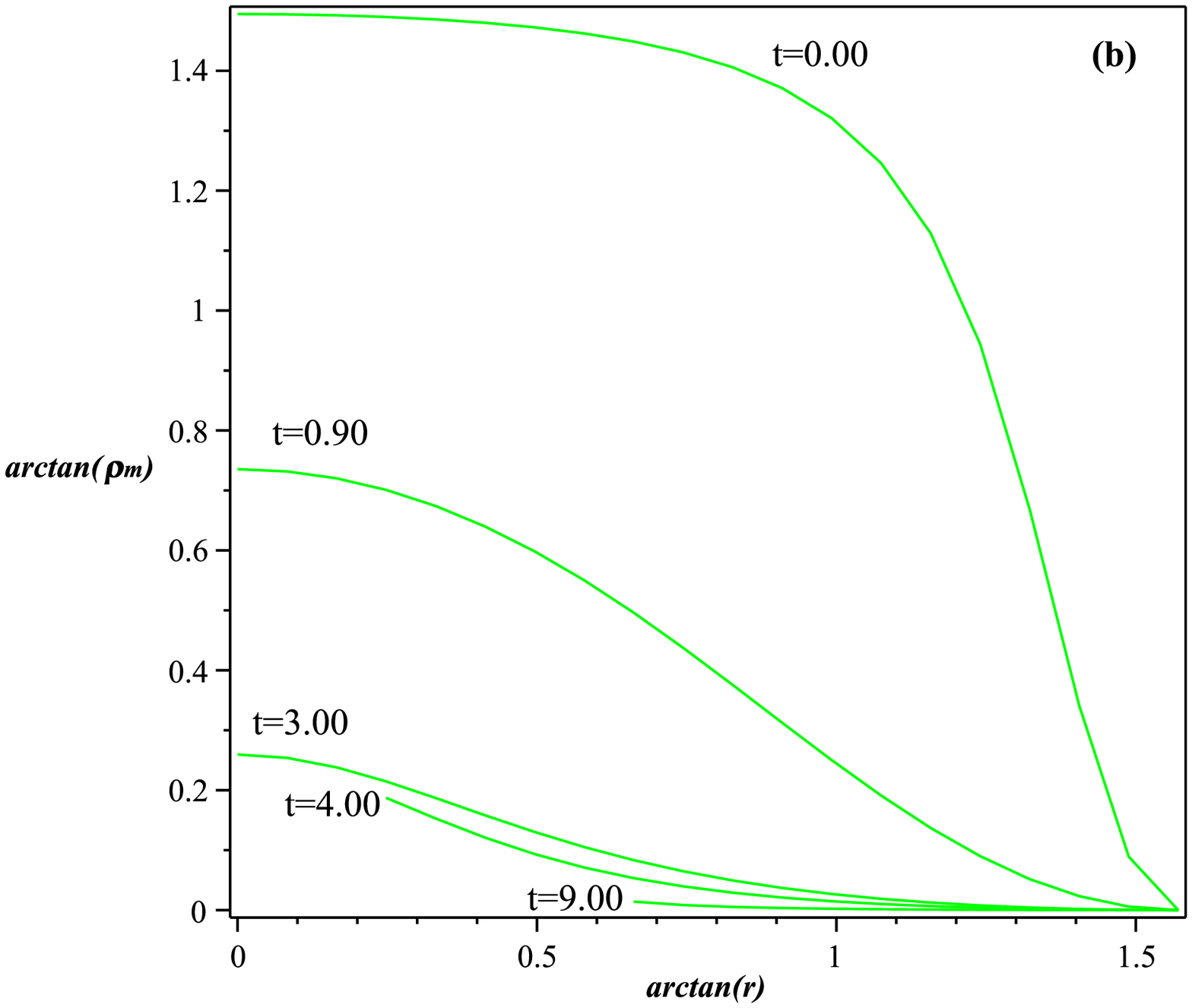}
\includegraphics*[scale=0.25]{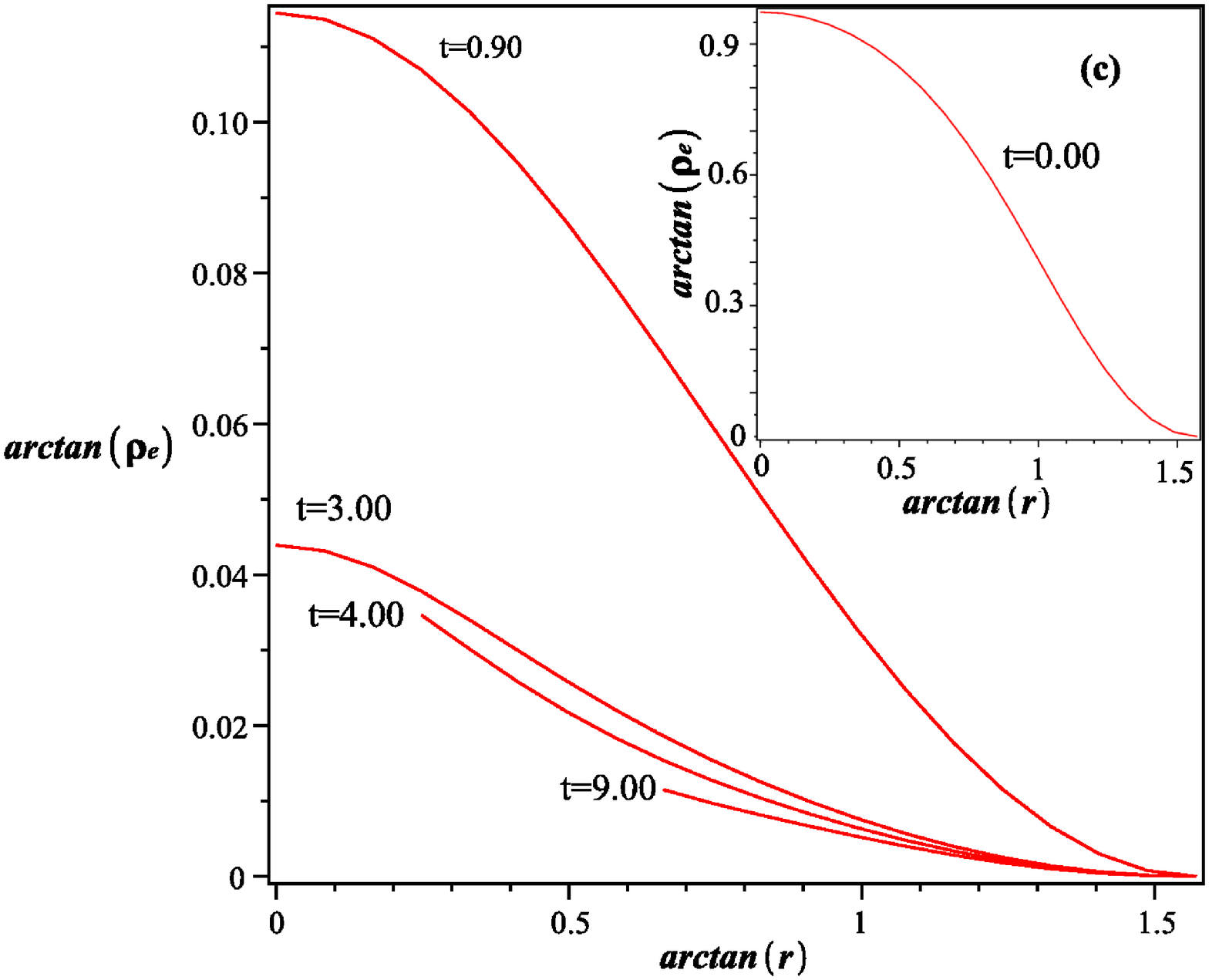}
\includegraphics*[scale=0.25]{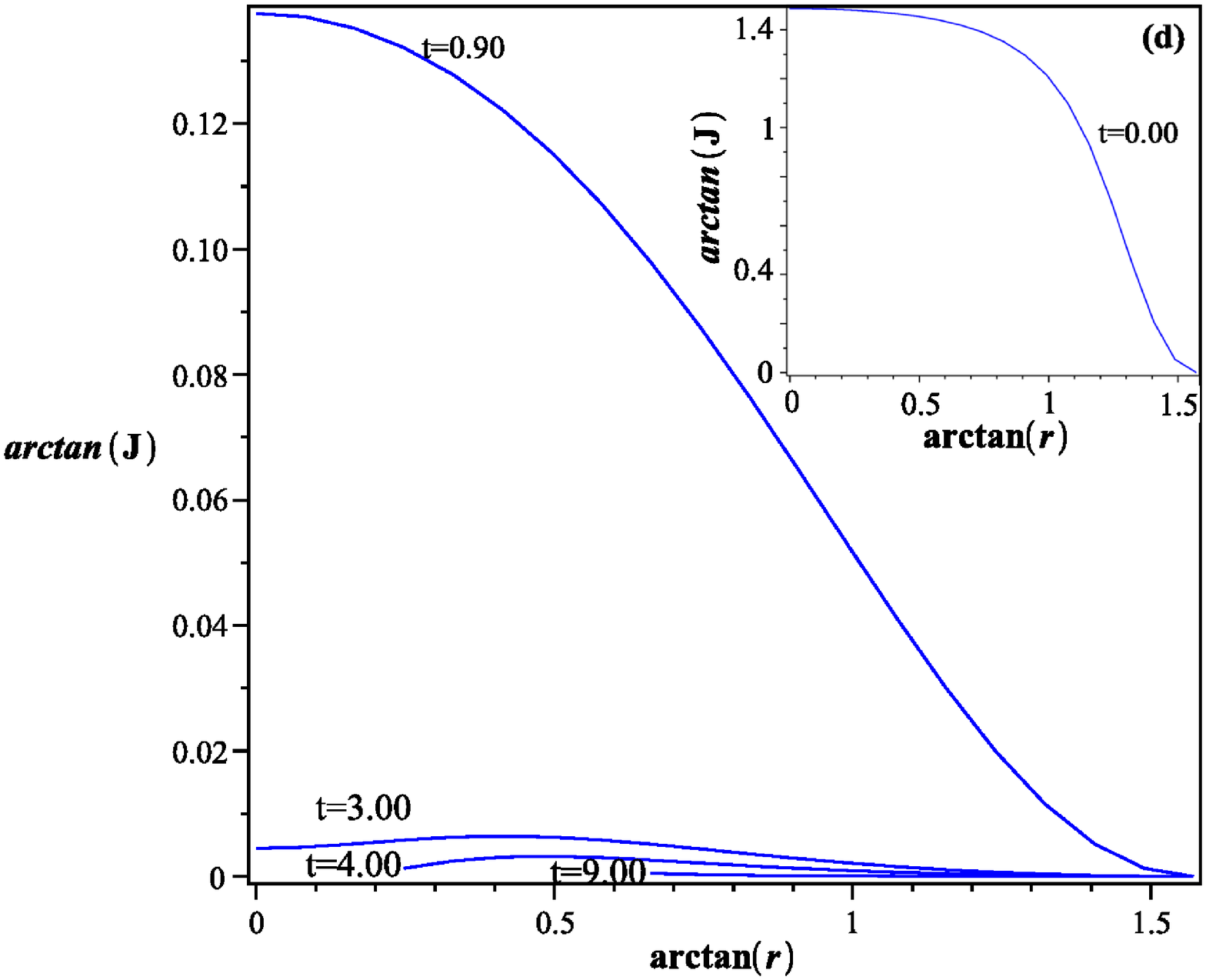}
\caption{Panel (a): Local profile of the scalar $\HH$ for different instants of time for the configuration with initial conditions given by (\ref{p14}) and $w=-1$, $\alpha=0.1$. Panel (b): Local profile of the scalar $\rho_m$ for different instants of time. Panel (c): Local profile of the scalar $\rho_e$ for different instants of time. Panel (d): Local profile of the scalar $J$ for different instants of time.} \label{fig12}\end{figure}

The local profiles of $\HH$, $\rho_m$, $\rho_e$ and $J$ are plotted in panels (a), (b), (c) and (d) of figure \ref{fig12} at different instants of time. Note that at the instant $t=3.00$ no shell has collapsed yet. At the instant $t=4.00$ some of the inner shells have collapsed but the other inner shells are still experimenting expansion behavior. Finally, at the instant $t=9.00$ only the outer shells are expanding and all the inner shells have already collapsed. From panel (c), we can also appreciate that $\rho_e$ decreases very fast with the expansion due to the coupling term (CDE is transferring its energy to the CDM given that $\alpha>0$). When the collapse of the inner shells starts, there is still a non null CDE present, so we can conclude that the CDE collapses with the CDM.

Changing the free parameters in this scenario will lead to more shells collapsing (if the slope of the invariant line is increased), or, on the other hand, to more shells evolving to the future attractor (if the slope of the invariant line is decreased). It is possible to reach a point where no inner shells collapse and all of them evolve to $PCA$ following a behavior phenomenologically identical to that of the previous ever-expanding scenario of figure \ref{fig5}.

\section{Conclusions.}\label{conclusions}
In this work we have extended the results of the $\Lambda$--CDM LTB model in \cite{izsuss10} by changing the cosmological constant source for a CDE source (a perfect fluid with constant local equation of state $p_e=w \rho_e$ coupled to the CDM through a local energy-momentum flux $J$). The former model is then a particular case of the CDE model for $w=-1$ and $J=0$. The Quasi-local scalars approach is used with the unknown scalar functions of the LTB metric and the evolution equations are transformed in a dynamical system of five autonomous non-linear first order time derivative differential equations at every shell of constant $r$ and a constraint Hubble-like equation. As in the $\Lambda$--CDM LTB case, we chose to represent the five-dimensional phase-space by means of an homogeneous projection $\Ome vs \Omm$ and an inhomogeneous three-dimensional projection.

We have chose an energy-momentum flux whose QL counterpart follows a relation as eq. (\ref{intj_m}) where $\alpha$ is a constant dimensionless free parameter. Given that QL scalars can be related to FLWR cosmological functions, we made that choice for the flux in order to represent a well-known in the literature coupling term. The system of autonomous equation reads, then, (\ref{sistdinint1a}-\ref{sistdinint1e}). Equations (\ref{sistdinint1a}) and (\ref{sistdinint1b}) are not coupled to the rest and form the homogeneous subspace. The critical points of the system are studied and classified in terms of the free parameters $w$ and $\alpha$ in the table \ref{criticalpoints}. Similarly to the $\Lambda$--CDM LTB model and for any choice of the free parameters $w$ and $\alpha$, the system has a future attractor and a past attractor. The past attractor only has physical meaning for $\alpha>0$, making any $\alpha<0$ scenario not viable by itself. This fact has been known in the literature in the FLRW scenario \cite{boe08,GZun14}. Additionally, the dynamical system has five saddle points.

In the homogeneous projection for the $\alpha>0$ case, the past attractor $PCR$ is displaced with respect to that of the $\Lambda$--CDM model (the latter lays on the $\Ome=0$ axis, while the former depends on the free parameters). As a consequence of that, an invariant line can be defined as (\ref{invline}) on the homogeneous subspace that separate it in two regions: the shells with initial conditions that lay under the invariant line evolve to the $\Ome=0$ axis; and the shells with initial condition on or over the invariant line evolve to infinity or to the future atractor. The trajectories that evolve to the $\Ome=0$ axis represent shells where the CDE cedes its energy to the CDM and disappears. The corresponding shell, from this point on, will evolve as a pure-dust LTB scenario. On the other hand, the trajectories that evolve to the infinity or to the future attractor can expand forever or experiment a collapse, but in any case the CDE source will be present. In figure \ref{fig1}a, an example of PCR, PCA and its invariant line in the homogeneous projection is shown, together with some simulated possible trajectories.

Also, when considering $\alpha>0$, the future attractor $PCA$ is identical to that of the $\Lambda$--CDM model in the homogeneous projection. On the other hand, in the inhomogeneous projection, the future attractor is a different point depending on the choice of parameter $w$: when $w>-1$ the attractor is PC1 (a line parallel to the $\Dm$ axis) independently of the choice of $\alpha$; when $w=-1$ the attractor is the point PC3 (and lays over the line PC1) and its coordinates are $\alpha$ dependent; and, finally, for $w<-1$ the attractor is PC3 whose position depends on both $w$ and $\alpha$ while Pc1 is a saddle point. Panels a,b, and c of figure \ref{fig2} represent the attractor and two saddle points in three examples of each of the cases mentioned above, while panel d represents the rest of saddle points in a different inhomogeneous projection.

Given the conformal invariance of the LTB metric, it is possible to find the initial conditions for every shell for the scalars of the dynamical system from a given set of densities and curvature profiles at an initial instant. A shell can also experiment a bounce or expand/contract forever depending on its initial values of $\Omm$ and $\Ome$. After setting the free parameters and from an initial set of profiles, it is posible to solve the evolution equations. Several examples are studied in this work.

In figures \ref{fig3} and \ref{fig4}, we represent the evolution of an ever expanding LTB metric with both CDE and CDM sources evolving together. In figures  \ref{fig5} and \ref{fig6}, we represent the evolution of a configuration where the CDE disappears from all the shells and the metric evolves as a pure dust LTB metric. In figures  \ref{fig7} and \ref{fig8}, we plot the evolution of a mixed configuration where the CDE of the inner shells disappears while CDE of the outer shells. In this case the resulting pure dust inner shells collapse after some instants leading to a shell cross singularity. In figures  \ref{fig9} and \ref{fig10}, we plot the evolution of a mixed configuration as well. In this case the resulting pure dust inner shells expand forever. The difference between those two configuration lay in the curvature of the inner shells. Finally, in figures  \ref{fig11} and \ref{fig12}, the evolution of a configuration leading to a structure formation toy model is shown. In the latter configuration, the CDE and the CDM collapse together in the inner shells while the outer shells keep their expanding evolution to the future atractor. This configuration is similar to that reported in the $\Lambda-CDM$ case \cite{izsuss10}. The dust structure generated in this example cannot be reproduced in a pure dust scenario as the collapsing shells present positive curvature.

Summarizing, in this work we have demonstrated that the study of the LTB metrics with CDE and CDM as sources is interesting, useful and necessary. The LTB structure formation scenarios are possible and the existence of some unique structures rise, which should motivate us for future works. Apart from structures where both sources collapse, as the one presented in this work, we can think of a initial configuration where the inner shells evolve to a pure dust LTB and collapse, while the outer shells keep the mixture of CDE and CDM sources in an perpetual expansion. The time span where those collapses takes place are closely related with the length scale of the perturbations, and should be studied in order to clarify whether, or under which conditions, it has sense to use LTB CDE metrics to explain the observed structures in the Universe. The addition of a non coupled dust term can be made in order to represent the ordinary matter present in Cosmology. Other coupling terms can be taken into account as well, such as the coupling proportional to CDE density (where no restrictions to $\alpha<0$ have been found in Cosmology to our knowledge).

\section*{Acknowledgments}

The authors would like to thank Dr. Fernando Ongay for the useful lessons on invariant lines of dynamical systems. RAS acknowledges support from CONACYT project number CONACYT 239639 and PAPIIT-DGAPA RR107015.

\end{document}